\documentclass[journal]{IEEEtran}

\ifCLASSINFOpdf
\else
\fi

\usepackage{graphicx,float, placeins}
\usepackage{xspace}
\usepackage{subfigure}

\usepackage{graphicx}
\usepackage{url}
\usepackage{bm}
\usepackage[comma,numbers,square,sort&compress]{natbib}
\usepackage{hyperref} 
\usepackage{amsmath,amssymb,amsthm}  
\usepackage{paralist}

\usepackage{algorithm}
\usepackage{algorithmic}      
\usepackage{multirow}
\renewcommand{\algorithmicrequire}{\textbf{Initialization:}}
\renewcommand{\algorithmicensure}{\textbf{Iteration:}}
\renewcommand{\algorithmiclastcon}{\textbf{Output:}} 

\newtheorem{theorem}{Theorem}
\newtheorem{definition}{Definition}

\newtheorem{remark}{Remark}
\newcommand{\e}{\begin{equation}}
\newcommand{\ee}{\end{equation}}
\newcommand{\en}{\begin{equation*}}
\newcommand{\een}{\end{equation*}}
\newcommand{\eqn}{\begin{eqnarray}}
\newcommand{\eeqn}{\end{eqnarray}}
\newcommand{\bmat}{\begin{bmatrix}}
\newcommand{\emat}{\end{bmatrix}}
\newcommand{\BIT}{\begin{itemize}}
\newcommand{\EIT}{\end{itemize}}

\newcommand{\mrm}{\mathrm}

\newcommand{\Trans}{{\mathrm{T}}}  %

\newcommand{\xmath}[1]{\ensuremath{#1}\xspace}
\newcommand{\blmath}[1]{\bm{\mathrm{#1}}}

\newcommand{\uvb}{\blmath{b}}

\newcommand{\uvq}{\blmath{q}}

\newcommand{\uvs}{\blmath{s}}

\newcommand{\uvu}{\blmath{u}}
\newcommand{\uvv}{\blmath{v}}

\newcommand{\uvx}{\blmath{x}}
\newcommand{\uvy}{\blmath{y}}
\newcommand{\uvz}{\blmath{z}}

\newcommand{\umA}{\blmath{A}}
\newcommand{\umB}{\blmath{B}}
\newcommand{\umC}{\blmath{C}}
\newcommand{\umD}{\blmath{D}}

\newcommand{\umI}{\blmath{I}}

\newcommand{\umL}{\blmath{L}}

\newcommand{\umP}{\blmath{P}}

\newcommand{\umS}{\blmath{S}}

\newcommand{\umU}{\blmath{U}}

\newcommand{\umW}{\blmath{W}}
\newcommand{\umX}{\blmath{X}}
\newcommand{\umY}{\blmath{Y}}

\newcommand{\mPhi}{\bm \Phi}

\newcommand{\Phik}{\xmath{\mPhi_k}}

\newcounter{oursection}

\newcommand{\LfP} {\xmath{L_f^{\umP}}}
\newcommand{\deff} {\xmath{d_{\mathrm{eff}}}}
\usepackage{color}
\usepackage{cleveref}
\usepackage{silence}
\usepackage{soul}
\theoremstyle{plain} 

\newtheorem{thm}{Theorem}[section] 
\newcommand{\thistheoremname}{}
\newtheorem{genericthm}[thm]{\thistheoremname}

\newtheorem*{genericthm*}{\thistheoremname}
\newenvironment{namedthm*}[1]
  {\renewcommand{\thistheoremname}{#1}%
   \begin{genericthm*}}
  {\end{genericthm*}}

\usepackage{mathtools, cuted}
\usepackage{relsize}
\usepackage[export]{adjustbox}
\usepackage[dvipsnames]{xcolor}
\usepackage[skins]{tcolorbox}

\definecolor{darkred}{rgb}{0.6,0,0}
\definecolor{darkgreen}{rgb}{0,0.5,0}
\definecolor{darkblue}{rgb}{0,0,0.5}
\definecolor{goldenrod}{rgb}{0.85, 0.65, 0.13}
\definecolor{goldenbrown}{rgb}{0.6, 0.4, 0.08}
\usepackage{tikz,pgfplots}
\pgfplotsset{compat=1.5.1}
\usepgfplotslibrary{groupplots}

\usepackage{multicol}
\usepackage{slashbox}
\usepackage{diagbox}
\usepackage{enumitem,kantlipsum}
\usepackage{xspace}

\newcommand{\MRcb}[1]{{\color{black}#1}}
\newcommand{\MRcbT}[1]{{\color{black}#1}}

\newcommand{\NY}{Nystr\"{o}m\xspace}
\newcommand{\IRM}{IRM\xspace}
\newcommand{\RNP}{RNP\xspace}
\newcommand{\RNA}{RNA\xspace}
\newcommand{\lplq}{$\ell_p-\ell_q$\xspace}
\long\def\red#1{\bgroup\color{red}#1\egroup}

\newcommand{\stepsize}{\alpha}

\usepackage{listings}
\begin{document}

\title{
\MRcbT{Using Randomized \NY Preconditioners
\\
to Accelerate Variational Image Reconstruction}}
\author{ 
Tao Hong, \IEEEmembership{Member, IEEE},
Zhaoyi Xu, Jason Hu, \IEEEmembership{Student Member, IEEE} 
and Jeffrey A. Fessler, \IEEEmembership{Fellow, IEEE} \\


\thanks{T. Hong is with the Oden Institute for Computational Engineering and Sciences,
University of Texas at Austin, Austin, TX 78712, USA (Email: \texttt{tao.hong@austin.utexas.edu}). TH was with the Department of Radiology, University of Michigan, Ann Arbor, MI 48109, USA. TH was partially supported by NIH grant R01NS112233.}
\thanks{Z. Xu is with the Department of Mechanical Engineering,
University of Michigan, Ann Arbor, MI 48109,  
(Email: \texttt{zhaoyix@umich.edu}).}

\thanks{J. Hu and J. Fessler
are with the Department of Electrical and Computer Engineering,
University of Michigan, Ann Arbor, MI 48109, USA 
(Email: \texttt{jashu,fessler@umich.edu}). JF was supported in part by NIH grants
R01EB035618 and R21EB034344.}

}


\markboth{}{}

\maketitle
\begin{abstract}
Model-based iterative reconstruction plays a key role in solving inverse problems.
However, the \MRcb{associated minimization problems are generally large-scale, nonsmooth,
and sometimes even nonconvex}, which present challenges in designing efficient iterative solvers.
Preconditioning methods can significantly accelerate the convergence of iterative methods.
In some applications, computing preconditioners on-the-fly is beneficial.
Moreover, forward models in image reconstruction are typically represented as operators,
and the corresponding explicit matrices are often unavailable,
which brings additional challenges in designing preconditioners.
Therefore, for practical use, computing and applying preconditioners
should be computationally inexpensive.
This paper adapts the randomized \NY approximation
to compute effective preconditioners that accelerate image reconstruction
without requiring an explicit matrix for the forward model.
We leverage modern GPU computational platforms to compute the preconditioner on-the-fly.
Moreover, we propose efficient approaches for applying the preconditioners
to problems with classical nonsmooth regularizers,
\MRcbT{i.e., wavelet, total variation, and Hessian Schatten-norm}.
Our numerical results on image deblurring, super-resolution with impulsive noise,
and 2D computed tomography reconstruction
illustrate the efficiency and effectiveness of the proposed preconditioner.
\end{abstract}

\IEEEpeerreviewmaketitle

\section{Introduction}

\IEEEPARstart{T}{HE} task of image reconstruction
is to recover a clean image from degraded measurements.
Model-based iterative reconstruction recovers a clean image $\uvx$
by solving the following type of minimization problem:
\begin{equation}
    \label{eq:RecoProbCM}
    \min_{\uvx\in\mathbb R^N} f(\uvx) + \lambda \, g(\uvx),
\end{equation}
where $f(\uvx)$ denotes a data-fidelity
specifying the discrepancy between the model's predictions and the degraded measurements, $g(\uvx)$ denotes a regularizer imposing prior information on $\uvx$, and $\lambda>0$ is a trade-off parameter to balance $f(\uvx)$ and $g(\uvx)$.
Both $f(\uvx)$ and $g(\uvx)$ can be nonsmooth and nonconvex.
This paper focuses on linear inverse problems of the form
\begin{equation}
	\label{eq:lpReg}             
	\min_{\uvx\in\mathbb R^N} \frac{1}{p}\underbrace{\|\umA\uvx-\uvy\|_p^p}_{f(\uvx)}
 + \lambda \, g(\uvx),
\end{equation}
where $\umA\in\mathbb R^{M\times N}$ with $M\leq N$ denotes the forward operator, 
which maps the unknown image $\uvx$ to the measurements $\uvy$.
In this paper, we assume $p\in(0,2]$. 

In practice,
the matrix $\umA$ is under-determined or ill-conditioned
and the measurements are noisy,
so one uses a regularizer $g(\uvx)$ to stabilize the solution.
Typical \MRcb{regularizer} choices that are effective include
wavelet sparsity \cite{starck2001very},
total variation (TV) \cite{rudin1992nonlinear},
dictionary learning \cite{aharon2006k},
and low-rank models \cite{dong2014compressive}.
When using such hand-crafted regularizers,
we refer to \eqref{eq:RecoProbCM} as variational image reconstruction.
Over the past decades, deep learning (DL) \cite{lecun2015deep}
has received significant attention
in the field of image reconstruction.
Instead of using hand-crafted priors,
DL-based methods learn implicit priors from massive training data.
This includes end-to-end learning \cite{mccann2017convolutional}
and physics-informed deep unrolling \cite{chen2022learning,monga2021algorithm}.
Alternatively, one can use plug-and-play (PnP)~\cite{venkatakrishnan2013plug,sun2021scalable,hong2024provable}
or regularization by denoising (RED)~\cite{romano2017little,hong2020solving}
where one trains a deep denoiser \cite{zhang2017beyond} first
and then uses it as an implicit prior for general inverse problems.
Recently, building priors with generative models, such as diffusion models,
has also attracted much attention.
We refer the reader to \cite{daras2024survey} for a survey of those directions.

While deep learning methods often outperform variational methods in many applications,
variational image reconstruction
\cite{rodriguez2008efficient,guerquin2011fast,wang2017reweighted,fessler2024linear}
offers significant advantages in terms of interpretability, theoretical guarantees,
and robustness to noise or distribution shifts.
Variational methods do not require large training datasets
and provide predictable, stable behavior.
Moreover, they allow for easier customization of priors based on domain knowledge,
making them particularly suitable for applications
where interpretability and reliability are critical, such as medical imaging.
Recent work has also shown that novel neural network structures
can be derived from variational methods, yielding state-of-the-art performance
\cite{cherkaoui2020learning,zheng2022learn,gu2022revisiting,mai2022deep,soh2022variational}.
Therefore, this paper considers hand-crafted regularizers for \eqref{eq:lpReg}.
Specifically, we consider $g(\uvx)=\Psi(\umL\uvx)$
where $\umL\in\mathbb R^{L\times N}$ refers to the regularization operator
and $\Psi(\cdot)$ denotes an energy function.
A typical choice of $\Psi(\cdot)$ is a norm or quasi-norm. 



\subsection{\texorpdfstring{\lplq}{} Image Reconstruction}

For $p\in(0,2]$, we consider
$\Psi(\cdot)=\frac{1}{q}\|\cdot\|_q^q$ with $q\in(0,2]$
and \eqref{eq:lpReg} becomes
\begin{equation}
	\label{eq:lplqReg}
	\min_{\uvx\in\mathbb R^N} \underbrace{\frac{1}{p} \|\umA\uvx-\uvy\|_p^p}_{f(\uvx)}+ \lambda \underbrace{\frac{1}{q}\|\umL\uvx\|_q^q}_{g(\uvx)},
\end{equation}
which is known as the \lplq problem \cite{rodriguez2008efficient}.
For $p,q \leq 1$, both $f(\uvx)$ and $g(\uvx)$ are nonsmooth.
To address the nonsmooth challenge,
Clason et al. considered \eqref{eq:lplqReg} in its dual formulation \cite{clason2010duality},
where the objective function becomes differentiable,
at the cost of additional box and linear equality constraints.
For general $p,\,q\in(0,2)$,
Huang et al. \cite{huang2017majorization} first smoothed $f$ and $g$
and then applied the majorization-minimization for the smoothed problem,
leading to an inner least-squares problem.
\MRcbT{Then, \cite{huang2017majorization} used the Krylov subspace to represent the image, thereby reducing the dimension of the associated least-squares problem to that of the subspace. Consequently, it can be solved efficiently using a direct method.} Following the same strategy, Buccini et al. \cite{buccini2020modulus}
applied modulus-based iterative methods \cite{bai2017modulus}
to the smoothed problem to address the nonegativity constraint.
Alternatively, Chan et al. \cite{chan2014half} solved \eqref{eq:lplqReg}
with a half-quadratic algorithm.
Lanza et al. \cite{lanza2015generalized} proposed a \MRcb{generalized Krylov subspace (GKS)} method
to reduce computation of the method in \cite{chan2014half}
by assuming that the signal can be represented in a subspace. \MRcb{Gazzola et al. \cite{gazzola2021iteratively} presented two flexible Krylov subspace methods for sparse reconstruction.}

The solvers discussed above for \eqref{eq:lplqReg}
can be considered as iteratively reweighted methods (IRMs) \cite{rodriguez2008efficient}
because we eventually need to solve a linear equation
\eqref{eq:lplq:sub:updateX:linearEq}
with updating of the associated weightings at each iteration.
For completeness,
\Cref{sec:AdaptPreImageRes:sub:lplq}
describes one type \IRM for \eqref{eq:lplqReg}.
Typically the appealing conjugate gradient (CG) method is chosen
to address the associated linear equation.
However, the convergence of CG can be extremely slow for some problems,
especially for $p,\,q<1$, as the problem becomes nonconvex.
Preconditioning methods \cite{benzi2002preconditioning,bertaccini2018iterative}
are widely used to accelerate the convergence of CG.
Because the weightings $\umW_f^k$ and $\umW_g^f$ (defined in \eqref{eq:WfWg:IRM}) change at each iteration,
an effective preconditioner should change at each iteration,
so computing such a preconditioner must be computationally inexpensive. 

\subsection{\texorpdfstring{$\ell_2$}{} with Mixed Norm Image Reconstruction}
\label{sec:intro:sub:l2mixed}

When $p=2$, we consider $\Psi(\cdot)$ to be an $\ell_{1,\phi}$ mixed norm
(i.e., $\|\mathcal V\|_{1,\phi} = \sum_{l=1}^G \|\uvv_l\|_\phi$
with $\phi\geq 1$ and $\uvv_l$ denoting the $l$th group of
$\mathcal V=\{\uvv_1,\uvv_2,\cdots,\uvv_G\}$),
for which we can develop more efficient approaches than \IRM to solve \eqref{eq:lplqReg}
and a broader range of choices for $\umL$.
Moreover, we can also address convex constraints $\mathcal C$ easily.
Dividing the components of $\mathcal V$ into $G$ groups, \MRcb{ the mixed norm $\ell_{1,\phi}$ is defined by applying an $\ell_\phi$ norm to each group
and then applying an $\ell_1$ norm to the entire set of groups.} Specifically, we address
\begin{equation}
	\label{eq:lpl1qMixedNorm}
	\min_{\uvx\in\mathcal C} \underbrace{\frac{1}{2} \|\umA\uvx-\uvy\|_2^2}_{f(\uvx)}
 + \lambda \underbrace{\|\umL\uvx\|_{1,\phi}}_{g(\uvx)}.
\end{equation}
When $\umL$ is a first-order differential operator
and $\phi=1$ (respectively, $\phi=2$),
then \eqref{eq:lpl1qMixedNorm} represents the anisotropic (respectively, isotropic)
TV-based reconstruction.
When $\umL$ denotes the second-order finite-difference operator,
then \eqref{eq:lpl1qMixedNorm} becomes
the Hessian Schatten (HS) norm based reconstruction \cite{lefkimmiatis2013hessian}.
\Cref{app:sec:examplesMixedNorm} presents an explicit form for $\|\umL\uvx\|_{1,\phi}$
for various $\umL$ to clarify the definition of mixed norm.

Since \eqref{eq:lpl1qMixedNorm} is a standard composite minimization problem,
an appealing algorithm for solving it 
is the accelerated proximal gradient (APG) method \cite{beck2009fast}.
Alternatively, one could use a primal-dual method \cite{chambolle2011first,komodakis2015playing}
or ADMM \cite{boyd2011distributed} etc.
To accelerate the convergence of solving \eqref{eq:lpl1qMixedNorm},
one can adoapt a quasi-Newton proximal method
that uses second-order information
\cite{repetti2021variable,becker2019quasi,hong2024complex}.
However, the efficiency of using a quasi-Newton proximal method
depends highly on the accuracy of approximating the Hessian matrix of $f(\uvx)$
and the efficiency of solving the associated weighted proximal mapping
(WPM, defined in \eqref{eq:def:WeightedProximal}). \MRcbT{For $\phi=2$, $g(\uvx)$ reduces to the group sparsity case, for which one can adapt a flexible Krylov subspace method~\cite{chung2024flexible} to solve \eqref{eq:lpl1qMixedNorm}.}

\subsubsection{Previous Work on Preconditioning Methods in Image Reconstruction}
\label{sec:intro:PreviousPreMethods}

Preconditioning methods are used widely
to accelerate the convergence of iterative methods \cite{chen2005matrix}
in scientific computing
and have been extended to accelerate the convergence of image reconstruction.
Clinthorne et al. \cite{clinthorne1993preconditioning}
applied a spatially invariant preconditioner
to accelerate the convergence of reconstructing
single-photon emission computed tomography images with quadratic $g(\uvx)$.
Fessler et al. \cite{fessler1999conjugate} proposed a shift-variant preconditioner
for accelerating positron emission tomography (PET) reconstruction
with non-quadratic smooth $g(\uvx)$.
Lin et al. \cite{fu2012preliminary} derived cosine transform
and incomplete factorization based preconditioners 
for image super-resolution task with smooth $g(\uvx)$.
Pelletier et al. \cite{pelletier2011preconditioning}
developed a block circulant with circulant block preconditioner
for edge-preserving image super-resolution.
The authors in \cite{tsai2017fast,ong2019accelerating}
considered diagonal preconditioners for accelerating PET
and magnetic resonance imaging reconstruction.
To reduce the computation of computing WPM,
\cite{savanier2022unmatched} discussed an unmatched preconditioner method for image reconstruction
by computing WPM approximately.
To address the nonsmoothness of $g(\uvx)$,
Koolstra et al. \cite{koolstra2019accelerating} adapted the preconditioner to ADMM
to accelerate the convergence of solving the associated least-squares problem.
By leveraging  deep learning,
\cite{koolstra2022learning} proposed an approach
to learn preconditioners to accelerate the convergence of image reconstruction. 

\subsection{Our Contribution}

The  preconditioners discussed in \Cref{sec:intro:PreviousPreMethods}
were either used for smooth $g(\uvx)$
\cite{clinthorne1993preconditioning,fessler1999conjugate,fu2012preliminary,tsai2017fast,ong2019accelerating}
or applied to a nonsmooth $g(\uvx)$ after smoothing
\cite{pelletier2011preconditioning}.
Moreover, the previous methods require domain knowledge
because they were developed for specific applications,
making them difficult to generalize to other applications.
Although the preconditioners proposed in
\cite{koolstra2019accelerating,koolstra2022learning}
can be used to address nonsmooth $g(\uvx)$,
the convergence rate of ADMM is only $\mathcal O(1/k)$.
To address these challenges,
we adapt the randomized \NY approximation (\RNA) \MRcb{\cite{martinsson2020randomized,frangella2023randomized}} to design an effective preconditioner,
called randomized \NY preconditioner (\RNP),
to accelerate variational image reconstruction.
Our main contributions are summarized as follows.
\begin{itemize}   

\item
To the best of our knowledge,
this is the first work to adapt \RNP to accelerate image reconstruction with nonsmooth regularizers.
Computing \RNP requires only the evaluation of $\umA\uvx$
and $\umA^\Trans \uvx$ ($\Trans$ denotes the transpose operator),
and does not require to access the explicit matrix $\umA$
or knowledge of any specific structure of $\umA$.
Therefore, \RNP can generalize to a wide range of applications.

\item
By leveraging modern high-performance GPU computational platforms,
we compute \RNP on-the-fly, allowing it to accelerate \IRM for \eqref{eq:lplqReg}
in a computationally efficient way.
Our numerical experiments on image deblurring and super-resolution with impulsive noise
showed that, by using \RNP, we can reduce wall time by more than $90$\%.
To the best of our knowledge, this is the first work to consider the use of hardware platform
in conjunction with computing preconditioners on-the-fly in image reconstruction.

\item
We show how \RNP can be adapted to directly address nonsmooth $g(\uvx)$.
Specifically, we propose efficient approaches for integrating \RNP with APG
to solve \eqref{eq:lpl1qMixedNorm}.
Our experiments on computed tomography (CT) reconstruction
with wavelet, TV, and HS norm regularizers
demonstrate that \RNP can significantly accelerate the convergence of APG. 
\end{itemize}

\subsection{Preliminaries and Roadmap}

This subsection introduces some definitions and a theorem
that will be used frequently in the subsequent discussion,
and then presents a roadmap of this paper.

\begin{definition}[Weighted proximal mapping (WPM)]
    Given a proper closed convex function $h(\uvx)$,
    a symmetric positive-definite matrix
    $\umW\succ0\in\mathbb{R}^{N\times N}$, and $\lambda>0$,
    the WPM associated with $h$ is defined as
\begin{equation}
\mrm{prox}_{\lambda  h}^{\umW}(\uvx)=\arg\min\limits_{\uvu\in\mathbb R^N}\left(\lambda  h(\uvu)+\frac{1}{2}\|\uvu-\uvx\|^2_{\umW}\right),
	\label{eq:def:WeightedProximal}
\end{equation} 
where $\|\uvx\|_{\umW}\triangleq{\sqrt{\uvx^\Trans\umW\uvx}}$ denotes the $\umW$-norm.
\end{definition}
For $\umW=\umI$, \eqref{eq:def:WeightedProximal} simplifies
to the proximal mapping:
\begin{equation}
	\label{eq:proximal}
	\mathrm{prox}_{\lambda h}(\uvx)=\arg\min_{\uvu\in\mathbb R^N}\lambda h(\uvu)+ \frac{1}{2}\|\uvu-\uvx\|_2^2.
\end{equation}
If $h(\uvx)$ is a characteristic function
$$
\delta_{\mathcal C}(\uvx) = \left \{ \begin{array}{rl}
     0 &  \uvx\in\mathcal C\\
     \infty & \uvx\notin\mathcal C,
\end{array}\right.
$$
then the proximal mapping becomes
projection onto the set $\mathcal C$.

In general, computing a WPM
\eqref{eq:def:WeightedProximal}
is computationally expensive.
However, when $\umW$ is represented by a diagonal matrix
plus a rank-$r$ correction,
then we can use \Cref{them:structuredWPM:evaluation} to compute WPM relatively cheaply.
\begin{theorem}[\cite{becker2019quasi}, Theorem 3.4]
\label{them:structuredWPM:evaluation}
Let $\umW=\umD\pm {\umU}{\umU}^\mathsf{T}$, $\umW\succ 0\in\mathbb R^{N\times N}$,
and $\umU \in\mathbb R^{N\times r}$.
Then, it holds that 
\begin{equation}
\label{eq:WPM:structure:equiv}
	\mrm{prox}_{\lambda h}^{\umW}(\uvx)
    = \mrm{prox}_{\lambda h}^{\umD}(\uvx\mp{\umD}^{-1}{\umU}\bm \gamma^*),
\end{equation}
where $\bm\gamma^*\in\mathbb R^r$
is the unique solution of 
\begin{equation}
\label{eq:WPM:structure:equiv:nonlinearSystems}
	{\umU}^\mathsf{T}\left(\uvx-\mrm{prox}_{\lambda h}^{\umD}
    \left(\uvx\mp{\umD}^{-1} {\umU}\bm\gamma\right)\right) + \bm\gamma = \bm 0.
\end{equation}
\end{theorem}
\noindent
Semi-smooth Newton methods can solve
\eqref{eq:WPM:structure:equiv:nonlinearSystems}
efficiently \cite{eberhard1999survey,becker2019quasi,hong2023mini}.

The rest of this paper is organized as follows.
\Cref{sec:RNPandRelated} describes how RNA can be a preconditioner for linear equations.
\Cref{sec:AdaptPreImageRes} shows how to adapt \RNP to accelerate \IRM and APG
for solving \eqref{eq:lplqReg} and \eqref{eq:lpl1qMixedNorm}, respectively.
\Cref{sec:NumericalExp} presents
numerical experiments on image deblurring and super-resolution with impulsive noise
and computed tomography reconstruction
to demonstrate the effectiveness and efficiency of using \RNP for acceleration.
\Cref{sec:Conclusion} \MRcb{presents} the conclusion and future work. \MRcbT{Code to reproduce the results in the paper
is available at
\url{https://github.com/hongtao-argmin/RNP-AccImageRecon}.}

\section{Randomized \NY Approximation}
\label{sec:RNPandRelated}

The \NY method provides a way to build a low-rank approximation
for a symmetric positive semidefinite (SPSD) matrix $\bm\Phi\succeq 0\in\mathbb R^{N\times N}$.
Let $\bm\Omega\in\mathbb R^{N\times K}$ be \MRcb{an i.i.d. Gaussian} random test matrix
\cite{martinsson2020randomized,frangella2023randomized} with sketch size $K\geq 1$.
The \NY method uses the following formula as a low-rank approximation of $\bm\Phi$,
called the \NY approximation
\begin{equation}
	\label{eq:NYApprox}
	\bm\Phi\left<\bm\Omega\right>
 = (\bm\Phi \bm\Omega)(\bm\Omega^\Trans \bm\Phi \bm\Omega)^{\dagger}(\bm\Phi \bm\Omega)^\Trans,
\end{equation} 
where $^{\dagger}$ denotes the pseudo-inverse.
Clearly $\bm\Phi\left<\bm\Omega\right>$ is also a SPSD matrix
and its rank is at most $K$ \cite[Lemma A.1]{zhao2022nysadmm}.
Because a direct implementation of \eqref{eq:NYApprox} is numerically unstable, 
we adopt the robust scheme proposed in \cite{martinsson2020randomized}, 
which is summarized in \Cref{alg:NyApprox}. 
This algorithm computes $\umU$ and $\hat{\umS}$ such that 
$\bm\Phi\langle \bm\Omega \rangle = \umU \hat{\umS}\umU^\Trans$, 
although this product is not formed explicitly in practice.

Many scientific computing applications \cite{saad2003iterative}
require solving the following linear equation: 
\begin{equation}	
\label{eq:linearEq}
(\bm\Phi+\mu\umI)\uvx=\uvb,~\mu>0.
\end{equation} 
Using \RNA to formulate a preconditioner to address \eqref{eq:linearEq},
the associated formulations of \RNP and its inverse \cite{frangella2023randomized} are as follows:
\begin{equation}
\label{eq:RYPreconditioner}
	\begin{array}{ll}
	\umP \quad = & \frac{1}{{\hat s}_K+\mu}\umU(\hat{\umS}+\mu\umI)\umU^\Trans
 + (\umI-\umU\umU^\Trans), \\
	\umP^{-1} = & ({\hat s}_K+\mu) \umU(\hat{\umS}+\mu\umI)^{-1}\umU^\Trans+(\umI-\umU\umU^\Trans),
	\end{array}
\end{equation}
where ${\hat s}_K$ is the $K$th eigenvalue of $\bm\Phi\left<\bm\Omega\right>$.
$\umU$ and $\hat{\umS}$ are defined in \Cref{alg:NyApprox}.
Let $\deff(\mu)=\mathrm{tr}\left(\bm\Phi(\bm\Phi+\mu\umI)^{-1}\right)$ be the effective dimension.
Then, we have the following theorem to describe the condition number of $(\bm\Phi+\mu\umI)$ 
after using \RNP, which reveals the effectiveness of \RNP.
\begin{theorem}[\cite{frangella2023randomized}, Theorem 5.1]
\label{them:NYPre:CN}
Suppose we build the \NY preconditioner through \eqref{eq:RYPreconditioner}
with sketch size $K=2\lceil 1.5 \deff(\mu)+1 \rceil$.
Then using $\umP$ as the preconditioner for $(\bm\Phi+\mu\umI)$
results in the expectation of condition number bound
$$
\mathbb E[\kappa (\umP^{-1/2}(\bm\Phi+\mu\umI)\umP^{-1/2})]<28,
$$
where $\kappa(\cdot)$ denotes the condition number function.
\end{theorem}
\Cref{them:NYPre:CN} shows that, by choosing an appropriate sketch size $K$,
\RNP can significantly reduce the condition number of $(\bm\Phi+\mu\umI)$.
Thus, one can significantly accelerate the convergence of iterative methods
for solving \eqref{eq:linearEq} with \RNP.
However, in practice, computing \deff is computationally expensive.
Moreover, the value of $K=2\lceil 1.5 \deff(\mu)+1 \rceil$ can be very large,
which will significantly \MRcb{increase} the computation of computing and applying \RNP.
In our experimental settings, we found that using a moderate value for $K$ is sufficient.
We refer the reader to \cite{frangella2023randomized,martinsson2020randomized},
where the authors discussed more theoretical properties of using \RNP
for considering \eqref{eq:linearEq}.
The main purpose of this paper is to study how to adapt \RNP
to accelerate variational image reconstruction,
focusing on evaluating its effectiveness and efficiency.
We also refer the reader to \cite{frangella2023randomized,zhao2022nysadmm,chu2024randomized}
and the references therein that discuss using \RNP for other applications.
\MRcb{Although \Cref{them:NYPre:CN} presents the split preconditioner $\umP^{-1/2}$,
we use $\umP^{-1}$ in practice for simplicity.}

\begin{algorithm}[t]       
\caption{\NY approximation.}           
\label{alg:NyApprox}                
\begin{algorithmic}[1]
\REQUIRE Sketch size $K$, machine accuracy $\epsilon$,
and symmetric positive semidefinite operator
$\bm\Phi\in\mathbb R^{N\times N}$.
\lastcon       
$\umU\in\mathbb R^{N\times K},~\hat{\umS}\in\mathbb R^{K\times K}$.
\STATE Generate a random matrix $\bm\Omega\in \mathbb R^{N\times K}$
\STATE $\umY=\bm\Phi\bm\Omega \in \mathbb R^{N \times K}$ \label{alg:NyApprox:phix}
\STATE $v=\epsilon\|\bm \Omega\|_F$
\STATE $\umY_v = \umY + v \bm\Omega$
\STATE $\umC=\mathrm{chol}({\mathbf \Omega}^\Trans\umY_v) \in \mathbb R^{K \times K}$
\% Cholesky decomp. \label{alg:NyApprox:CD}
\STATE $\umB=\umY_v\umC^{-\Trans} \in \mathbb R^{N \times K} $ \label{alg:NyApprox:GetB}
\STATE $[\umU,\umS,\sim]=\mathrm{SVD}(\umB)$ \label{alg:NyApprox:SVD}
\STATE $\hat{\umS}=\max\{\bm 0,\umS^2-v\umI\}$
\end{algorithmic}
\end{algorithm}



\section{On Adapting \NY Preconditioners in Image Reconstruction}
\label{sec:AdaptPreImageRes}

\MRcb{For impulsive noise, a natural data-fidelity term is the $\ell_p$ quasi-norm with $p \leq 1$, allowing image reconstruction by solving the $\ell_p-\ell_q$ problem. When the noise is assumed to be i.i.d. Gaussian, we use the $\ell_2$ norm as the metric in the data-fidelity term, resulting in the \MRcbT{$\ell_2-\ell_{1,\phi}$} problem.
This section presents two baseline algorithms
for solving $\ell_p-\ell_q$ and \MRcbT{$\ell_2-\ell_{1,\phi}$},
and then study how they can be accelerated using \RNP.}

Following we first describe \IRM in \Cref{sec:AdaptPreImageRes:sub:lplq}
and then show how \RNP can be adapted to accelerate \IRM for solving \eqref{eq:lplqReg} efficiently. Because the weightings in \IRM change at each iteration,
computing \RNP must be computationally inexpensive.
We leverage the modern GPU computational platform and the PyTorch library
to present an on-the-fly implementation for computing \RNP at the end of this section.
Additionally, we show how \RNP can be incorporated with APG \cite{parikh2014proximal}
to efficiently handle \eqref{eq:lpl1qMixedNorm}.
Furthermore, we propose efficient approaches to solve the associated WPM
to reduce the computational cost involved in applying \RNP.
We consider separately the \lplq and $\ell_2-\ell_{1,\phi}$ cases.

\subsection{\texorpdfstring{\lplq}{} Image Reconstruction Acceleration}
\label{sec:AdaptPreImageRes:sub:lplq}

Following \cite[Lemma 1]{chan2014half},
we represent the $p$th power of $r\in\mathbb R\setminus \{0\}$ with $0<p<2$ as
\begin{equation}
	\label{eq:halfQuad}
	|r|^p =\min_{\beta >0} \left\{ \beta r^2+\frac{1}{b_p \beta ^{a_p}} \right\},
\end{equation}
where $a_p=\frac{p}{2-p}$ and \MRcb{$b_p=\frac{2^{\frac{2}{2-p}}}{(2-p)\cdot p^{\frac{p}{2-p}}}$} are two positive scalars.
The minimizer of \eqref{eq:halfQuad} is 
\begin{equation}
\label{eq:halfQuadMin}
	\beta ^*=\frac{p}{2}|r|^{p-2}.
\end{equation} 
For the singular case, $r=0$,
we simply set $|r|_\epsilon=\sqrt{r^2+\epsilon}$ with $\epsilon>0$,
following \cite{chan2014half}.

Representing \eqref{eq:lplqReg} in a component-wise form, we get 
\begin{equation}
	\label{eq:lplq:componentwise}
	\min_{\uvx} \frac{1}{p}\sum_{m=1}^M |(\umA\uvx)_m - y_m|^p
 + \frac{\lambda}{q} \sum_{l=1}^L\sum_{\zeta=1}^\pi |(\umL\uvx)_l^\zeta|^q,
\end{equation}
where $(\umA\uvx)_m$ and $y_m$ represent the $m$th element 
of $\umA\uvx$, $\uvy$, respectively. $(\umL\uvx)_l^\zeta$ denotes the $\zeta$th element in the $l$th group of $\umL\uvx$. Here, we assume each group has the same number of elements and consider cases where each group represents a scalar or vector. \MRcb{Here, $\pi$ denotes the number of elements in each group.} 
Using \eqref{eq:halfQuad}, we rewrite \eqref{eq:lplq:componentwise} as
\begin{equation}
	\label{eq:lplq:constrained}
	\begin{array}{ll}
		\min\limits_{\substack{\uvx, \\ \uvv>\bm 0, \\ \uvz>\bm 0}} &
  F(\uvx,\uvv,\uvz)\equiv\frac{1}{p}\sum_{m=1}^M\Big(v_m |(\umA\uvx)_m - y_m|^2+
  \\[6pt]
  &\quad \quad \frac{1}{b_p v_m^{a_p}} \Big)+
  \frac{\lambda}{q} \sum_{l=1}^L \sum_\zeta \Big(z_l^\zeta |(\umL\uvx)_l^\zeta|^2  +\frac{1}{b_q (z_l^\zeta)^{a_q}}\Big),
	\end{array}
\end{equation}
where $v_m$ and $z_l^\zeta$ denote the $m$ and $(l-1)\pi+\zeta$th element of $\uvv$ and $\uvz$, respectively.
By using an alternating minimization algorithm,
\IRM \cite{rodriguez2008efficient} solves \eqref{eq:lplq:constrained}
by successively updating $\uvv$, $\uvz$, and $\uvx$.
\Cref{alg:alternating:lplq} summarizes the \IRM steps.
Using \eqref{eq:halfQuadMin},
we derive closed-form solutions for $\uvv^k$ and $\uvz^k$
at steps \ref{alg:alternating:lplq:updateV} and \ref{alg:alternating:lplq:updateZ}
of \Cref{alg:alternating:lplq}, i.e.,
\begin{equation}
	\label{eq:v:z:updating}
			\uvv^k=\frac{p}{2}|(\umA\uvx^k)-\uvy|^{p-2}~~\text{and}~~ 
		\uvz^k=\frac{q}{2}|\umL\uvx^k|^{q-2},
\end{equation}
where the powers $(p-2)$ and $(q-2)$
are element-wise.

By defining 
\begin{equation}
\label{eq:WfWg:IRM}
    \umW_f^k=\mathrm{diag}\left(\frac{2}{p}\uvv^k\right)
~~\text{and}~~
\umW_g^k=\mathrm{diag}\left(\frac{2}{q}\uvz^k\right),
\end{equation}
the minimization problem at step \ref{alg:alternating:lplq:updateX}
of \Cref{alg:alternating:lplq}
becomes
\begin{equation}
	\label{eq:lplq:sub:updateX}
	\uvx^k=\arg\min_{\uvx\in\mathbb R^N} \|\umA\uvx-\uvy\|^2_{\umW_f^k}
 +\lambda \|\umL\uvx\|_{\umW_g^k}^2.
\end{equation}
The first-order optimality condition of \eqref{eq:lplq:sub:updateX} is given by 
\begin{equation}
	\label{eq:lplq:sub:updateX:linearEq}
\Phik \uvx=\umA^\Trans\umW_f^k\uvy
, \quad
\Phik = \umA^\Trans \umW_f^k \umA +\lambda \umL^\Trans \umW_g^k \umL.
\end{equation}
Since the Hessian \Phik
is a SPSD matrix,
we use CG to solve \eqref{eq:lplq:sub:updateX:linearEq}.
However, solving \eqref{eq:lplq:sub:updateX:linearEq} is time-consuming
when \Phik
is ill-conditioned.
\MRcbT{We} use \eqref{eq:RYPreconditioner} to construct a preconditioner
for \Phik
and apply the preconditioned CG (PCG) method to solve \eqref{eq:lplq:sub:updateX:linearEq}.
Because $\umW_f^k$ and $\umW_g^k$ change at each iteration\footnote{\MRcb{Note that $\umW_f^k$ and $\umW_g^k$ are changed at each outer iteration but remain unchanged within CG iterations when solving \eqref{eq:lplq:sub:updateX:linearEq}.}},
computing $\umP$ in \eqref{eq:RYPreconditioner} must be computationally inexpensive.
\Cref{sec:OTF:RNP} discusses an on-the-fly implementation of \RNP to address this computational issue.
If $p,q=2$, then \eqref{eq:lplqReg} is a quadratic minimization problem,
allowing us to directly apply PCG with \RNP.
When $p=2$ and $q\neq 1$, \IRM\ is still applicable, but APG is more efficient, as discussed in the following subsection.

\begin{algorithm}[t]       
\caption{Alternating minimization method for solving \eqref{eq:lplq:constrained}.}           
\label{alg:alternating:lplq}                
\begin{algorithmic}[1]
\REQUIRE Initialization $\uvx^0$, tolerance $\varepsilon$,
and maximum number of iterations Max\_Iter.
\lastcon  $\uvx^*$.
 \FOR{$k=1,2,\cdots,$ Max\_Iter} 
 \STATE \MRcb{$\uvv^k=\arg\min_{\uvv>0}F(\uvx^{k-1},\uvv,\uvz^{k-1})$}\label{alg:alternating:lplq:updateV}
 \STATE \MRcb{$\uvz^k=\arg\min_{\uvz>0}F(\uvx^{k-1},\uvv^k,\uvz)$}\label{alg:alternating:lplq:updateZ}
 \STATE $\uvx^k=\arg\min_{\uvx} F(\uvx,\uvv^k,\uvz^k)$\label{alg:alternating:lplq:updateX}
 \IF {$\frac{\|\uvx^k-\uvx^{k-1}\|}{\|\uvx^k\|}<\varepsilon$}
\STATE break
\ENDIF
 \ENDFOR
 \STATE $\uvx^*=\uvx^{k}$
\end{algorithmic}
\end{algorithm}


\subsection{\texorpdfstring{$\ell_2-\ell_{1,\phi}$}{} Image Reconstruction Acceleration}
\label{sec:AdaptPreImageRes:sub:l2l1} 

APG, which only uses first-order information
and achieves the optimal worst-case convergence rate of $\mathcal O(1/k^2)$ \cite{beck2009fastTV},
is an appealing algorithm for addressing \eqref{eq:lpl1qMixedNorm}.
To further accelerate the convergence of APG, we adapt \RNP to APG,
resulting in the weighted accelerated proximal gradient (WAPG) algorithm
that is summarized in \Cref{alg:WeightedacceleratedProximal}.
For any $\uvx_1,\uvx_2\in\mathbb R^N$, the Lipschitz constant \LfP of $f$
with respect to the $\umP$-norm satisfies the following inequality
\begin{equation}
	\label{eq:LipPNorm}
	\|\umP^{-1}(\nabla f(\uvx_1)-\nabla f(\uvx_2))\|_{\umP}\leq \LfP\|\uvx_1-\uvx_2\|_{\umP}.
\end{equation}
If $f,g$ are convex,
the worst-case convergence rate
of the cost function sequence
generated by \Cref{alg:WeightedacceleratedProximal}
is
$\frac{2\LfP\|\uvx^1-\uvx^*\|_{\umP}^2}{(k+1)^2}$
\cite[Chapter 10.7.5]{beck2017first}
where $\uvx^1$ and $\uvx^*$ are the initial and optimal values, respectively.
Clearly, for $\umP=\umI$, WAPG simplifies to APG.
By choosing a suitable $\umP$, one can expect a lower bound than $\umP=\umI$
yielding faster convergence.
Indeed, our experimental results in CT reconstruction
illustrate the faster convergence of WAPG with \RNP,
\MRcb{i.e., experiments in \Cref{sec:numericalExp:sub:CTReco,sec:numericalExp:sub:RealCTReco}}. 

For noninvertible $\umL$,
the associated WPM in \Cref{alg:WeightedacceleratedProximal}
generally does not have a closed-form solution, even with $\umP=\umI$,
so WPM requires iterative methods.
Following \cite{hong2023mini}, we solve WPM using its dual problem
to address the nonsmoothness of $g(\uvx)$.
Moreover, compared with computing WPM with $\umP=\umI$,
we show that solving WPM with \RNP
negligibly increases computation.
Whereas \cite{hong2023mini} only discussed constrained TV regularization,
we generalize their derivation to an abstract $\umL$ with mixed norm in this work.

Computing WPM at step \ref{alg:WeightedacceleratedProximal:WPM}
of \Cref{alg:WeightedacceleratedProximal} at $k$th iteration is equivalent to solving
\begin{equation}
\label{eq:WPM:Primal}
    \min_{\uvx\in\mathcal C}\frac{1}{2}\|\uvx-\uvs^k\|_{\umP}^2 + \bar{\lambda} \|\umL\uvx\|_{1,\phi},
\end{equation}
where $\uvs^k=\uvu^{k-1}-\alpha \umP^{-1}\nabla f(\uvu^{k-1})$ and $\bar{\lambda}=\alpha\lambda$.
\MRcbT{By using the dual representation
\begin{equation}
\label{eq:dualMixedNormRep}
  \|\umL\uvx\|_{1,\phi}
  = \max_{\mathcal Q \in B_{\infty,\psi}}
  \left\langle \mathcal Q, \umL\uvx \right\rangle,
\end{equation}
where $B_{\infty,\psi}$ denotes the $\ell_{\infty,\psi}$ mixed unit-norm ball, i.e.,
\begin{equation}
    \label{eq:mixedUnitNormBall}    
     B_{\infty,\psi}= \Big\{ \mathcal Q=\{\uvq_1,\uvq_2,\ldots,\uvq_G\}: \|\uvq_l\|_\psi \leq 1,~\forall l\Big\},
\end{equation}
and $\psi \geq 1$ satisfies $\tfrac{1}{\phi} + \tfrac{1}{\psi} = 1$,
we rewrite \eqref{eq:WPM:Primal} as
\begin{equation}
\label{eq:dualMixedNormDualRep}
  \min_{\uvx \in \mathcal C}
  \max_{\mathcal Q \in B_{\infty,\psi}}
  \frac{1}{2}\|\uvx - \uvs^k\|_{\umP}^2
  + \bar{\lambda}\,\left\langle \umL^\mathcal T \mathcal Q, \uvx \right\rangle,
\end{equation}
where $\umL^\mathcal T$ denotes the adjoint of $\umL$. Denote by $\mathcal V = \umL\uvx$ with $\mathcal V = \{\uvv_1,\uvv_2,\ldots,\uvv_G\}$.
The inner product in \eqref{eq:dualMixedNormRep}
is defined as $\sum_{l=1}^G \langle \uvq_l,\uvv_l\rangle$.}
Reorganizing \eqref{eq:dualMixedNormDualRep}, we get
\begin{equation}
\label{eq:dualMixedNormDualRep:Quad}
\begin{array}{rl}
     \max\limits_{\mathcal Q\in B_{\infty,\psi}} \min\limits_{\uvx\in\mathcal C}
     & \|\uvx-(\uvs^k-\bar{\lambda} \umP^{-1}\umL^\mathcal T\mathcal Q)\|_{\umP}^2 \\
     & \quad -\|\uvs^k-\bar{\lambda} \umP^{-1}\umL^\mathcal T\mathcal Q\|_{\umP}^2. 
\end{array}
\end{equation}
Because $\uvx$ appears only in the first term of \eqref{eq:dualMixedNormDualRep:Quad},
the minimizer of $\uvx$ is exactly the WPM associated with $\delta_\mathcal C$:
\begin{equation}
    \label{eq:OptimalxWPM}
    \bar{\uvx}^* = \mathrm{prox}^{\umP}_{\delta_\mathcal C}
    (\uvs^k-\bar{\lambda} \umP^{-1}\umL^\mathcal T\mathcal Q).
\end{equation}
Substituting \eqref{eq:OptimalxWPM} into \eqref{eq:dualMixedNormDualRep:Quad}, we get
\begin{equation}
\label{eq:dualMixedNormDualRep:QuadOnlyV}
\begin{array}{rl}
     \mathcal Q^*=\arg\min\limits_{\mathcal Q\in B_{\infty,\psi}}
     & \|\uvs^k-\bar{\lambda}\umP^{-1}\umL^\mathcal T\mathcal Q\|_{\umP}^2\\
     &\,\,-\Big\|\mathrm{prox}^{\umP}_{\delta_\mathcal C}
     (\uvs^k-\bar{\lambda}\umP^{-1}\umL^\mathcal T\mathcal Q) \\
     &\quad \quad \quad -(\uvs^k-\bar{\lambda} \umP^{-1}\umL^\mathcal T\mathcal Q)\Big\|_{\umP}^2.
\end{array}
\end{equation} 
Following \cite{hong2023mini}, the gradient and the corresponding Lipschitz constant of the objective function in \eqref{eq:dualMixedNormDualRep:QuadOnlyV} are
\begin{equation}
    \label{eq:dualMixedNormDualRep:QuadOnlyVgradient}
    -2\bar{\lambda} \umL\,\mathrm{prox}_{\delta_\mathcal C}^{\umP}
    (\uvs^k-\bar{\lambda}\umP^{-1}\umL^\mathcal T\mathcal Q),
\end{equation}
and $2\sigma^{\umP}_{\mathrm{min}}\bar{\lambda}^2\|\umL\|^2$, respectively.
$\sigma^{\umP}_{\mathrm{min}}$ is the smallest eigenvalue of $\umP$.
Since the objective function in \eqref{eq:dualMixedNormDualRep:QuadOnlyV} is differentiable,
we apply APG to solve \eqref{eq:dualMixedNormDualRep:QuadOnlyV}.
After getting $\mathcal Q^*$, we compute
\begin{equation}
\label{eq:OptimalxWPM:k}
\uvx^{k+1} = \mathrm{prox}^{\umP}_{\delta_\mathcal C}
(\uvs^k-\bar{\lambda} \umP^{-1}\umL^\mathcal T\mathcal Q^*).    
\end{equation}

Both \eqref{eq:dualMixedNormDualRep:QuadOnlyVgradient} and \eqref{eq:OptimalxWPM:k}
require computing
$\mathrm{prox}^{\umP}_{\delta_\mathcal C}(\cdot)$ and $\umP^{-1}$,
which can be computationally expensive for a general $\umP$.
A natural way to choose $\umP$ is to approximate
the Hessian $\umA^\Trans\umA$.
By using \RNP, $\umP$ (see \eqref{eq:RYPreconditioner})
can be rewritten as
$\umI+\bar{\umU}\bar{\umU}^\Trans$
with $\bar{\umU}=\umU\sqrt{\frac{1}{\hat{s}_K+\mu}(\hat {\umS}+\mu\umI)-\umI}$,
where the square root here is applied \MRcbT{elementwise}.
Clearly, $\umP$ has the same structure as $\umW$ proposed in \Cref{them:structuredWPM:evaluation},
so getting
$\mathrm{prox}^{\umP}_{\delta_\mathcal C}(\cdot)$
reduces to addressing a nonlinear equation \eqref{eq:WPM:structure:equiv:nonlinearSystems}.
Thus, the total computation for $\mathrm{prox}^{\umP}_{\delta_\mathcal C}(\cdot)$
can be dramatically reduced,
as demonstrated by our numerical experiments in \Cref{sec:numericalExp:sub:CTReco}.
Note that ${\hat s}_K$ is the $K$th eigenvalue of $\bm\Phi\left<\bm\Omega\right>$,
so the last column in $\bar{\umU}$ will be a zero vector.
In practice, we found that using $\sqrt{{\hat s}_K}$
instead of ${\hat s}_K$ resulted in faster convergence.
So, in this paper, we use ${\hat s}_K\leftarrow \sqrt{{\hat s}_K}$ in \eqref{eq:RYPreconditioner}.
Since $\umU$ is an orthogonal matrix,
we have $\sigma^{\umP}_{\mathrm{min}}=1$,
so the Lipschitz constant of the cost function
in \eqref{eq:dualMixedNormDualRep:QuadOnlyV} becomes $2\bar{\lambda}^2\|\umL\|^2$.



When $\umL$ is invertible and $\mathcal C=\mathbb R^N$,
we can rewrite \eqref{eq:lpl1qMixedNorm} as
\begin{equation}
	\label{eq:lplqMixedNorm:invertibleL}
	\min_{\bar{\uvx}\in\mathbb R^{\bar{N}}} \frac{1}{2}\|\umA{\umL}^{-1} \bar{\uvx}-\uvy\|_2^2
    +\lambda \|\bar{\uvx}\|_{1,\phi},
\end{equation}
where $\uvx=\umL^{-1}\bar{\uvx}$.
If $\phi=1$ and $\umL$ refers to a wavelet transform,
\eqref{eq:lplqMixedNorm:invertibleL} represents wavelet-regularized reconstruction.
In such cases,
we do not need to compute WPM iteratively, saving further computation.
By setting $\umP$ to approximate $(\umL^{-1})^\mathcal T \umA^\Trans\umA \umL^{-1}$ with \RNP,
we only need to solve a small  nonlinear equation to compute the associated WPM.
We did not observe any degradation in our subsequent CT reconstruction
when using \eqref{eq:lplqMixedNorm:invertibleL},
even though it does not account for the constraint $\mathcal C$.



\begin{algorithm}[t]       
\caption{Weighted accelerated proximal gradient (WAPG) method.}
\label{alg:WeightedacceleratedProximal}                
\begin{algorithmic}[1]
\REQUIRE Initialization $\uvu^1,~\uvx^1$, stepsize $\alpha$, $t^0=1$, $\umP$,
and maximum number of iterations Max\_Iter.
\lastcon  $\uvx^{\mathrm{Max\_Iter}}$.
 \FOR{$k=1,2,\cdots,$ Max\_Iter} 
 \STATE $\uvx^{k+1}=\mathrm{prox}_{(\alpha \lambda) g+\delta_{\mathcal C}}^{\umP}
 (\uvu^k-\stepsize\umP^{-1}\nabla f(\uvu^k))$\label{alg:WeightedacceleratedProximal:WPM}
 \STATE $t^k = \frac{1+\sqrt{1+4(t^{k-1})^2}}{2}$
 \STATE $\uvu^{k+1} = \uvx^{k+1}+\frac{t^{k-1}-1}{t^k}(\uvx^{k+1}-\uvx^k)$
 \ENDFOR
\end{algorithmic}
\end{algorithm}

\subsection{\MRcb{Implementation Details of On-the-Fly \RNP Method}}
\label{sec:OTF:RNP}

\Cref{sec:AdaptPreImageRes:sub:lplq,sec:AdaptPreImageRes:sub:l2l1}
show how to adapt \RNP to accelerate \IRM and APG
for \eqref{eq:lplqReg} and \eqref{eq:lpl1qMixedNorm}, respectively.
However, if computing \RNP is expensive,
then using \RNP may lose its practical utility.
Indeed, in \eqref{eq:lplq:sub:updateX:linearEq}, $\umW_f^k$ and $\umW_g^k$
change at each iteration,
so the preconditioner should be recomputed.
Therefore, the entire algorithm may be slower in terms of wall time
than one without using \RNP if obtaining \RNP is computationally expensive. 

\Cref{alg:NyApprox} shows that, to compute \RNP, one must evaluate $\bm\Phi\bm\Omega$.
Each column of $\bm\Omega$ represents an image.
In image reconstruction $\umA$ and $\umL$ are represented as operators,
so the associated $\bm\Phi$ is also an operator,
such as $\bm\Phi=\umA^\Trans\umA$ in \eqref{eq:lpl1qMixedNorm}
and $\bm\Phi=(\umL^{-1})^\mathcal T \umA^\Trans\umA \umL^{-1}$
in \eqref{eq:lplqMixedNorm:invertibleL}.
For some $\umA$ and $\umL$, evaluating $\umA\uvx$ and $\umL\uvx$ many times
could be computationally expensive.
Moreover, in practice,
one computes $\bm\Phi\bm\Omega$
by applying $\bm\Phi$ to each column of $\bm\Omega$ sequentially,
which could be slow when $K$ is moderately large.
So step \ref{alg:NyApprox:phix} in \Cref{alg:NyApprox} will dominate the overall computation,
making it crucial to significantly reduce this computational cost.

On classical CPU computational platforms,
a parallel scheme may be used to evaluate $\bm\Phi\bm\Omega$.
However, a central server still must collect the results to formulate $\umY$,
making communication time a potential bottleneck.
Modern machine learning libraries
(e.g., PyTorch \cite{paszke2019pytorch} and TensorFlow \cite{abadi2016tensorflow})
support a batch mode, allowing efficient parallel computation of $\bm\Phi\bm\Omega$.
Additionally, these libraries are highly optimized for modern GPU computational platforms,
and the batch mode is easy to implement with basic Python knowledge.
These features enable us to compute \RNP on-the-fly.
\Cref{tab:SKTime:CT:20} presents the wall time of computing \RNP
for CT reconstruction with different acquisitions on CPU and GPU computational platforms,
clearly showing the advantage of using GPU. \MRcbT{However, \RNP requires storing an additional $K$ images. As a result, it may encounter memory limitations when handling high-dimensional images on a GPU platform. Therefore, developing a memory-efficient approach to apply \RNP to high-dimensional images would be an interesting future direction.}
\begin{remark}
\MRcb{
In this section, we introduce \IRM (respectively, WAPG) for solving \eqref{eq:lplqReg} (respectively, \eqref{eq:lpl1qMixedNorm}) and discuss how \RNP can be incorporated for acceleration. However, in some special cases, e.g., $ p = 2 $ and $ \mathcal{C} = \mathbb{R}^N $, \eqref{eq:lplqReg} is identical to \eqref{eq:lpl1qMixedNorm}. This naturally raises the question of how to choose between \IRM and WAPG. From our experiments, when both \IRM and WAPG are applicable, we find that \IRM converges faster than WAPG in terms of iterations. However, since \IRM requires more computation per iteration than WAPG, we consistently observe that WAPG converges faster than \IRM in terms of wall time. We validate this observation in \Cref{sec:NumericalExp:sub:Comp} and \Cref{sec:numericalExps:sub:CT:sub:CompIRM_GKS}.  In this paper, we only consider linear inverse problems.  However, we believe that \RNP also has the potential to accelerate numerical solvers for nonlinear inverse problems. One example is the full-waveform inversion \cite{metivier2013full} in geophysics, where truncated Newton methods are among the appealing algorithms. However, one needs to solve a linear equation at each iteration, and the related coefficient matrix (i.e., the Hessian matrix) changes at every iteration.
So using effective on-the-fly preconditioners could significantly accelerate the numerical efficiency. We leave the investigation of this direction for future work, as it is beyond the scope of this paper.
}

\end{remark}

\section{Numerical Experiments}
\label{sec:NumericalExp}

\begin{table*}[t]
\centering
\caption{The associated minimization problems and the used priors for different imaging tasks.
$\nabla$ and $\nabla^2$ refer to the first- and second-difference operators, respectively.
}
\setlength\tabcolsep{15pt}
\begin{tabular}{l || c | c | c}
\hline
\hline 
Image Task & Problem & $q/\phi$ & $\umL$\\
\hline
Debluring/Super-Resolution & \eqref{eq:lplqReg} with $p\in(0, 1]$ & $q=1$ & $\nabla$, TV\\ 
\hline 
\multirow{3}{*}{Computed Tomography}&  \eqref{eq:lplqMixedNorm:invertibleL}  & $\phi=1$  & wavelet\\
\cline{2-4}
& \eqref{eq:lpl1qMixedNorm} & $\phi=1$ & $\nabla$, TV\\
\cline{2-4}
&  \eqref{eq:lpl1qMixedNorm} & $\phi=1,2,\infty$ & $\nabla^2$, HS
\end{tabular}
\label{tab:ImageTaskVSProblemPriors}
\end{table*}

\begin{table*}[t]
    \centering
     \caption{Comparison of wall time for obtaining \RNP in computed tomography reconstruction on CPU and GPU.}
     \setlength\tabcolsep{15pt}
\begin{tabular}{l||rrr||rrr}
 \hline
  \hline
 &\multicolumn{3}{c||}{Parallel-beam}  & \multicolumn{3}{c}{Fan-beam}\\

  & Wavelet & TV & HS & Wavelet & TV & HS\\
  \hline
$K$  & $20$ & $20$ &$100$  &$20$  &$20$ & $100$\\
\hline

GPU&  $0.11$s  & $0.10$s &  $0.40$s &  $0.11$s  & $0.09$s & $0.46$s \\

\hline
CPU &$0.48$s  & $0.40$s & $4.20$s  & $0.48$s  & $0.52$s & $4.25$s \\
\end{tabular}
    \label{tab:SKTime:CT:20}
\end{table*}

This section studies the performance of using \RNP on three different image reconstruction tasks:
image deblurring and super-resolution with impulsive noise, and CT.
Following \cite{rodriguez2008efficient},
we use a TV regularizer for image deblurring and super-resolution tasks. For CT reconstruction,
we consider wavelet, TV, and HS norm regularizers.
\Cref{tab:ImageTaskVSProblemPriors} presents the associated minimization problems
and regularizers for different image tasks in this part. \MRcb{The trade-off parameter $\lambda$ is selected through a grid search to maximize the PSNR.}
All experiments were implemented with PyTorch and run on an NVIDIA GeForce RTX 3090. 

\subsection{Image Deblurring and Super-Resolution}

For the image deblurring task,
we first degraded the test images by convolving them with two different $9\times 9$ kernels:
a uniform blur and a Gaussian blur with a standard deviation of $1.6$.
In the image super-resolution task,
we generated the low-resolution image by convolving the high-resolution
with a $7\times7$ Gaussian blur with a standard deviation of $1.6$,
followed by downsampling with a factor of $2$.
We added $5\%$ salt-and-pepper noise to the degraded images.\footnote{We first randomly chose $5\%$ of the pixels and set them to $1$, and then randomly selected another $5\%$ of the pixels and set them to $0$.} 
The supplementary material
also reports the performance of using \RNP with different noise levels.
We set the sketch size $K=100$
and tolerance $\epsilon = 10^{-4}$
in \Cref{alg:NyApprox} and PCG, respectively.
Similar to \cite{romano2017little},
the restoration of an RGB image was conducted by converting it to the YCbCr color-space first.
Next, we applied the reconstruction algorithm to the luminance channel only
and then transformed the result back to the RGB domain.
For the deblurring task, we simply applied a median filter to the chroma channels for denoising.
For the super-resolution task,
we first used a median filter to denoise the chroma channels
and then applied bicubic interpolation to upscale them to the desired resolution.
Moreover, the peak signal-to-noise ratio (PSNR) values of RGB images
were evaluated on the luminance channel only.
We used $\ell_p-\ell_1$ to denote the method with different $p$.
For the one using \RNP, we added ``P-'', i.e., $\umP-\ell_p-\ell_1$.

\subsubsection{Image Deblurring}
\label{sec:NumericalExp:ImageDeblurring}
\begin{figure}[ht]
    \centering
    \subfigure[Uniform blur.]{\includegraphics[width=0.49\linewidth]{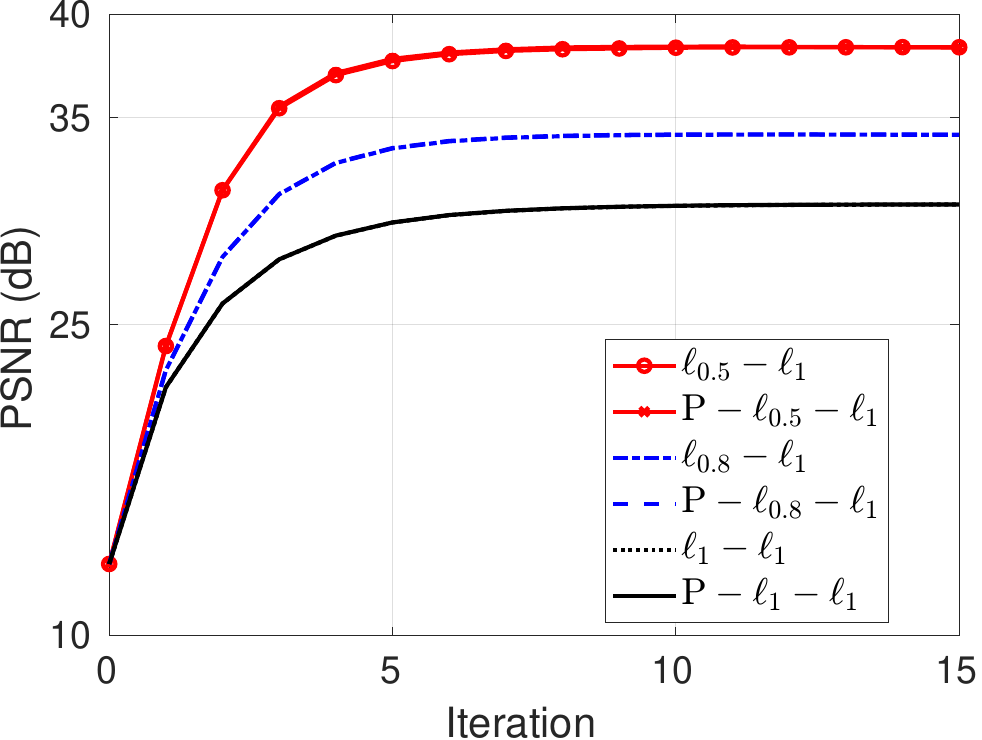}\label{fig:UniformDeblur:PSNRIterandTime:uniformIter}}
    \subfigure[Uniform blur.]{\includegraphics[width=0.475\linewidth]{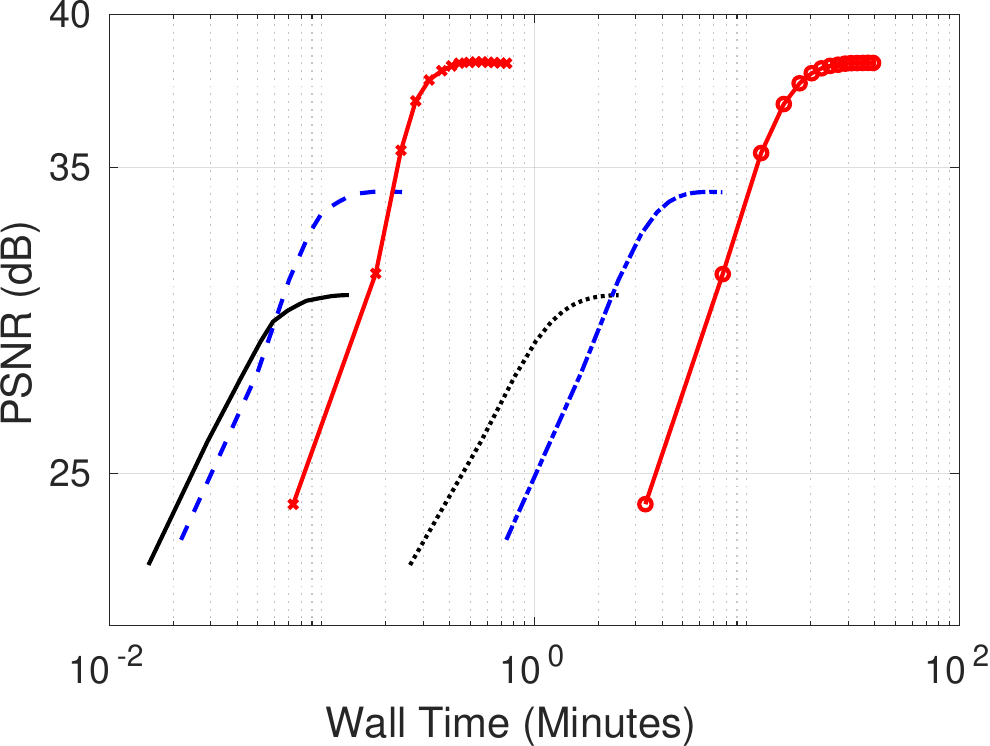}\label{fig:UniformDeblur:PSNRIterandTime:uniformTime}}

    \subfigure[Gaussian blur.]{\includegraphics[width=0.478\linewidth]{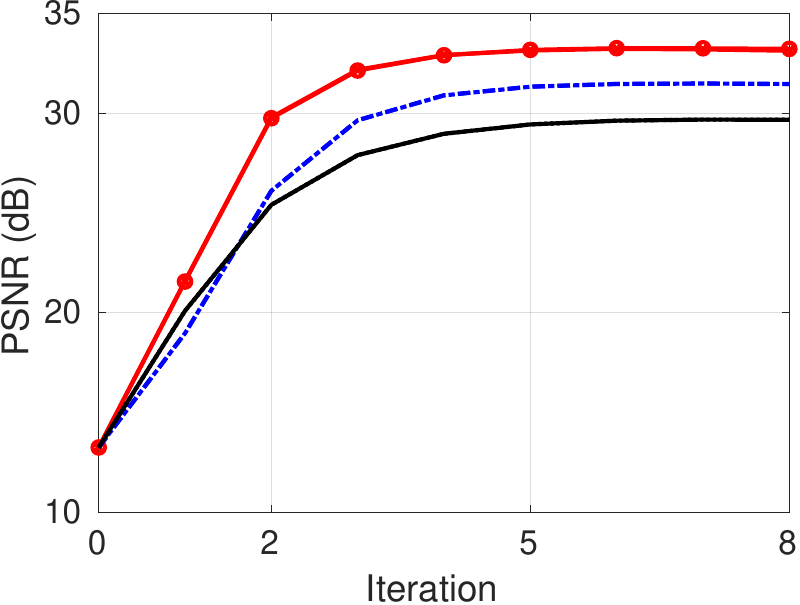}\label{fig:UniformDeblur:PSNRIterandTime:GaussIter}}
    \subfigure[Gaussian blur.]{\includegraphics[width=0.48\linewidth]{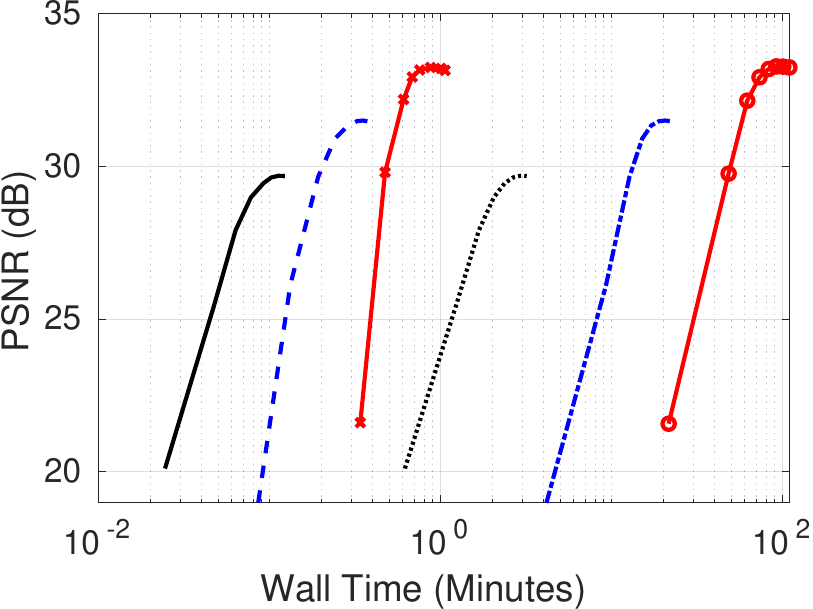}\label{fig:UniformDeblur:PSNRIterandTime:GaussTime}}
    \caption{PSNR values versus iteration and wall time for different $p$ in the image deblurring task. First (respectively, second)  row is tested on the starfish (respectively, leaves) image.}
    \label{fig:UniformDeblur:PSNRIterandTime}
\end{figure}

\begin{figure}[ht]
    \centering
   \subfigure[Starfish \& uniform blur.]{\includegraphics[width=0.48\linewidth]{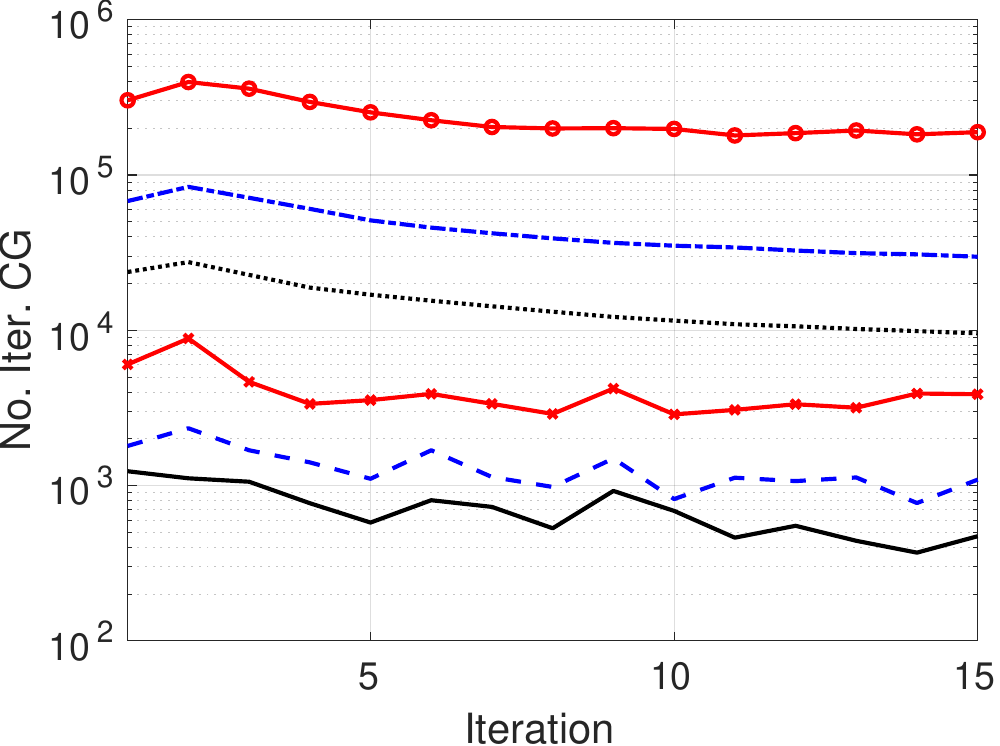}}
    \subfigure[Leaves \& Gaussian blur.]{\includegraphics[width=0.47\linewidth]{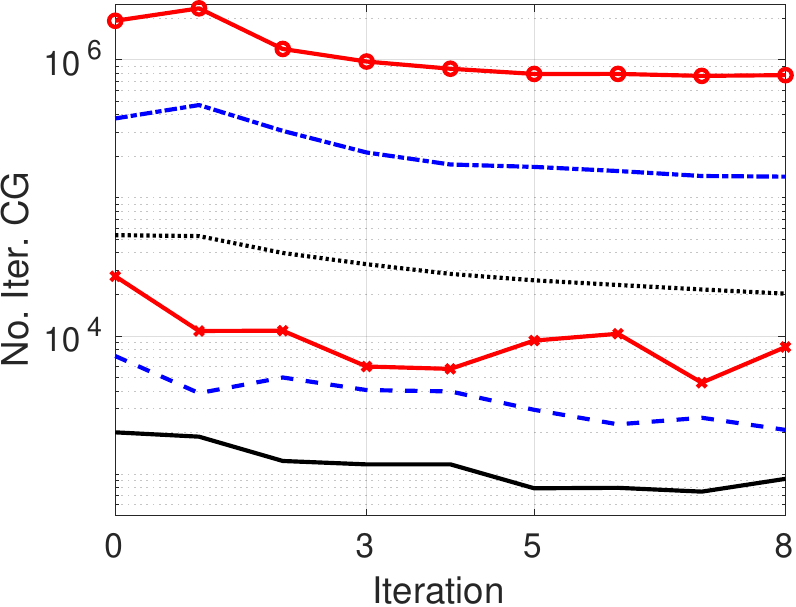}}
    \caption{Number of iterations within CG for different $p$ in the image deblurring task. The legend is identical to \Cref{fig:UniformDeblur:PSNRIterandTime}.}
    \label{fig:Deblur:CGIter}
\end{figure}

\begin{figure*}[ht]
\vspace{-2cm}
	\centering
\begin{tikzpicture}
    \begin{axis}[at={(0,0)},anchor = north west,
    xmin = 0,xmax = 250,ymin = 0,ymax = 70, width=0.95\textwidth,
        scale only axis,
        enlargelimits=false,
       axis line style={draw=none},
       tick style={draw=none},
        axis equal image,
        xticklabels={,,},yticklabels={,,},
       ]

    \node[inner sep=0pt, anchor = south west] (p1_1) at (0,0) {\includegraphics[ width=0.18\textwidth]{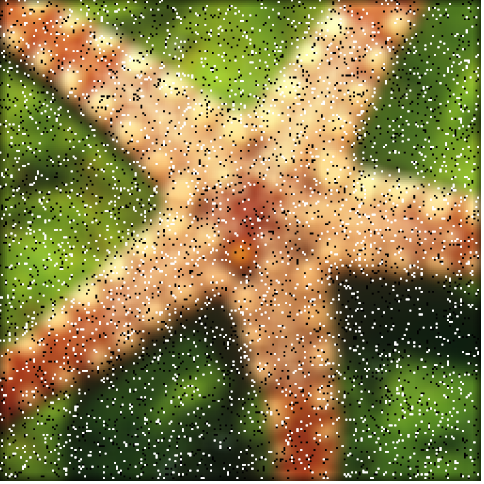}};
    
    \node at (8,3) {\color{white} Noisy};
    
     \node at (22,44) {\color{white} Deblurring, Uniform};
     
    \node[inner sep=2pt, anchor = west] (p1_2) at (p1_1.east) {\includegraphics[ width=0.18\textwidth]{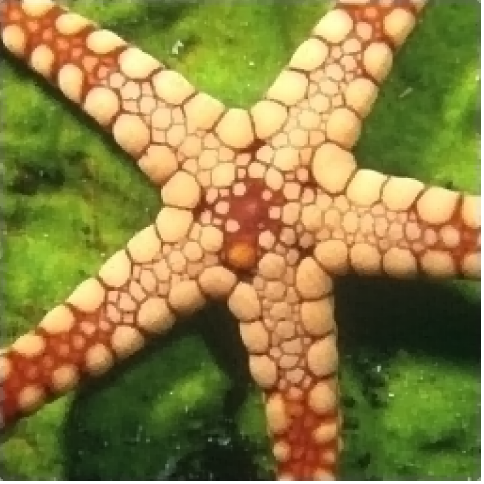}};

    \node at (57,3) {\color{white} $\ell_1-\ell_1$};
    \node at (88,3) {\color{white} $30.7$dB};
    
    
    \node[inner sep=1pt, anchor = west] (p1_3) at (p1_2.east) {\includegraphics[ width=0.18\textwidth]{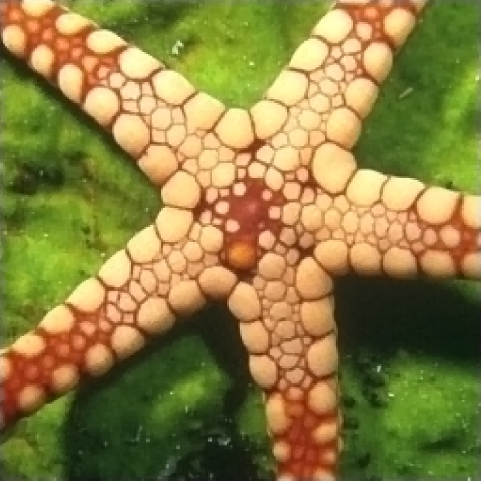}};
    
   \node at (107,3) {\color{white} $\ell_{0.8}-\ell_1$};
    \node at (137,3) {\color{white} $34.2$dB};
 \node[inner sep=1pt, anchor = west] (p1_4) at (p1_3.east) {\includegraphics[ width=0.18\textwidth]{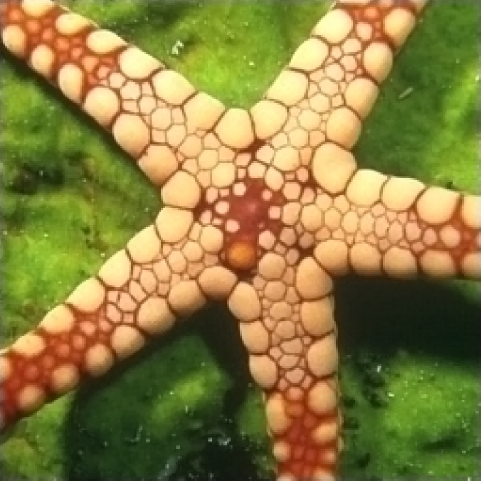}};
 
  \node at (156,3) {\color{white} $\ell_{0.5}-\ell_1$};
  \node at (186,3) {\color{white} $38.4$dB};
 \node[inner sep=1pt, anchor = west] (p1_5) at (p1_4.east) {\includegraphics[ width=0.18\textwidth]{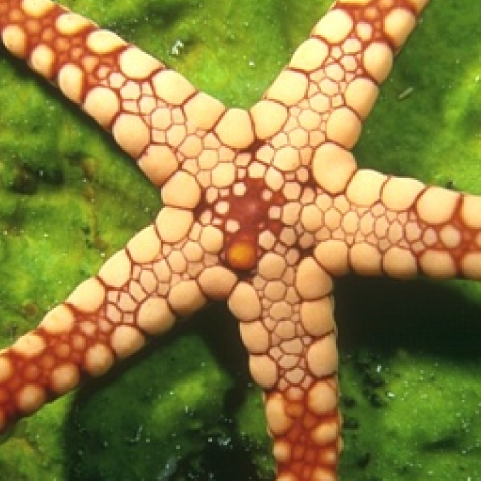}};
\node at (198.5,3) {\color{white} GT};
  
 \end{axis}

\begin{axis}[at={(p1_1.south west)},anchor = north west,
    xmin = 0,xmax = 250,ymin = 0,ymax = 70, width=0.95\textwidth,
        scale only axis,
        enlargelimits=false,
        yshift=1.5cm,
       axis line style={draw=none},
       tick style={draw=none},
        axis equal image,
        xticklabels={,,},yticklabels={,,},
       ]

\node[inner sep=0pt, anchor = south west] (Gaussp1_1) at (0,0) {\includegraphics[ width=0.18\textwidth]{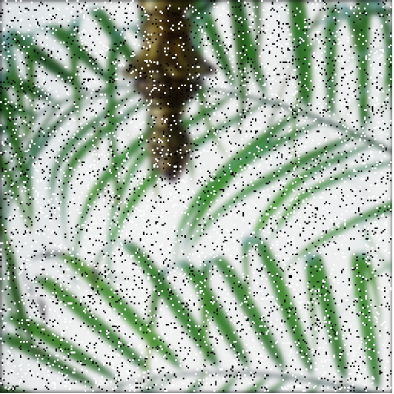}};
    
    \node at (22,44) {\color{black} Deblurring, Gaussian};
 \node[inner sep=2pt, anchor = west] (Gaussp1_2) at (Gaussp1_1.east) {\includegraphics[ width=0.18\textwidth]{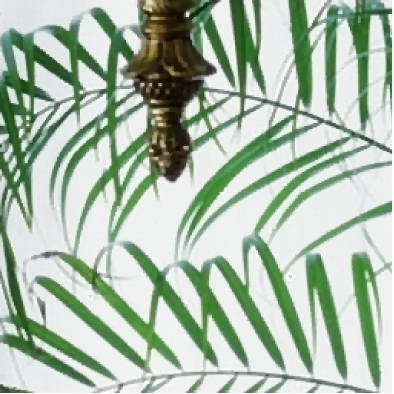}};

\node at (57,3) {\color{black} $\ell_1-\ell_1$};
\node at (88,3) {\color{black} $29.6$dB};
    
 \node[inner sep=1pt, anchor = west] (Gaussp1_3) at (Gaussp1_2.east) {\includegraphics[ width=0.18\textwidth]{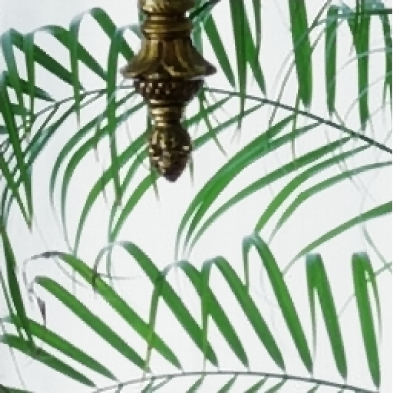}};
    
  \node at (107,3) {\color{black} $\ell_{0.8}-\ell_1$};

  \node at (137,3) {\color{black} $31.5$dB};
 
 \node[inner sep=1pt, anchor = west] (Gaussp1_4) at (Gaussp1_3.east) {\includegraphics[ width=0.18\textwidth]{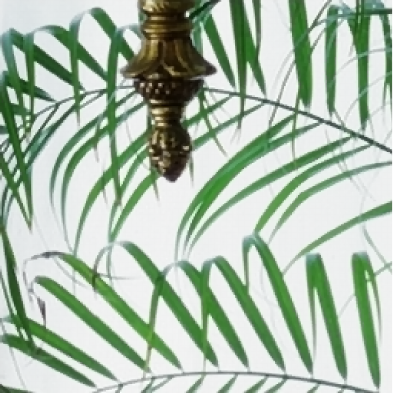}};
 
  \node at (156,3) {\color{black} $\ell_{0.5}-\ell_1$};
  \node at (186,3) {\color{black} $33.3$dB};

 \node[inner sep=1pt, anchor = west] (Gaussp1_5) at (Gaussp1_4.east) {\includegraphics[ width=0.18\textwidth]{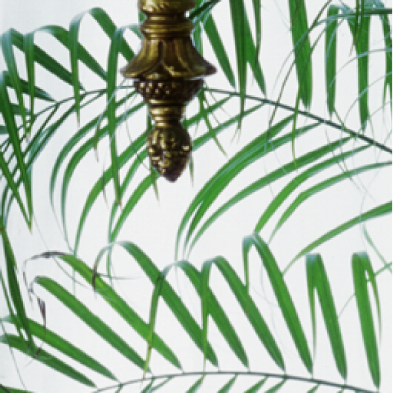}};
\end{axis}

\begin{axis}[at={(Gaussp1_1.south west)},anchor = north west,
    xmin = 0,xmax = 250,ymin = 0,ymax = 70, width=0.95\textwidth,
        scale only axis,
        enlargelimits=false,
        yshift=1.5cm,
       axis line style={draw=none},
       tick style={draw=none},
        axis equal image,
        xticklabels={,,},yticklabels={,,},
       ]

\node[inner sep=0pt, anchor = south west] (p1_1) at (0,0) {\includegraphics[ width=0.18\textwidth]{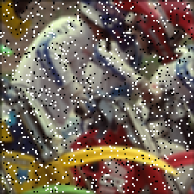}};
    
   \node at (18,44) {\color{white} Super-Resolution};
 \node[inner sep=2pt, anchor = west] (p1_2) at (p1_1.east) {\includegraphics[ width=0.18\textwidth]{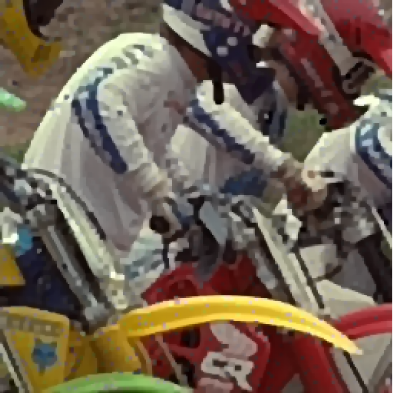}};

\node at (57,3) {\color{white} $\ell_1-\ell_1$};
\node at (88,3) {\color{white} $22.9$dB};
    
 \node[inner sep=1pt, anchor = west] (p1_3) at (p1_2.east) {\includegraphics[ width=0.18\textwidth]{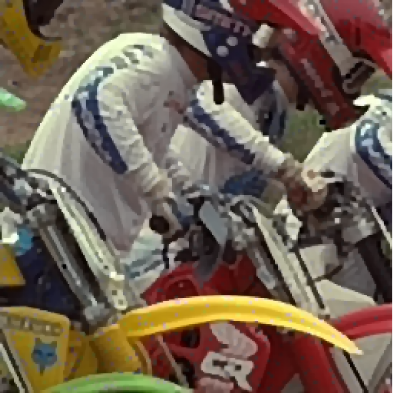}};
    
 \node at (107,3) {\color{white} $\ell_{0.8}-\ell_1$};

  \node at (137,3) {\color{white} $24.1$dB};

 
 \node[inner sep=1pt, anchor = west] (p1_4) at (p1_3.east) {\includegraphics[ width=0.18\textwidth]{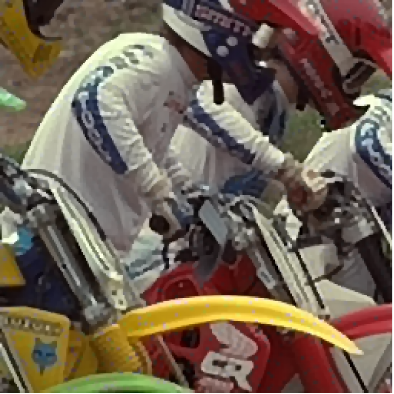}};
 
  \node at (156,3) {\color{white} $\ell_{0.5}-\ell_1$};
  \node at (186,3) {\color{white} $25.3$dB};

 \node[inner sep=1pt, anchor = west] (p1_5) at (p1_4.east) {\includegraphics[ width=0.18\textwidth]{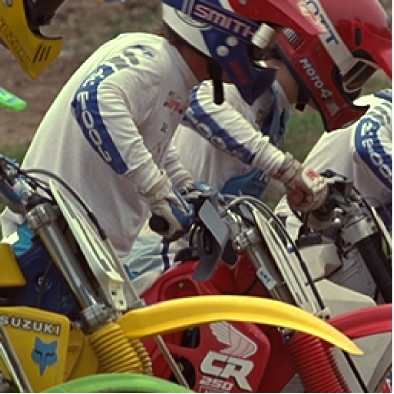}};
  
 \end{axis}

\end{tikzpicture} 

\caption{
The reconstructed images for different $p$ on the image deblurring (\MRcb{first and second rows}) and super-resolution (\MRcb{third row}) tasks. The PSNR value is labeled at the right bottom corner of each image. The first and fifth columns are the noisy measurement and the ground truth, respectively. The first (respectively, second) row is the reconstructed starfish (respectively, leaves) image at $10$th (respectively, $6$th)  iteration for the image deblurring task with uniform (respectively, Gaussian) blur. The third row is the reconstructed bike image at $7$th iteration for the image super-resolution task with a downsampling factor $2$
and $5$\% salt-and-pepper noise.
}
\label{fig:UniformGaussianDeblurSR:RecoImages}
\end{figure*}

\begin{table*}[t]
    \centering
     \caption{Comparison of PSNR and ST for different $p$ in the image deblurring task with uniform and Gaussian blurs. We ran  \IRM  $20$ iterations on each image and presented the highest PSNR values within these $20$ iterations along with the associated ST values. The highest PSNR and ST values for each image are marked in bold.}
 \setlength\tabcolsep{3.35pt}
\begin{tabular}{p{1cm}||rr|rr|rr|rr|rr|rr|rr|rr|rr}
 
 \hline
 \hline
Image&\multicolumn{2}{c|}{Butterfly}  & \multicolumn{2}{c|}{Boats}&\multicolumn{2}{c|}{C. Man} &\multicolumn{2}{c|}{House} &\multicolumn{2}{c|}{Parrot}  & \multicolumn{2}{c|}{Barbara} & \multicolumn{2}{c|}{Starfish} & \multicolumn{2}{c|}{Peppers}& \multicolumn{2}{c}{Leaves} \\

\hline

\multicolumn{19}{c}{Deblurring: Uniform blur, 5\% salt-and-pepper noise}
\\
\hline

& PSNR & ST  & PSNR & ST  &PSNR & ST &PSNR & ST  &PSNR & ST  &PSNR & ST  &PSNR & ST  & PSNR& ST 
&PSNR & ST 

 \\
 \cline{2-19}
 $p=1$  &  $31.6$&$0.94$&
        $32.9$  & $0.94$&
        $31.8$&$0.95$  &
        $35.9$ & $0.93$&
        
        $32.3$&$0.95$&
        $27.9$&$0.95$ &
        
        $30.8$&$0.94$&
        $32.2$ &$0.94$&
         $30.7$ &$0.94$

        
        
 \\

 \hline

 $p=0.8$  &  $35.3$&$0.96$&  
        $35.9$  & $0.95$&
        $35.1$&$0.97$  &
        $38.7$ &  $0.96$&
             
        $35.7$&$0.97$&
        $31.1$&$0.97$ &
        
        $34.2$&$0.96$&
        $35.2$ &$0.96$&
         $34.7$ &$0.96$

             
        
 \\       
 \hline

 $p=0.5$  &  $\bm{39.9}$&$\bm{0.98}$& 
        $\bm{39.9}$  & $\bm{0.97}$&
        $\bm{38.8}$&$\bm{0.98}$  &  
         $\bm{42.5}$ &  $\bm{0.98}$&

        $\bm{39.2}$&$\bm{0.98}$&
        $\bm{34.9}$&$\bm{0.98}$ &
        
        $\bm{38.4}$&$\bm{0.98}$&
        $\bm{38.8}$ &$\bm{0.98}$&
         $\bm{39.7}$ &$\bm{0.98}$


        
 \\
 \hline

 \multicolumn{19}{c}{Deblurring: Gaussian blur, 5\% salt-and-pepper noise}

 \\
 \hline
 
 $p=1$  &  $29.5$&$0.95$&   
        $30.9$  & $0.96$&
        $29.4$&$0.97$  &
        $34.1$ & $0.96$&
        $31.9$&$0.96$&
        $28.1$&$0.97$ &
        $29.9$&$0.96$&
        $31.3$ &$0.95$&
         $29.7$ &$0.96$

 \\

 \hline

 $p=0.8$  &  $31.5$  & $0.98$&  
        $32.9$  & $0.98$&
        $31.5$ & $0.98$&
        $35.6$&$0.98$  &
        $33.5$&$0.98$&
        $30.3$&$0.98$ &
        $31.9$&$0.98$&
        $32.5$ &$0.98$&
         $31.5$ &$0.98$

 \\      
 \hline

 $p=0.5$  &  $\bm{33.5}$&$\bm{0.99}$& 
        $\bm{35.2}$  & $\bm{0.99}$&
        $\bm{33.5}$&$\bm{0.99}$  &  
         $\bm{37.4}$ &  $\bm{0.99}$&
        $\bm{34.9}$&$\bm{0.99}$&
        $\bm{31.9}$&$\bm{0.99}$ &
        $\bm{33.7}$&$\bm{0.99}$&
        $\bm{34.1}$ &$\bm{0.99}$&
         $\bm{33.3}$ &$\bm{0.99}$


\end{tabular}
    \label{tab:DeblurTable}
\end{table*}

\begin{table*}[t]
    \centering
     \caption{Comparison of PSNR and ST for different $p$ in the image super-resolution task. We ran \IRM $20$ iterations on each image and presented the highest PSNR values within these $20$ iterations along with the associated ST values. The highest PSNR and ST values for each image are marked in bold.}
     \setlength\tabcolsep{3.35pt}
\begin{tabular}{p{1cm}||rr|rr|rr|rr|rr|rr|rr|rr|rr}
 
 \hline
  \hline
 \multicolumn{19}{c}{Super-resolution: scaling = 2, 5\% salt-and-pepper noise}
 \\
 \hline
\multirow{2}*{Image}&\multicolumn{2}{c|}{Butterfly}  & \multicolumn{2}{c|}{Flower}&\multicolumn{2}{c|}{Girl} &\multicolumn{2}{c|}{Parth.} &\multicolumn{2}{c|}{Parrot}  & \multicolumn{2}{c|}{Racoon} & \multicolumn{2}{c|}{Bike} & \multicolumn{2}{c|}{Hat}& \multicolumn{2}{c}{Plants} \\

\cline{2-19}

& PSNR & ST  & PSNR & ST  &PSNR & ST &PSNR & ST  &PSNR & ST  &PSNR & ST  &PSNR & ST  & PSNR& ST 
&PSNR & ST 

 \\
 \hline
 $p=1$  &  $25.6$&$0.74$&   
        $27.4$  & $0.74$&
        $31.6$&$0.78$  &
        $24.5$ & $0.43$&
        $27.5$&$0.76$&
        $27.3$&$0.68$ &
        $23.2$&$0.76$&
        $29.4$ &$0.76$&
        $31.9$ &$0.76$

 \\

 \hline

 $p=0.8$  &  $27.1$&$0.94$&
        $28.7$  & $0.93$&
        $32.5$&$0.91$  &
        $25.4$ &  $0.91$&
        $28.6$&$0.94$&
        $28.4$&$0.90$ &
        $24.3$&$0.93$&
        $30.3$ &$0.94$&
         $33.3$ &$0.92$

 \\       
 \hline

 $p=0.5$  &  $\bm{28.1}$&$\bm{0.96}$&
        $\bm{29.6}$  & $\bm{0.97}$&
        $\bm{33.1}$&$\bm{0.96}$  &  
         $\bm{25.9}$ &  $\bm{0.96}$& 
        $\bm{29.7}$&$\bm{0.97}$&
        $\bm{29.4}$&$\bm{0.97}$ &
        $\bm{25.3}$&$\bm{0.97}$&
        $\bm{31.4}$ &$\bm{0.96}$&
         $\bm{34.4}$ &$\bm{0.96}$

        
         \end{tabular}
    \label{tab:SRTable}
\end{table*}

\Cref{fig:UniformDeblur:PSNRIterandTime}
presents the PSNR values versus iteration and wall time for different $p$.
The first (respectively, second) row of \Cref{fig:UniformDeblur:PSNRIterandTime}
was tested on the starfish (respectively, leaves) image with uniform (respectively, Gaussian) blur.
\Cref{fig:UniformDeblur:PSNRIterandTime:uniformIter,fig:UniformDeblur:PSNRIterandTime:GaussIter}
show a small $p$ yielded a higher PSNR than a large $p$.
It is not \MRcb{surprising} because a small $p$ is more robust to outliers \cite{candes2008enhancing}.
However, \eqref{eq:lplqReg} becomes nonconvex for $p<1$,
so solving \eqref{eq:lplqReg} is more challenging.
Indeed,
\Cref{fig:UniformDeblur:PSNRIterandTime:uniformTime,fig:UniformDeblur:PSNRIterandTime:GaussTime}
illustrate that a small $p$ required much more wall time than a large one.
\Cref{fig:UniformDeblur:PSNRIterandTime} illustrates that using \RNP
significantly accelerated the convergence speed in terms of wall time.
The time for computing \RNP was included in the whole wall time.
From
\Cref{fig:UniformDeblur:PSNRIterandTime:uniformTime,fig:UniformDeblur:PSNRIterandTime:GaussTime},
we even saw $\mathrm{P}-\ell_{0.5}-\ell_1$ converged faster than $\ell_1-\ell_1$
in terms of the wall time illustrating the effectiveness of using \RNP.
\Cref{fig:Deblur:CGIter} describes the number of iterations within CG
at each iteration of \IRM for different $p$,
with the uniform and Gaussian kernels for the starfish and leaves images, respectively.
Using \RNP reduced the number of iterations required by CG more than $90$\%,
while achieving the same solution accuracy.
The first and second rows of \Cref{fig:UniformGaussianDeblurSR:RecoImages}
present the noisy, reconstructed, and ground truth images,
where the degradation is reduced significantly.

To quantitatively measure the performance of using \RNP,
we define a ``saved time'' (ST) criterion:
\begin{equation}
\label{eq:STDefine}
	\mathrm{ST} = \frac{\mathrm{Time}_\text{w/o}-\mathrm{Time}_\text{w}}{\mathrm{Time}_\text{w/o}},
\end{equation}
where $\mathrm{Time}_\text{w}$ and $\mathrm{Time}_\text{w/o}$
denote the wall time with and without using \RNP, respectively.
If $\mathrm{ST}$ is close to $1$,
then using \RNP significantly reduced the wall time.
To study the performance of \RNP further, we tested on additional $8$ images.
We ran \IRM $20$ iterations for different $p$. \Cref{tab:DeblurTable} presents the highest PNSR and the associated ST value for each image within $20$ iterations. Using \RNP saved more than $95\%$ wall time than without \RNP,
illustrating the effectiveness and efficiency of using \RNP.
Moreover, \Cref{tab:DeblurTable} also shows that a smaller $p$ yielded a higher PSNR.

\subsubsection{Image Super-Resolution}
\label{sec:NumericalExp:ImageSR}

\begin{figure}[ht]
    \centering
    \subfigure[Bike.]{\includegraphics[width=0.485\linewidth]{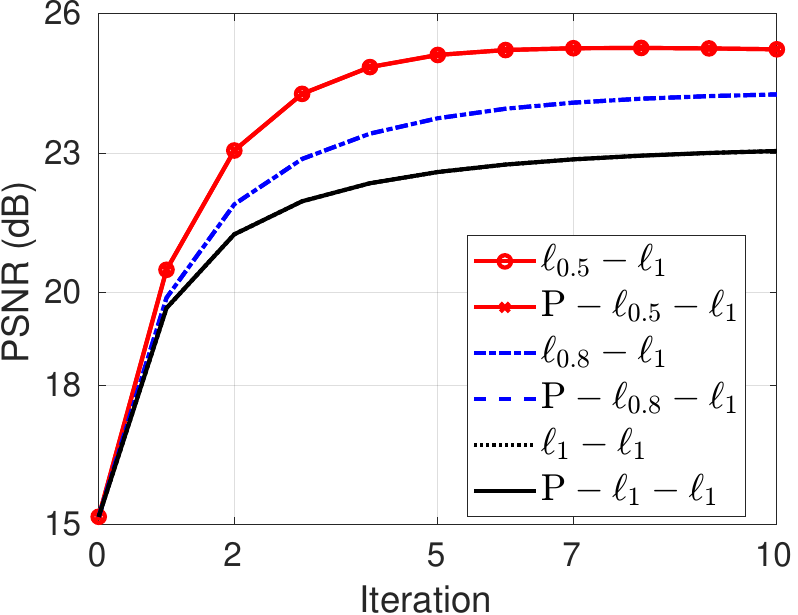}}
    \subfigure[Bike.]{\includegraphics[width=0.49\linewidth]{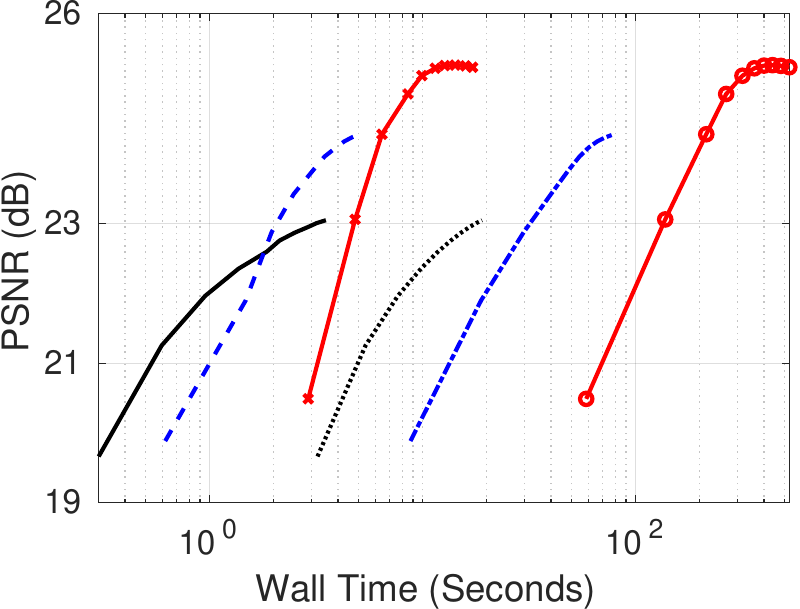}}
    \caption{PSNR values versus iteration and wall time for different $p$ in the image super-resolution task.}
    \label{fig:SRGauss:PSNRIterandTime}
\end{figure}

\begin{figure}[ht]
    \centering
    \includegraphics[width=0.8\linewidth]{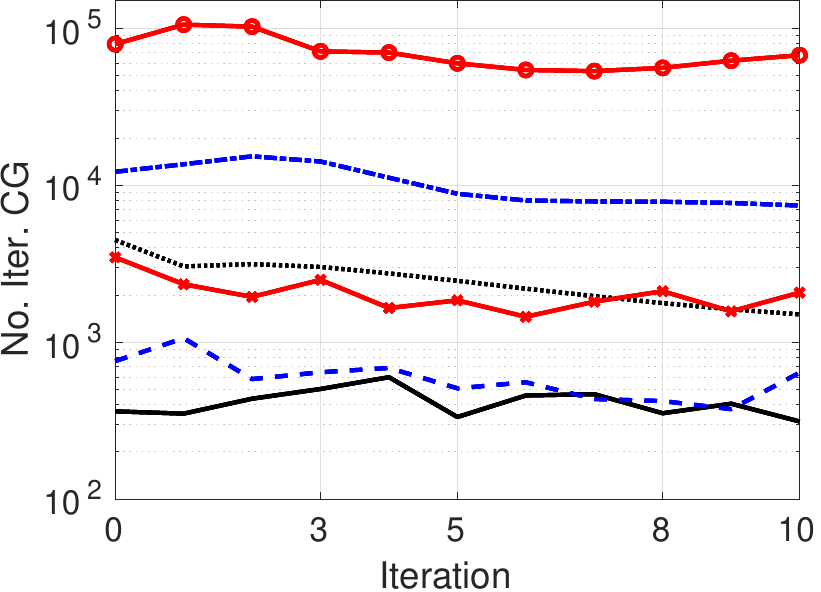}
    \caption{Number of iterations within CG for different $p$ in the super-resolution task. The downsampling factor is $2$. The legend is identical to \Cref{fig:SRGauss:PSNRIterandTime}.}
    \label{fig:SRGauss:CGIter}
\end{figure}

\Cref{fig:SRGauss:PSNRIterandTime}
shows the PSNR values for different $p$ in the image super-resolution of the bike image.
The results are consistent with the trends in \Cref{sec:NumericalExp:ImageDeblurring},
where a small $p$ yielded a higher PSNR than a larger one
and using \RNP significantly accelerated the convergence speed in terms of wall time.
Moreover,
\Cref{fig:SRGauss:CGIter}
presents the number of CG iterations for different $p$;
the preconditioned one required fewer iterations
illustrating the effectiveness of using \RNP.
The third row of \Cref{fig:UniformGaussianDeblurSR:RecoImages}
presents the reconstructed images,
where we observed that a small $p$ yielded a higher quality image.
We tested on additional $8$ images to study the performance of using \RNP further.
\Cref{tab:SRTable} shows using \RNP saved almost $70\%$ (respectively, $95\%$) time
for $p=1$ (respectively, $p=0.5$) 
in line with the observation in \Cref{sec:NumericalExp:ImageDeblurring}.

\subsubsection{The Choice of Sketch Size}
\label{sec:NumericalExp:ImageDeblurring:SketchSize}

\begin{figure}[ht]
    \centering
    \subfigure[$p=0.5.$]{\includegraphics[width=0.485\linewidth]{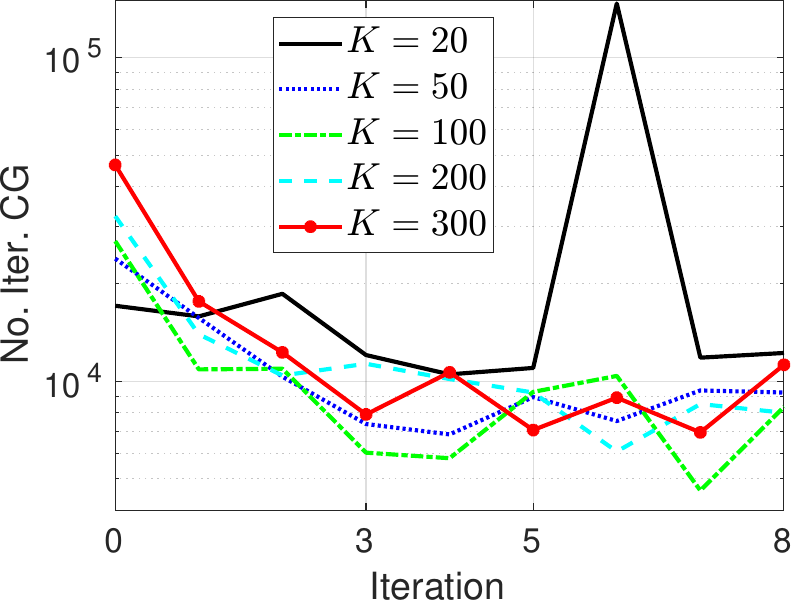}\label{fig:GaussDeblue:PSNRTimeandCGIter:sketchCGIter}}
    \subfigure[$p=0.5.$]{\includegraphics[width=0.46\linewidth]{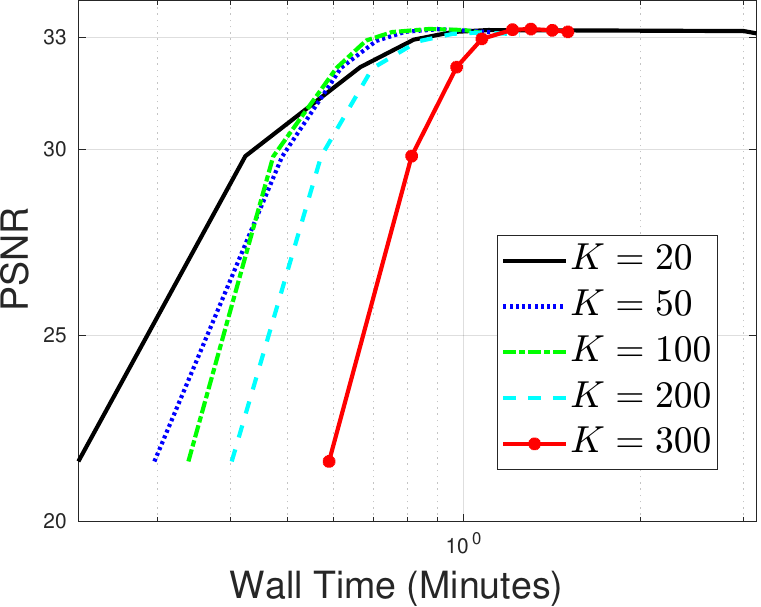}
    \label{fig:GaussDeblue:PSNRTimeandCGIter:PSNRTime}}
    \caption{Comparison of different $K$ in the image deblurring task with Gaussian kernel
    on the leaves image \MRcb{for $p=0.5$}. (a): the number of iterations within CG in each iteration;
    (b): PSNR values versus wall time.}
    \label{fig:GaussDeblue:PSNRTimeandCGIter}
\end{figure}

From \Cref{them:NYPre:CN},
one option for choosing the sketch size is to set
$K=2\lceil 1.5d_{\mathrm{eff}}(\mu)+1\rceil$.
However, $\deff(\mu)$ is difficult to compute in practice.
Moreover, $\deff(\mu)$ can be extremely large,
making computing \RNP time- and memory-intensive,
even with an on-the-fly implementation.
Indeed, for an extremely large $K$,
steps \ref{alg:NyApprox:CD}-\ref{alg:NyApprox:SVD} in \Cref{alg:NyApprox}
will dominate the computation instead of step \ref{alg:NyApprox:phix}.
Furthermore, applying \RNP can be also expensive for an extremely large $K$.
Murray et al. \cite{murray2023randomized}
presented an adaptive strategy to update $K$ at each iteration,
avoiding the need for $\deff(\mu)$.
However, the adaptive strategy requires calling \Cref{alg:NyApprox} multiple times
and executing many iterations of the power method to estimate the error
between $\bm\Phi$ and $\umU\hat{\umS}\umU^\Trans$.
Therefore, it is unsuitable for our experimental settings,
where \RNP must be computed on-the-fly.
We found that using a fixed, moderate $K$ works well in practice. 

\Cref{fig:GaussDeblue:PSNRTimeandCGIter:sketchCGIter}
presents the number of iterations within CG
in the image deblurring task with Gaussian blur \MRcb{and $p=0.5$} on the leaves image for different $K$.
For $K>20$, the number of iterations within CG did not change significantly. \MRcb{Notice that at the $7$th iteration, $K = 20$ requires significantly more CG iterations than other iterations, where the preconditioner is not very effective. However, the required number of CG iterations is still much lower than in the case without a preconditioner.}
In general, a larger $K$ would yield a more effective preconditioner,
but it also increases the computational cost to compute and apply.
\Cref{fig:GaussDeblue:PSNRTimeandCGIter:PSNRTime}
shows the reconstructed PSNR values versus wall time,
where clearly $K=300$ required more wall time than the others.
Moreover, $K=50$ and $K=100$ performed better than the others.
Since $K=100$ performed slightly better than $K=50$, we simply set $K=100$ in this subsection. 


\subsubsection{\MRcb{Comparison of GKS and ADMM}}
\label{sec:NumericalExp:sub:Comp}
\MRcb{
We compared the performance of \IRM with \RNP (w-$20$), GKS \cite{lanza2015generalized}, and ADMM for $p=0.5,1$ on the \MRcbT{super-resolution} task with the bike image. \Cref{fig:lplq:SR:bike:comp:GKSADMM} presents the cost values in terms of number of iterations and wall time. \Cref{fig:lplq:SR:bike:comp:GKSADMM:Iter:l1l1,fig:lplq:SR:bike:comp:GKSADMM:Iter:l05l1} show that w-$20$ is the fastest algorithm in terms of iterations. Compared to w-$20$, GKS requires significantly less computation per iteration. Indeed, \Cref{fig:lplq:SR:bike:comp:GKSADMM:Time:l1l1} indicates  GKS converged fastest than the other methods in terms of wall time for $p=1$. However, for $p=0.5$, we observed that  w-$20$ converged faster than GKS in terms of wall time illustrating the effectiveness of using \RNP for acceleration when the problem becomes more challenging.

\begin{figure}
\centering
	\subfigure[$p=1.$]{\includegraphics[width=0.49\linewidth]{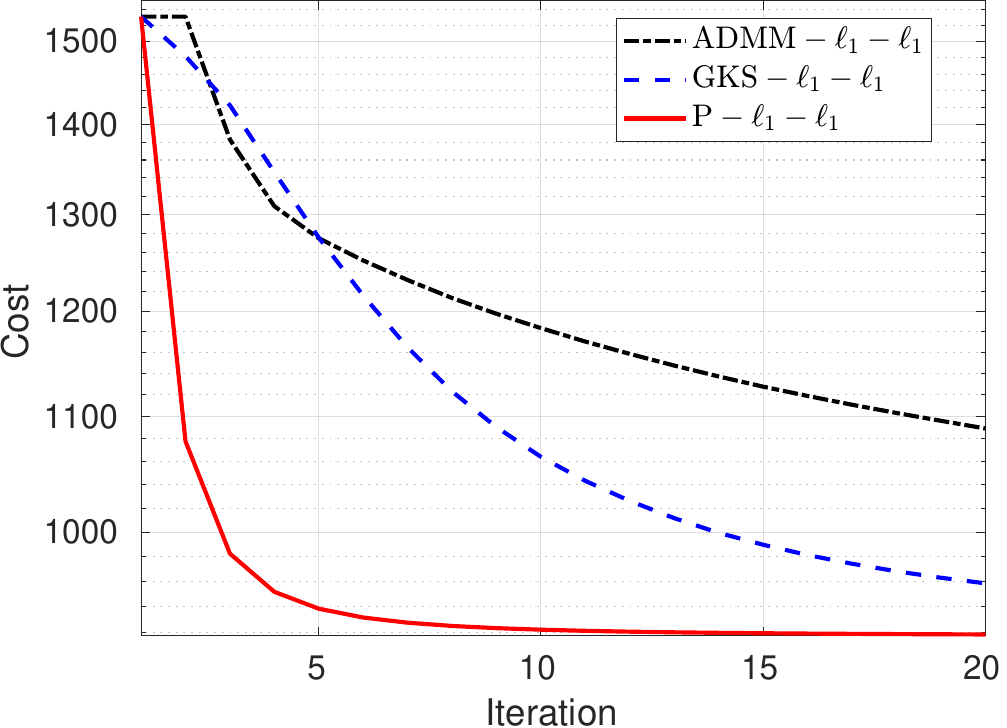}\label{fig:lplq:SR:bike:comp:GKSADMM:Iter:l1l1}}
		\subfigure[$p=1.$]{\includegraphics[width=0.48\linewidth]{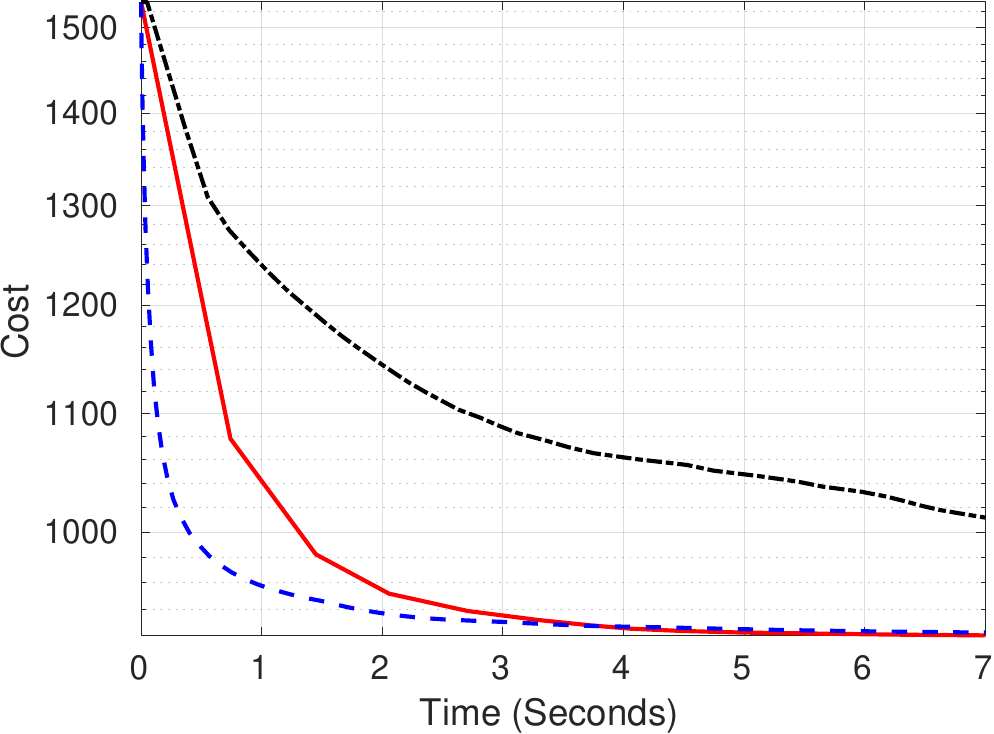}\label{fig:lplq:SR:bike:comp:GKSADMM:Time:l1l1}}
		
	\subfigure[$p=0.5.$]{\includegraphics[width=0.485\linewidth]{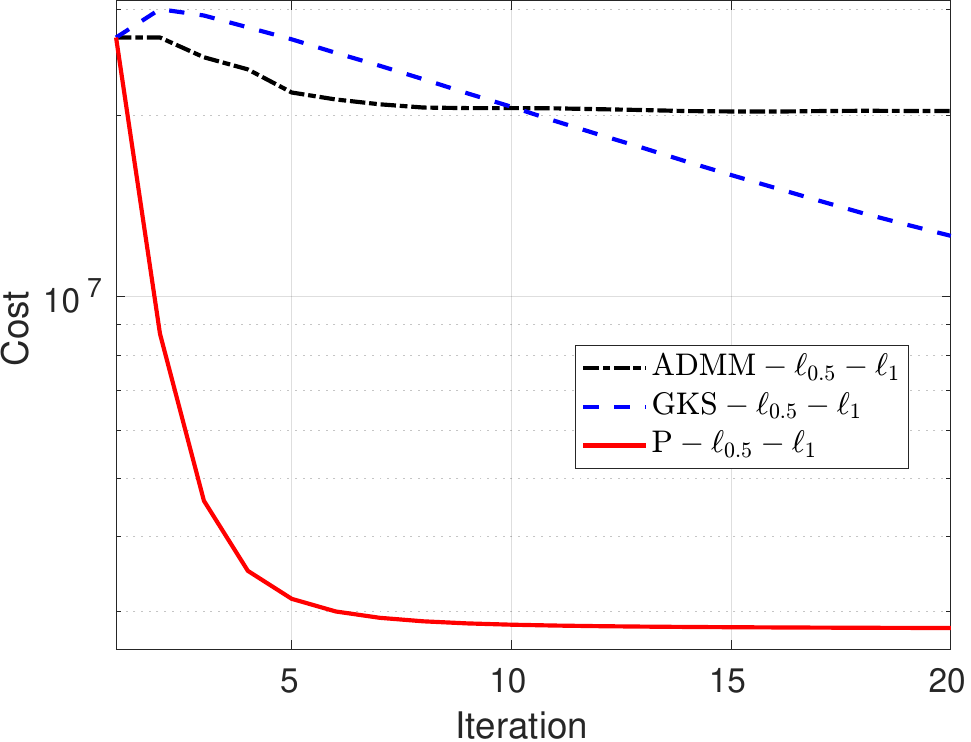}\label{fig:lplq:SR:bike:comp:GKSADMM:Iter:l05l1}}
	\subfigure[$p=0.5.$]{\includegraphics[width=0.485\linewidth]{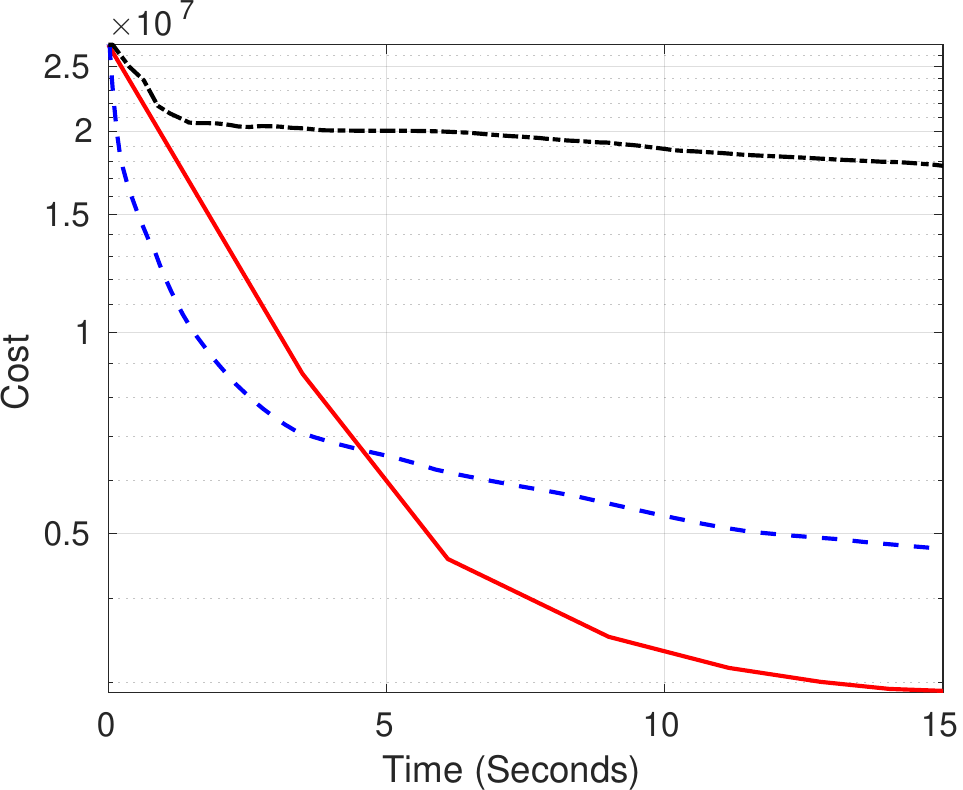}\label{fig:lplq:SR:bike:comp:GKSADMM:Time:l05l1}}	
	\caption{\MRcb{Comparison with GKS and ADMM on the image super-resolution task on the bike image with $p=0.5,1$.}}
	\label{fig:lplq:SR:bike:comp:GKSADMM}
\end{figure}
}

\subsection{Computed Tomography Reconstruction}
\label{sec:numericalExp:sub:CTReco}
\begin{figure}
    \centering
    \subfigure[]{\includegraphics[width=0.485\linewidth]{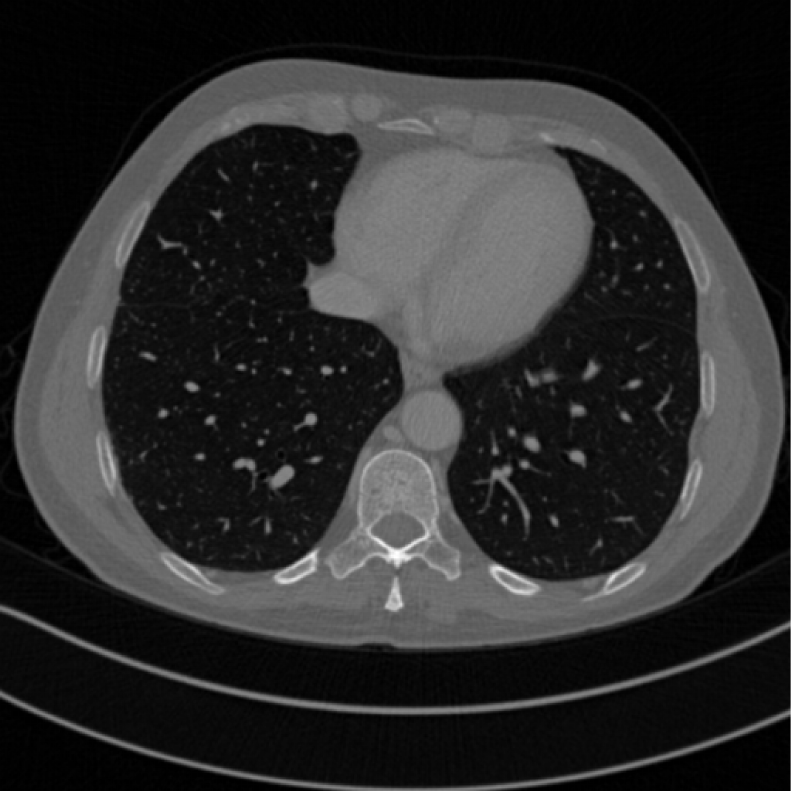}\label{fig:CT:GT:a}}
    \subfigure[]{\includegraphics[width=0.485\linewidth]{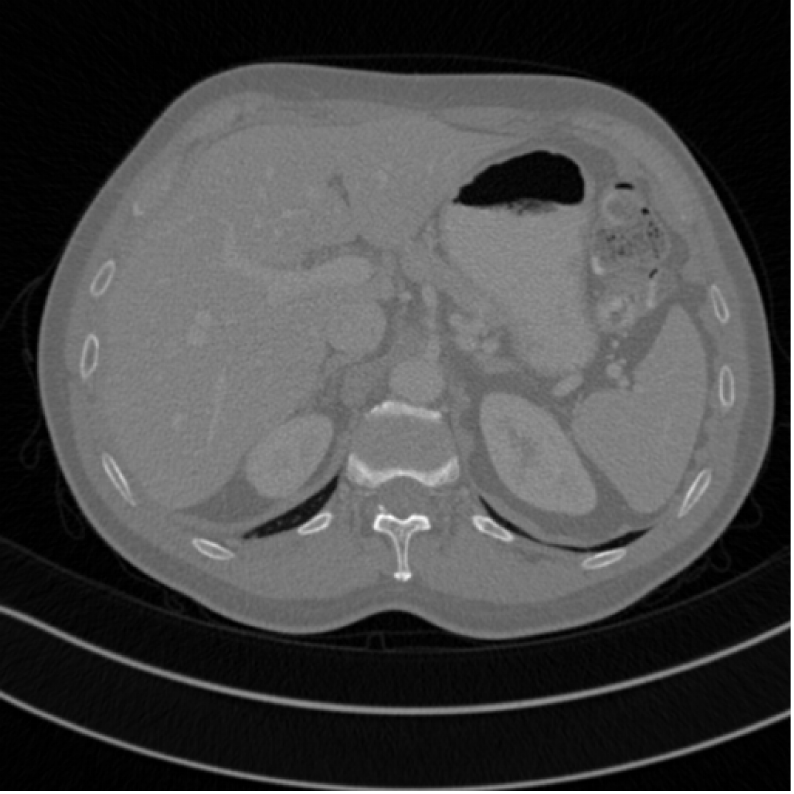}\label{fig:CT:GT:b}}
    \caption{Ground truth CT images.}
    \label{fig:CT:GT}
\end{figure}

We studied the performance of using \RNP for CT reconstruction
with three different regularizers: wavelet, TV with $\phi=1$, and HS$_\phi$ norm. \MRcb{Moreover, we also compared our method with the ADMM algorithm \cite{boyd2011distributed}.} For HS$_\phi$, we mainly studied $\phi=1$.
We used the ``daub4'' wavelet with $4$ levels
in the ``pywt'' \cite{wolter2024ptwt} toolbox.
The Operator Discretization Library (ODL) \cite{odl2017} was used for the CT forward model. 
\MRcbT{We investigated both parallel-beam and fan-beam acquisition geometries with $100$ projections. 
For both geometries, the reconstruction space was uniformly discretized on $[-20,20]\times[-20,20]\,\mathrm{cm}^2$ with a resolution of $512\times 512$. 
In the parallel-beam case, projection data were collected over uniformly spaced views in $[0,180^\circ]$, using a linear detector with $1024$ bins spanning $[-40,40]\,\mathrm{cm}$. 
In the fan-beam case, projection data were acquired over uniformly spaced views in $[0,360^\circ]$, with a linear detector of $1024$ bins covering $[-60,60]\,\mathrm{cm}$.} We scaled two slices from subject ``067'' in the AAPM CT Grand Challenge data \cite{AAPMct17}
to $[0,1]$ as the test images; see \Cref{fig:CT:GT}. \MRcbT{The projections were generated by applying the forward model and then adding Gaussian noise with zero mean and variance $10^{-4}$.} Due to the page limit,
the supplement shows the reconstruction results for \Cref{fig:CT:GT:b}
and for the fan-beam geometry.
Again we use ``w/o'' to denote results without using \RNP,
while ``w-$K$'' denotes \RNP with sketch size $K$.

\begin{figure}
	\centering
\includegraphics[width=0.45\textwidth]{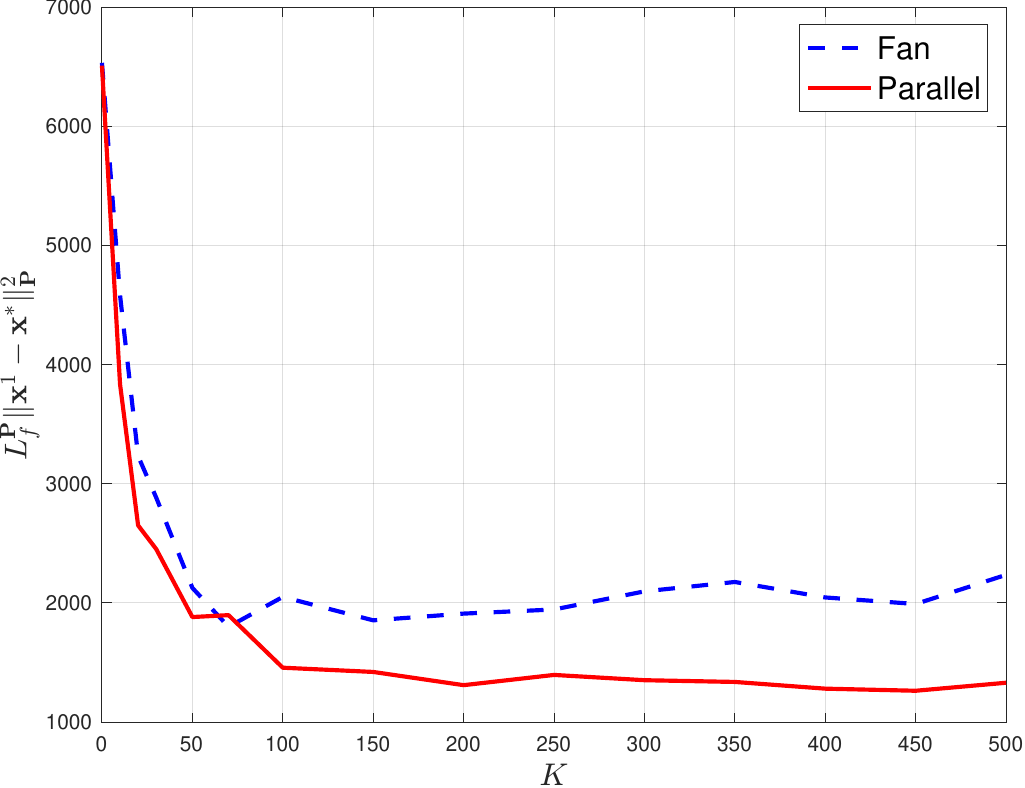}
    \caption{\MRcb{$\LfP\|\uvx^1-\uvx^*\|_{\umP}^2$ values versus different $K$ for fan-beam and parallel-beam acquisitions.}}
    \label{fig:CT:Lf:L}
\end{figure}

\Cref{tab:SKTime:CT:20} \MRcb{summarizes} the wall time for computing \RNP
for different acquisitions with various regularizers on both CPU and GPU computational platforms,
clearly highlighting the advantage of using the batch mode and the GPU computational platforms.
In general, a larger $K$ would yield a faster convergence in terms of iterations
than a small $K$. \MRcb{\Cref{fig:CT:Lf:L} presents the values of $\LfP\|\uvx^1-\uvx^*\|_{\umP}^2$ along the change of $K$ with fan-beam and parallel-beam acquisitions.  \Cref{fig:CT:Lf:L} shows that by using \RNP, we can significantly reduce $\LfP\|\uvx^1-\uvx^*\|_{\umP}^2$ that one can expect an acceleration. In this case, we observe that \RNP becomes less effective at reducing $\LfP\|\uvx^1-\uvx^*\|_{\umP}^2$ when $K>100$. Moreover, a larger $K$ may slightly increase $\LfP\|\uvx^1-\uvx^*\|_{\umP}^2$ but still remains effective at reducing it.}
In practice, large $K$ would also increase the computation of applying \RNP.
Here we set $K=20$ for wavelet and TV-based reconstruction,
and $K=100$ for HS$_1$-based reconstruction,
because these values worked well in practice.

\begin{figure*}[ht]
\vspace{-2.5cm}
	\centering
\begin{tikzpicture}
    \begin{axis}[at={(0,0)},anchor = north west,
    xmin = 0,xmax = 250,ymin = 0,ymax = 70,ylabel = Wavelet,width=0.95\textwidth,
        scale only axis,
        enlargelimits=false,
       axis line style={draw=none},
       tick style={draw=none},
        axis equal image,
        xticklabels={,,},yticklabels={,,},
        ylabel style={yshift=-0.3cm,xshift=-1.4cm},
       ]

    \node[inner sep=0.5pt, anchor = south west] (p1_1) at (0,0) {\includegraphics[ width=0.11\textwidth]{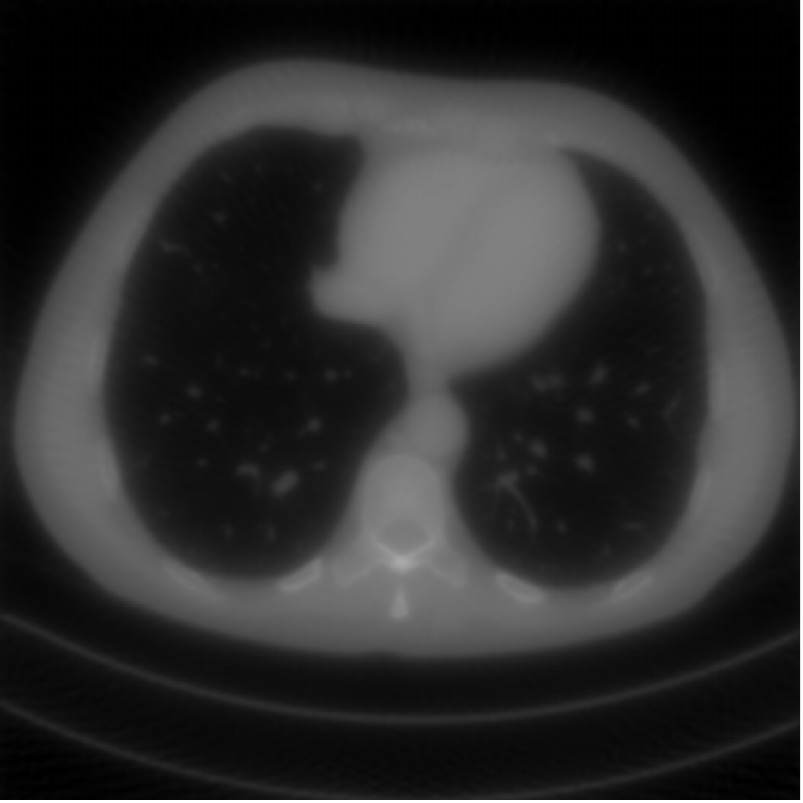}};
    
    \node at (26,26.5) {\color{white} $10$};
    
    \node at (5,3) {\color{white} w/o};
    \node at (24,3) {\color{white} $23.6$};
    
    
    \node[inner sep=0.5pt, anchor = west] (p1_2) at (p1_1.east) {\includegraphics[ width=0.11\textwidth]{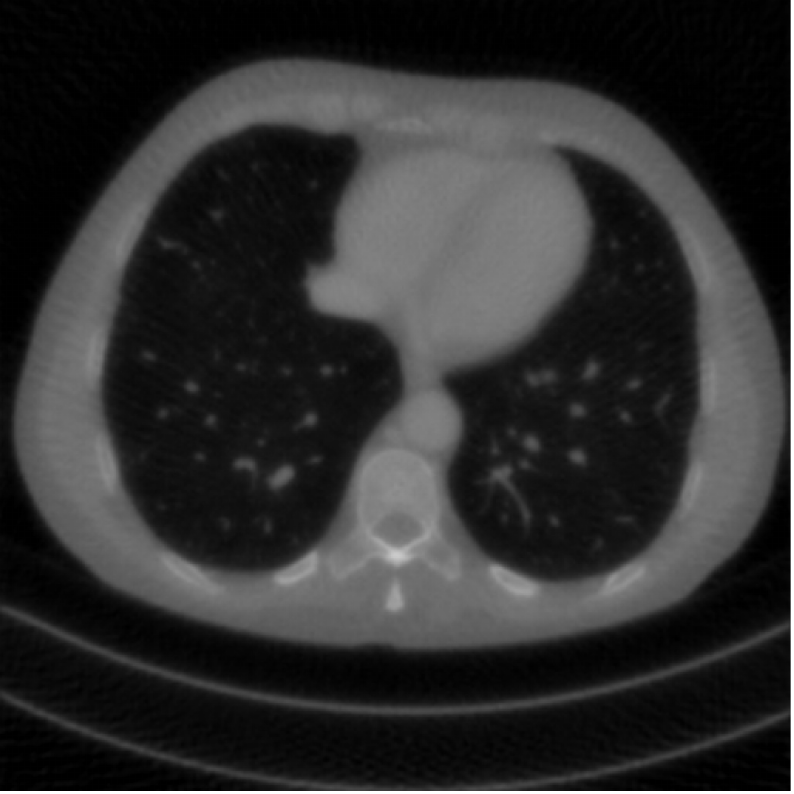}};

    
    \node at (54,3) {\color{white} $27.2$};
    \node at (33,3) {\color{white} w};
    
    \node[inner sep=0.5pt, anchor = west] (p1_3) at (p1_2.east) {\includegraphics[ width=0.11\textwidth]{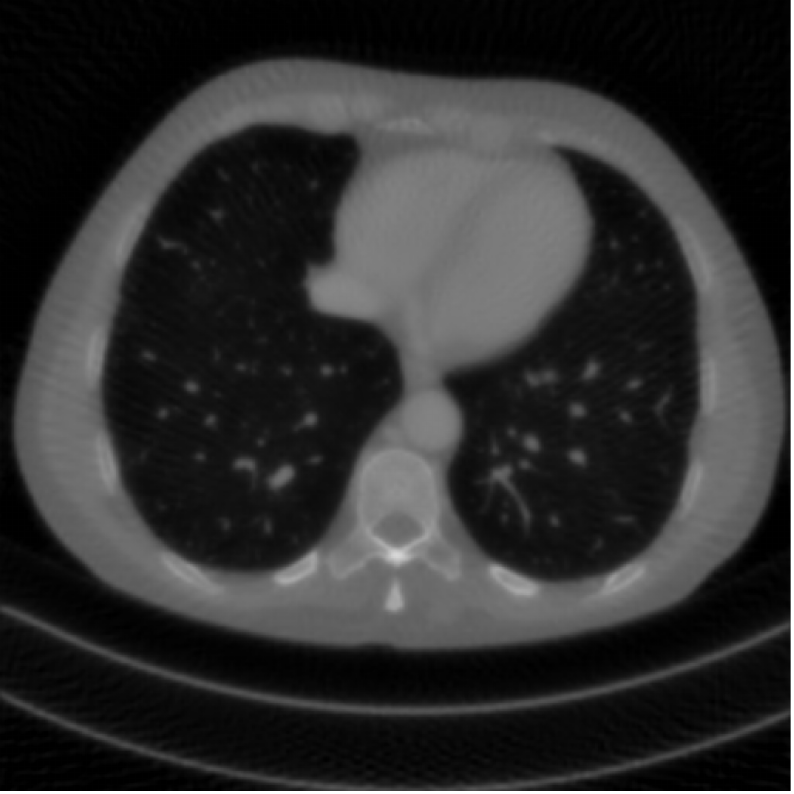}};

    \node at (85,26.5) {\color{white} $20$};
    
    \node at (64,3) {\color{white} w/o};
    
   \node at (83.5,3) {\color{white} $28.2$};
 \node[inner sep=0.5pt, anchor = west] (p1_4) at (p1_3.east) {\includegraphics[ width=0.11\textwidth]{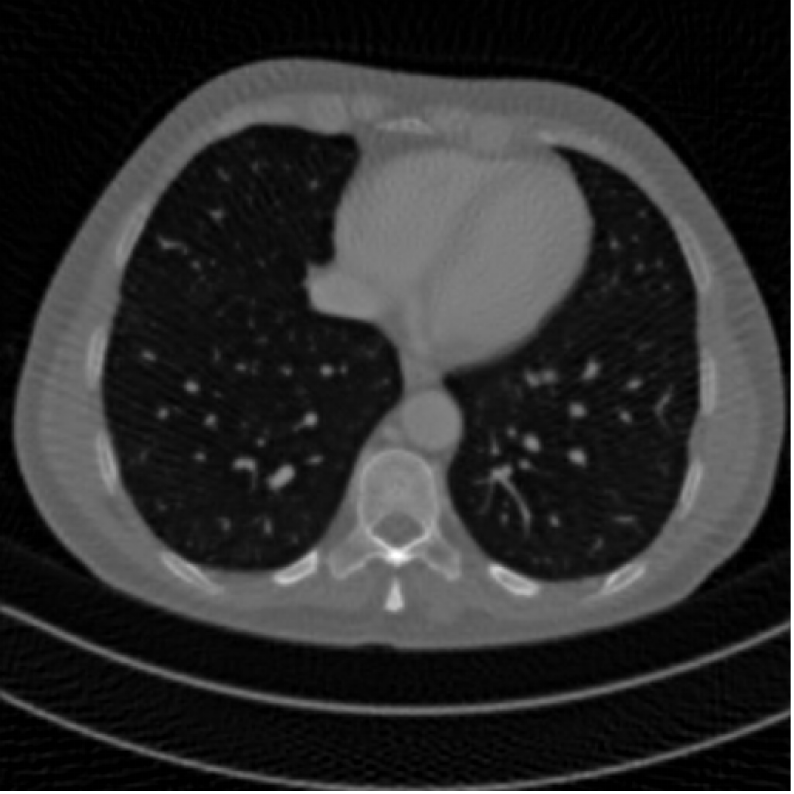}};

\node at (113,3) {\color{white} $32.3$};

  \node at (92,3) {\color{white} w};

 \node[inner sep=0.5pt, anchor = west] (p1_5) at (p1_4.east) {\includegraphics[ width=0.11\textwidth]{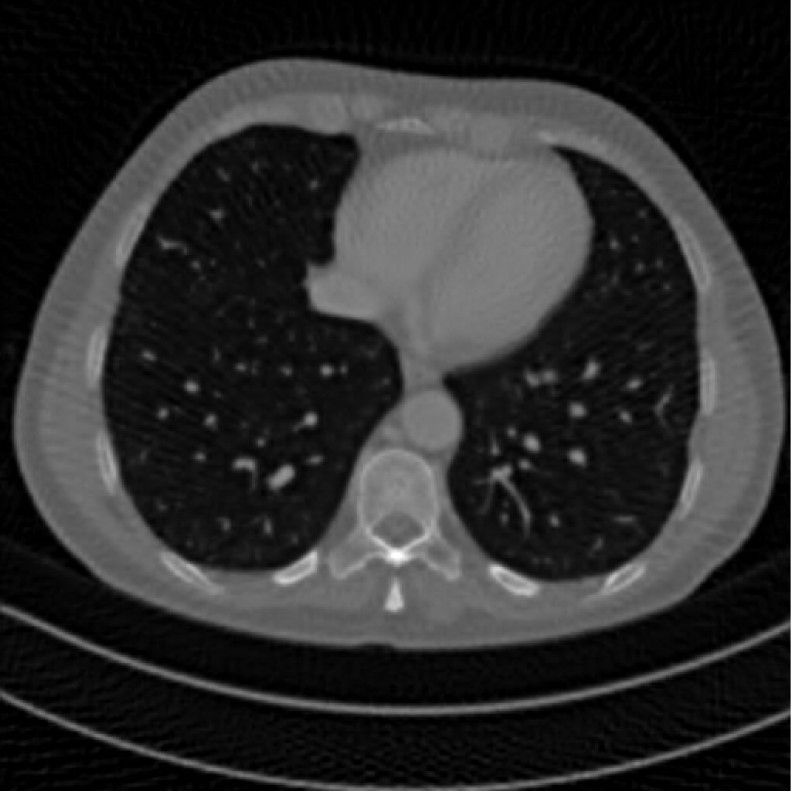}};

    \node at (145,26.5) {\color{white} $40$};
    
    \node at (123,3) {\color{white} w/o};
    
    \node at (143,3) {\color{white} $33.2$};
 \node[inner sep=0.5pt, anchor = west] (p1_6) at (p1_5.east) {\includegraphics[ width=0.11\textwidth]{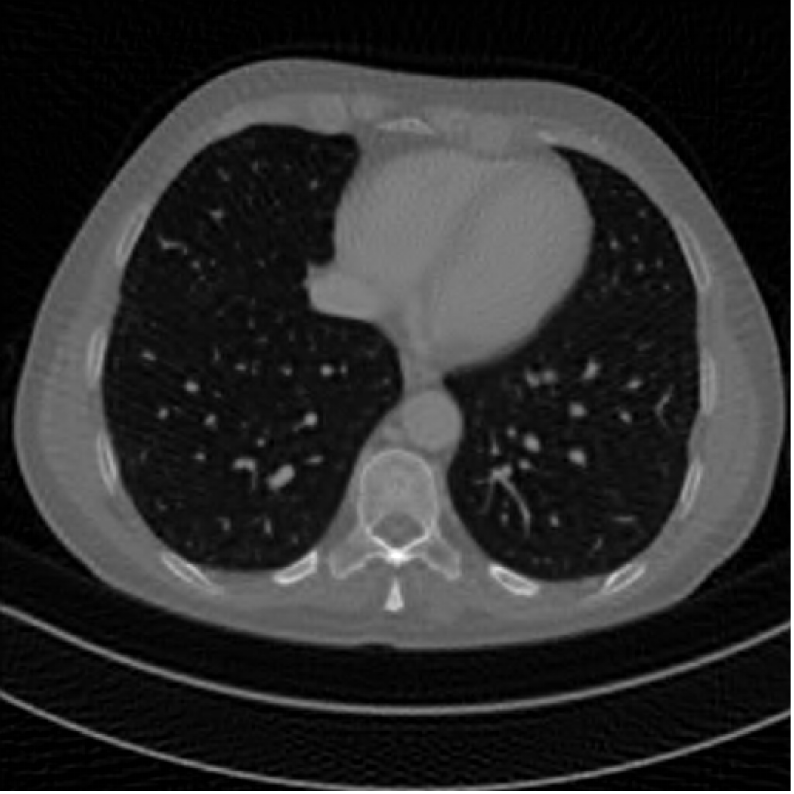}};

    \node at (151,3) {\color{white} w};
    
    \node at (172.5,3) {\color{white} $34.3$};

    
\node[inner sep=0.5pt, anchor = west] (p1_7) at (p1_6.east) {\includegraphics[ width=0.11\textwidth]{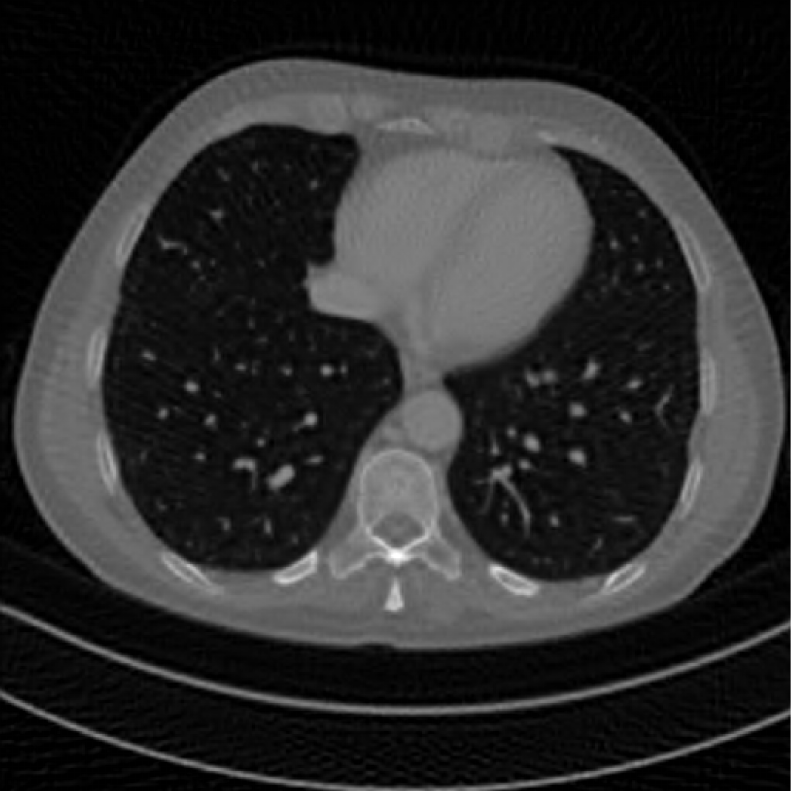}};

    \node at (204,26.5) {\color{white} $60$};
    
    \node at (183,3) {\color{white} w/o};
    
    \node at (201.5,3) {\color{white} $34.2$};

\node[inner sep=0.5pt, anchor = west] (p1_8) at (p1_7.east) {\includegraphics[ width=0.11\textwidth]{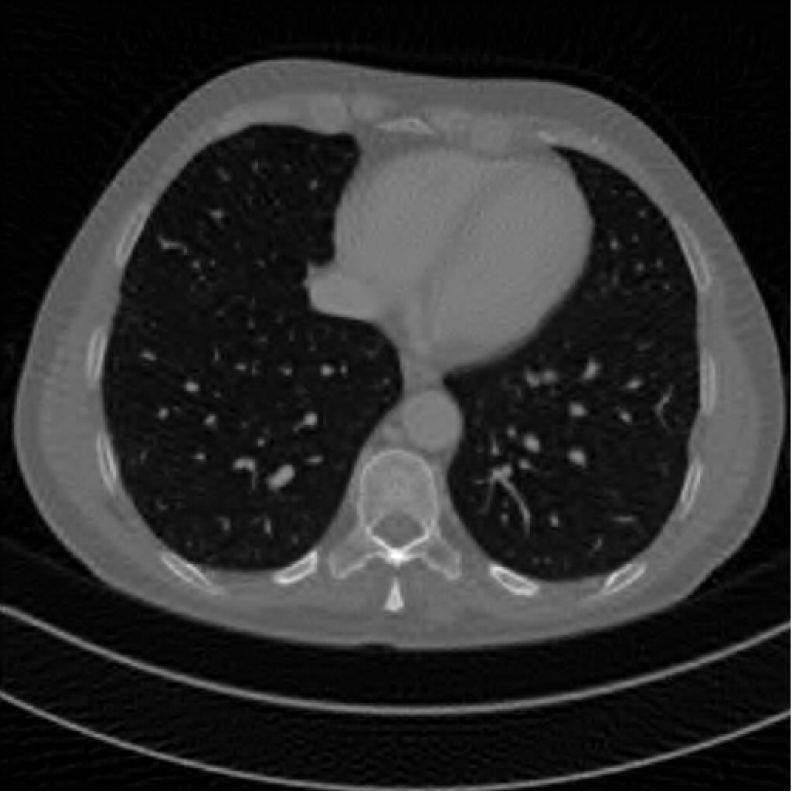}};

\node at (210,3) {\color{white} w};
    
\node at (231.5,3) {\color{white} $34.7$};


  

\end{axis}

 \begin{axis}[at={(p1_1.south west)},anchor = north west,
     xmin = 0,xmax = 250,ymin = 0,ymax = 70, width=0.95\textwidth,ylabel = TV,
         scale only axis,
         enlargelimits=false,
         yshift=2.8cm,
        axis line style={draw=none},
        tick style={draw=none},
         axis equal image,
         xticklabels={,,},yticklabels={,,},
         ylabel style={yshift=-0.3cm,xshift=-1.4cm},
        ]
        
   \node[inner sep=0.5pt, anchor = south west] (TVp1_1) at (0,0) {\includegraphics[width=0.11\textwidth]{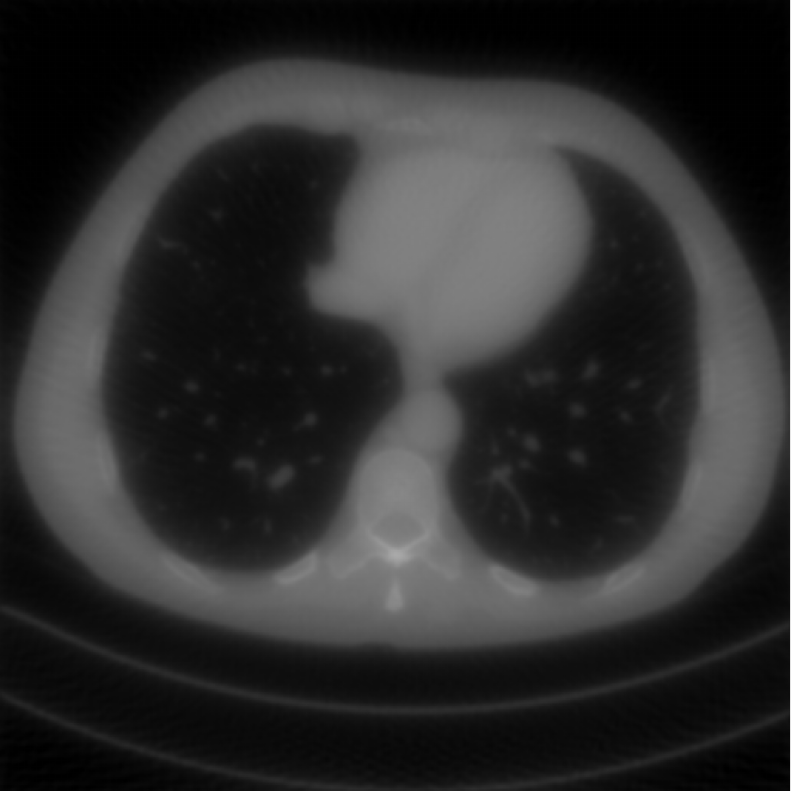}};

    \node at (24,3) {\color{white} $23.6$};
    
    \node[inner sep=0.5pt, anchor = west] (TVp1_2) at (TVp1_1.east) {\includegraphics[ width=0.11\textwidth]{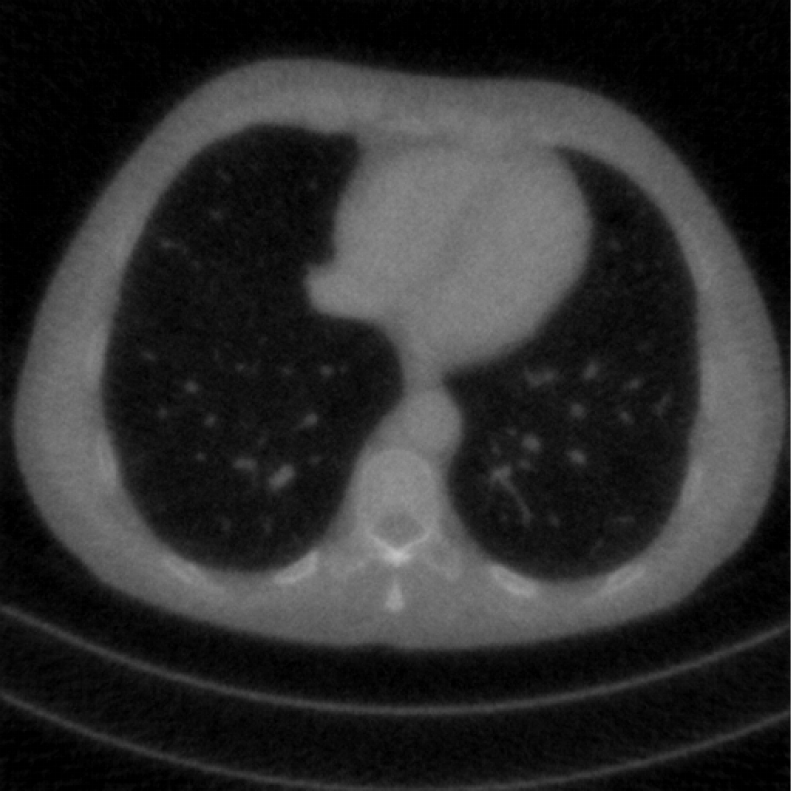}};
 
    \node at (54,3) {\color{white} $25.9$};
    
    \node[inner sep=0.5pt, anchor = west] (TVp1_3) at (TVp1_2.east) {\includegraphics[ width=0.11\textwidth]{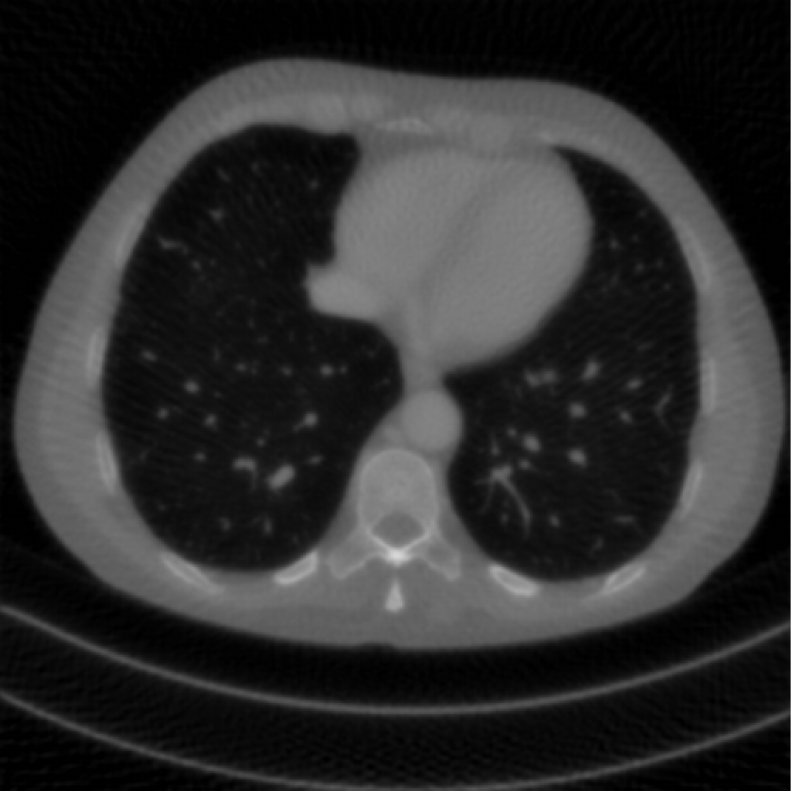}};

    \node at (83.5,3) {\color{white} $28.4$};
 \node[inner sep=0.5pt, anchor = west] (TVp1_4) at (TVp1_3.east) {\includegraphics[ width=0.11\textwidth]{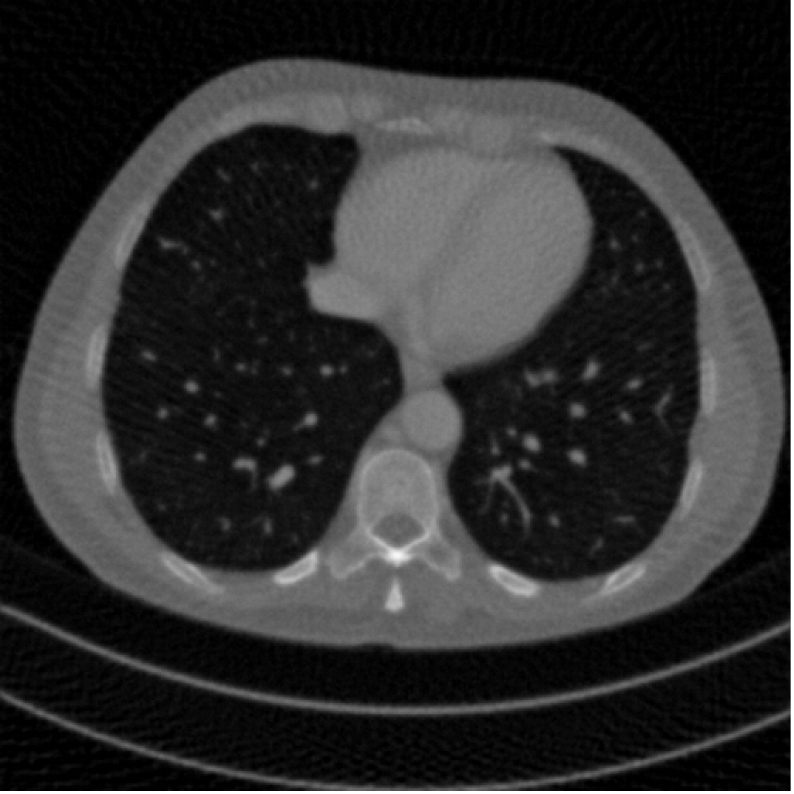}};

    
 \node at (113,3) {\color{white} $31.2$};

 \node[inner sep=0.5pt, anchor = west] (TVp1_5) at (TVp1_4.east) {\includegraphics[ width=0.11\textwidth]{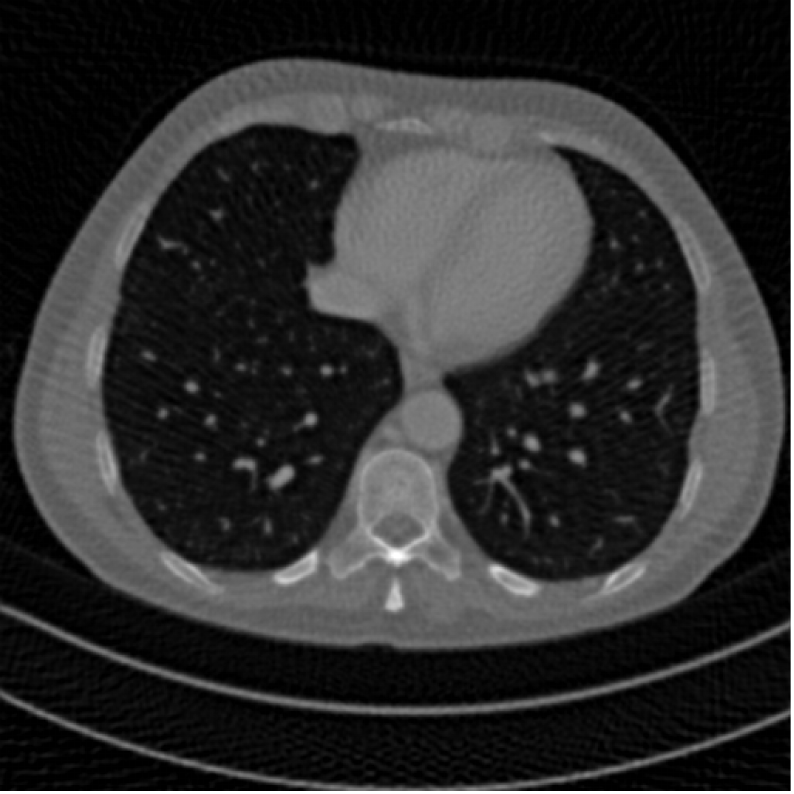}};


     \node at (143,3) {\color{white} $33.2$};
 \node[inner sep=0.5pt, anchor = west] (TVp1_6) at (TVp1_5.east) {\includegraphics[ width=0.11\textwidth]{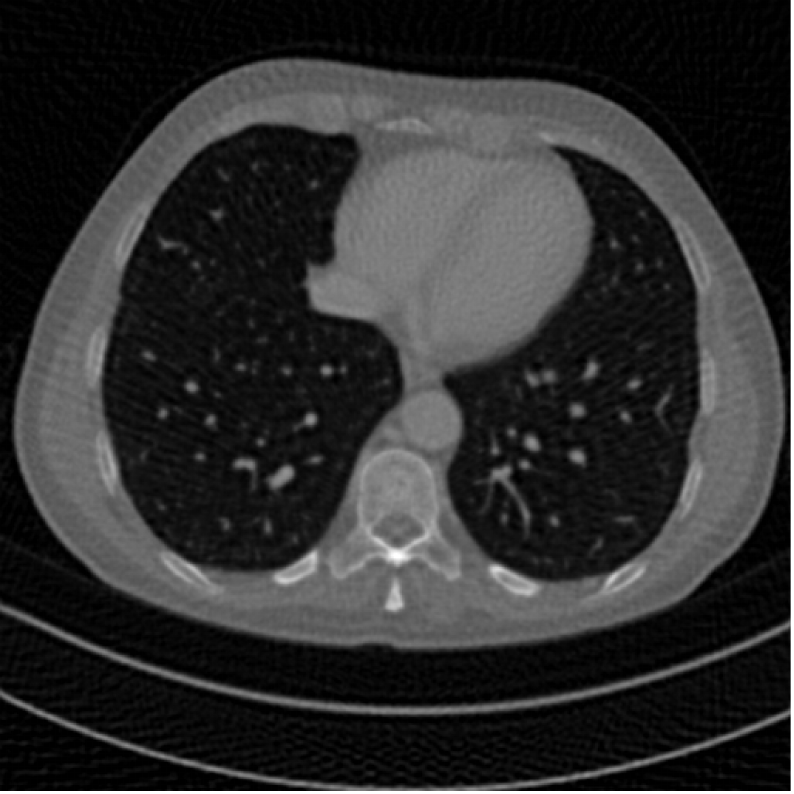}};


      \node at (171.5,3) {\color{white} $33.6$};
\node[inner sep=0.5pt, anchor = west] (TVp1_7) at (TVp1_6.east) {\includegraphics[ width=0.11\textwidth]{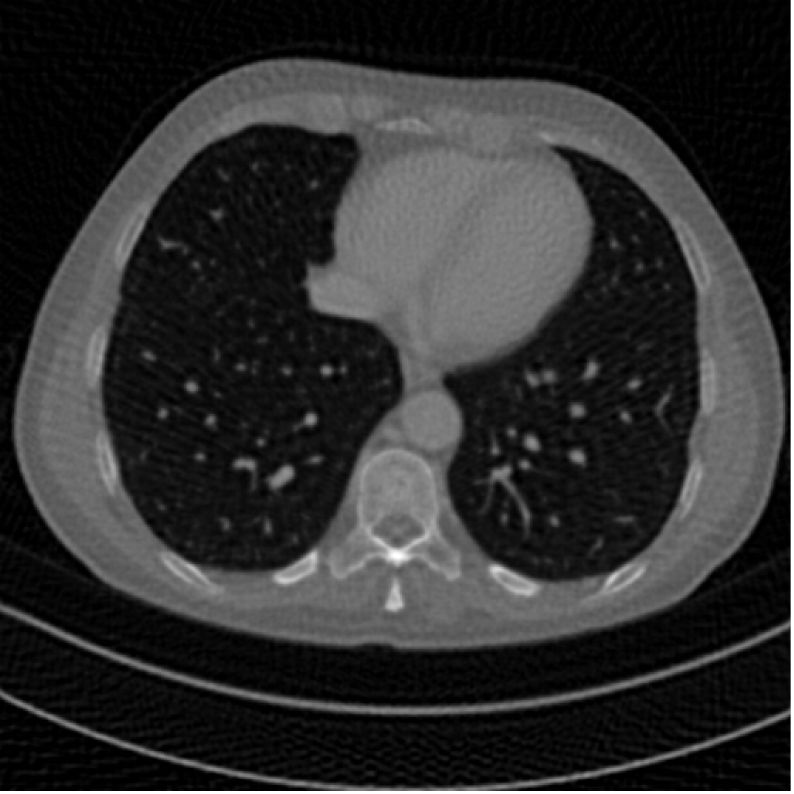}};

 \node at (201.5,3) {\color{white} $33.7$};
\node[inner sep=0.5pt, anchor = west] (TVp1_8) at (TVp1_7.east) {\includegraphics[ width=0.11\textwidth]{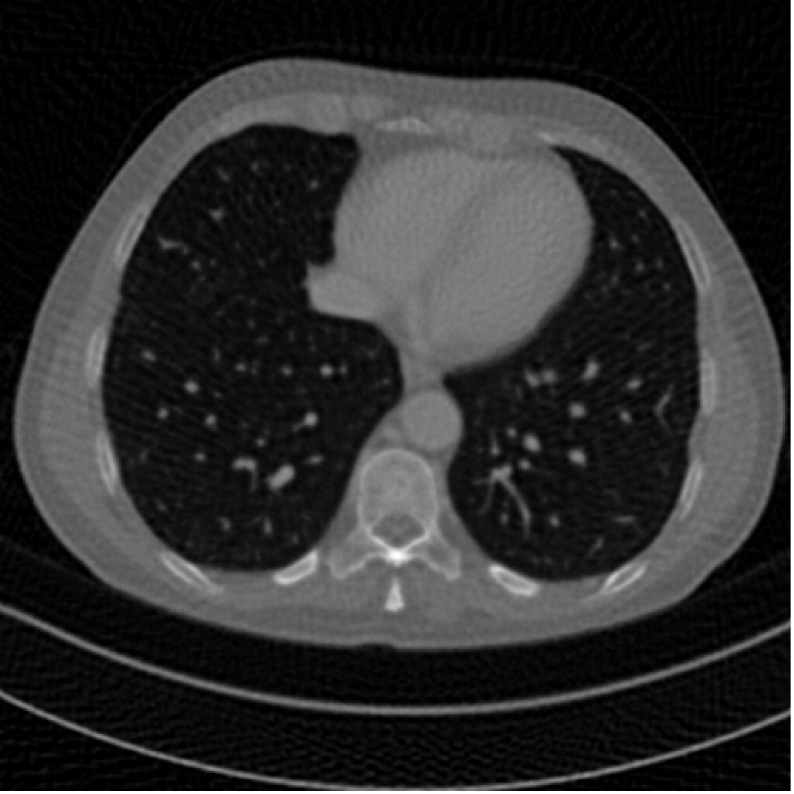}};

\node at (231.5,3) {\color{white} $33.7$};
 \end{axis}

 \begin{axis}[at={(TVp1_1.south west)},anchor = north west,
     xmin = 0,xmax = 250,ymin = 0,ymax = 70, width=0.95\textwidth,ylabel = HS$_1$,
         scale only axis,
         enlargelimits=false,
         yshift=2.8cm,
        axis line style={draw=none},
        tick style={draw=none},
         axis equal image,
         xticklabels={,,},yticklabels={,,},
         ylabel style={yshift=-0.3cm,xshift=-1.4cm},
        ]
        
   \node[inner sep=0.5pt, anchor = south west] (HSp1_1) at (0,0) {\includegraphics[width=0.11\textwidth]{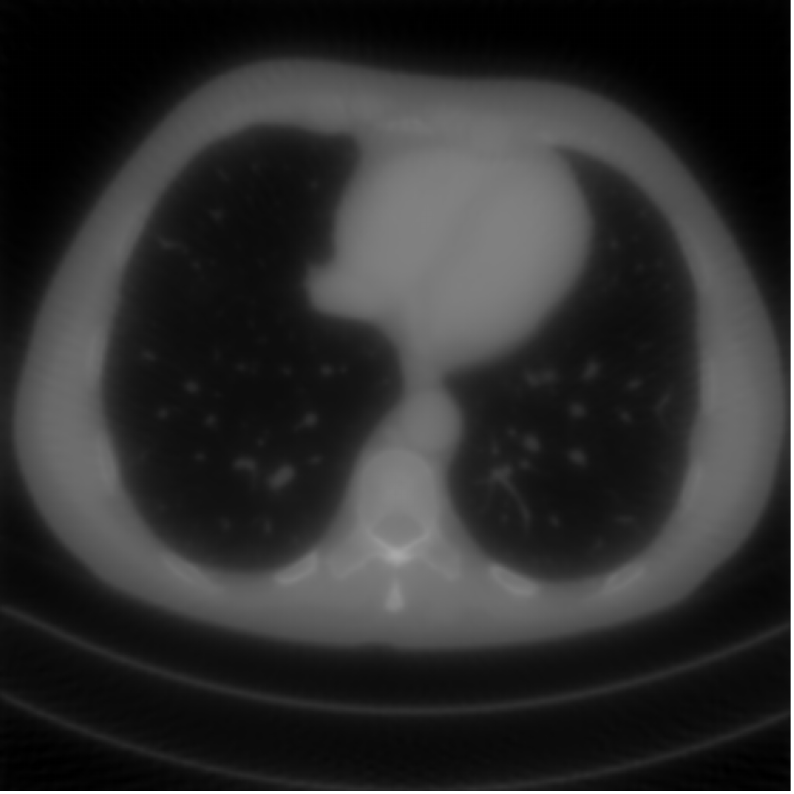}};

     \node at (24,3) {\color{white} $23.6$};
    
    \node[inner sep=0.5pt, anchor = west] (HSp1_2) at (HSp1_1.east) {\includegraphics[ width=0.11\textwidth]{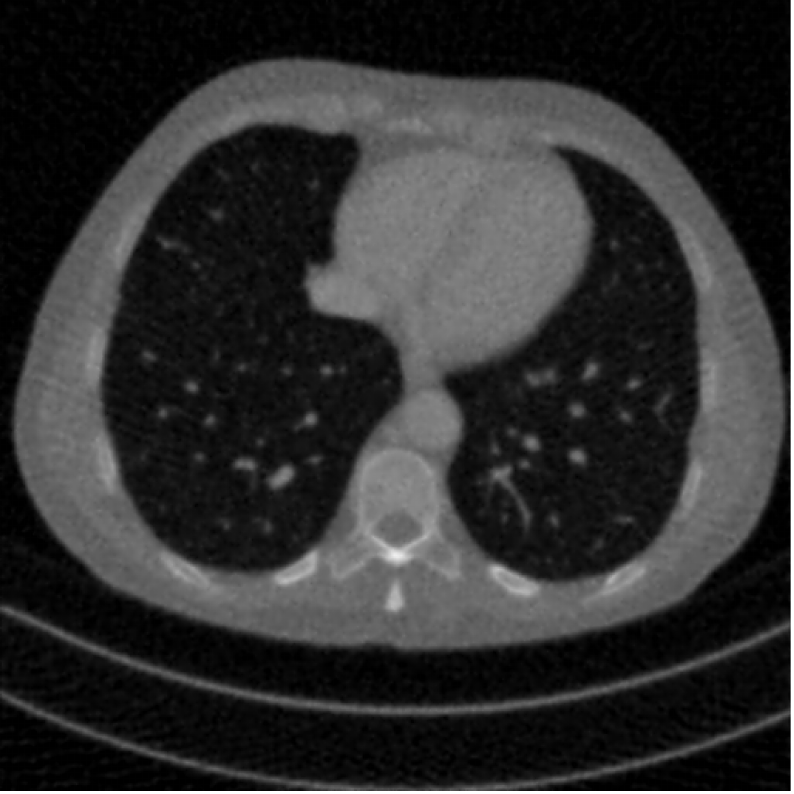}};
 
    \node at (53,3) {\color{white} $29.7$};
  
    \node[inner sep=0.5pt, anchor = west] (HSp1_3) at (HSp1_2.east) {\includegraphics[ width=0.11\textwidth]{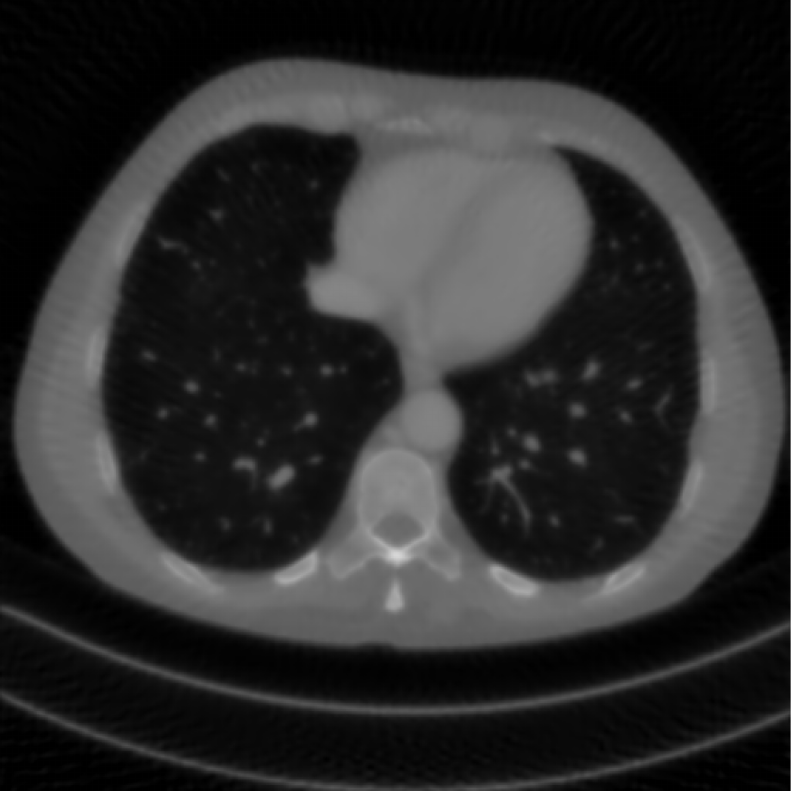}};

    \node at (82.5,3) {\color{white} $28.2$};

 \node[inner sep=0.5pt, anchor = west] (HSp1_4) at (HSp1_3.east) {\includegraphics[ width=0.11\textwidth]{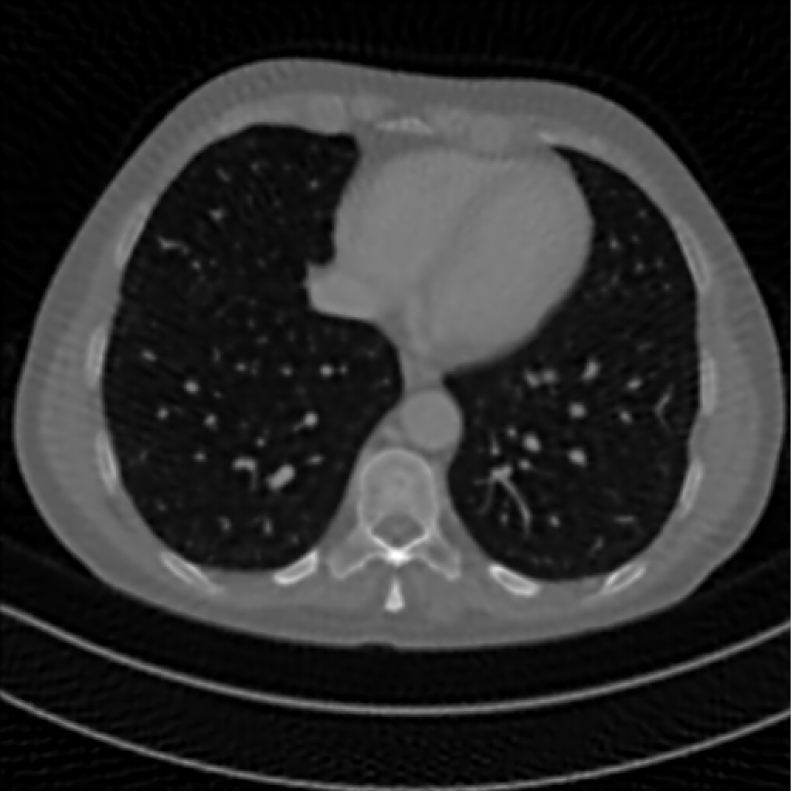}};

  \node at (113,3) {\color{white} $33.9$};
    

 \node[inner sep=0.5pt, anchor = west] (HSp1_5) at (HSp1_4.east) {\includegraphics[ width=0.11\textwidth]{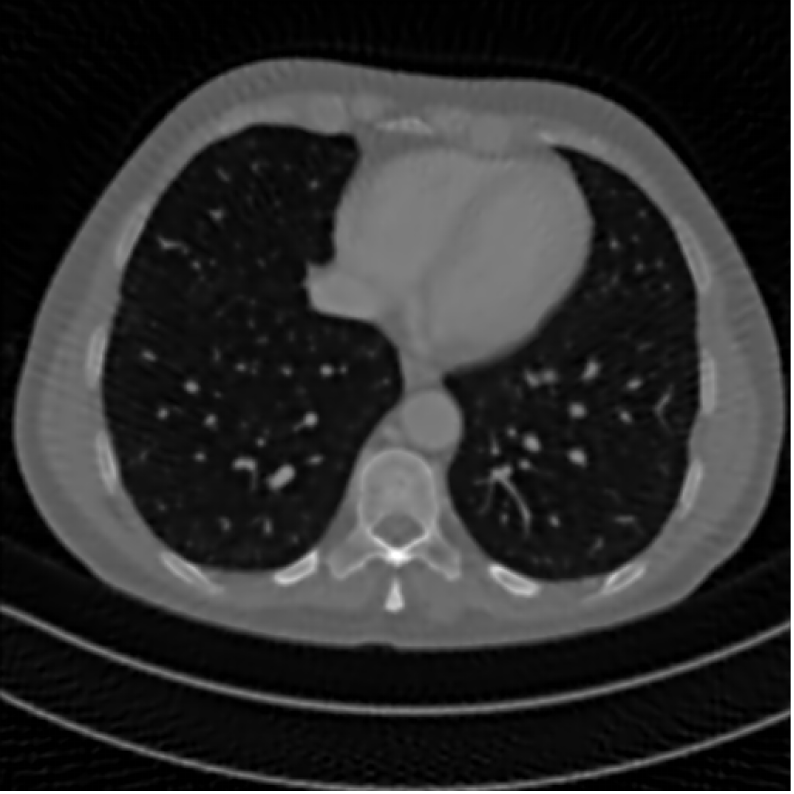}};

     \node at (142,3) {\color{white} $33.4$};
 \node[inner sep=0.5pt, anchor = west] (HSp1_6) at (HSp1_5.east) {\includegraphics[ width=0.11\textwidth]{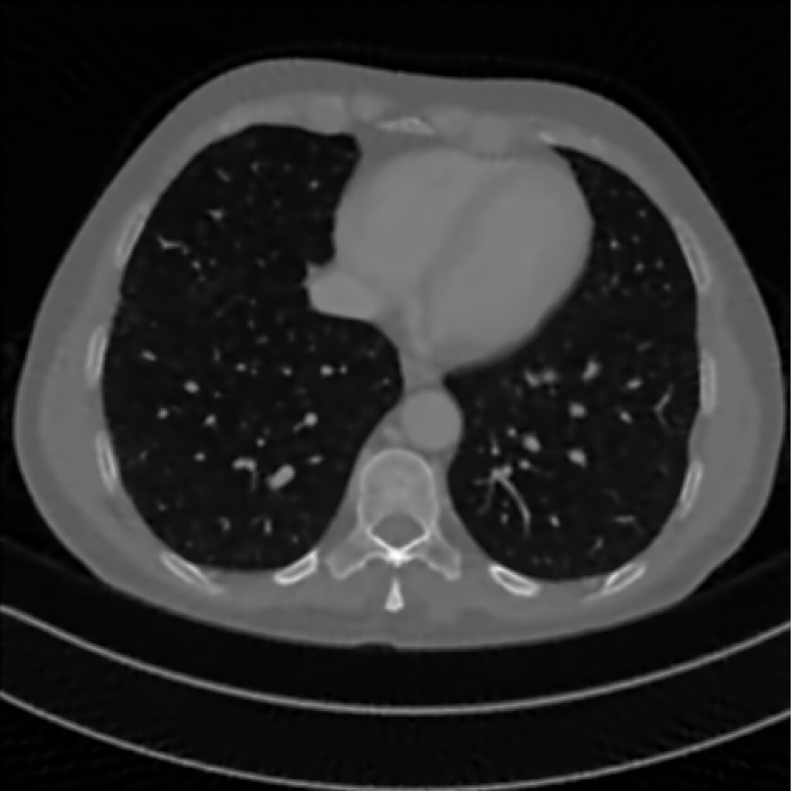}};

    \node at (171.5,3) {\color{white} $36.2$};

\node[inner sep=0.5pt, anchor = west] (HSp1_7) at (HSp1_6.east) {\includegraphics[ width=0.11\textwidth]{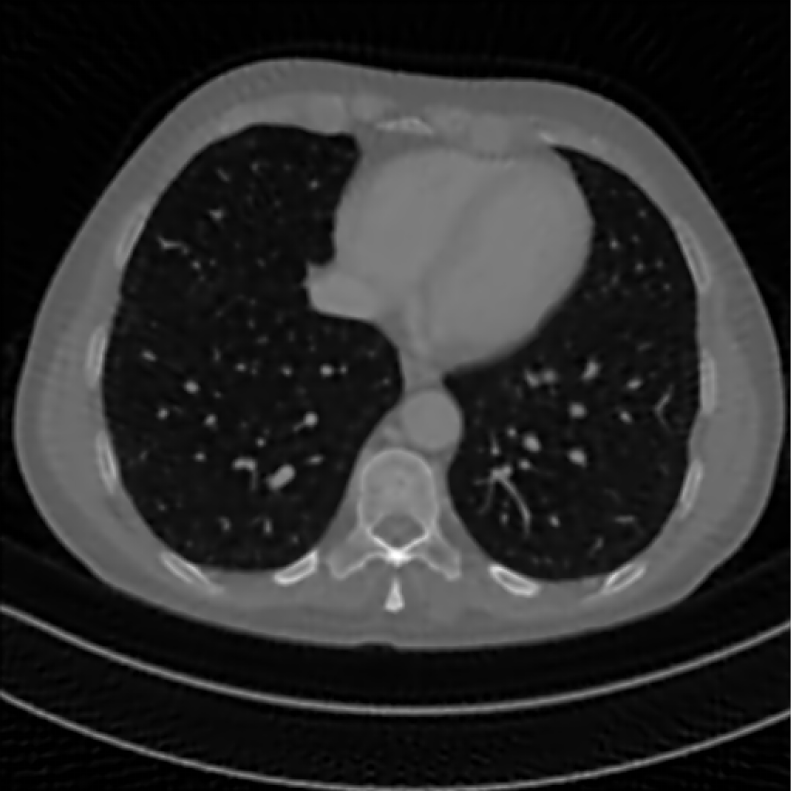}};

\node at (200.5,3) {\color{white} $35.1$};
\node[inner sep=0.5pt, anchor = west] (HSp1_8) at (HSp1_7.east) {\includegraphics[ width=0.11\textwidth]{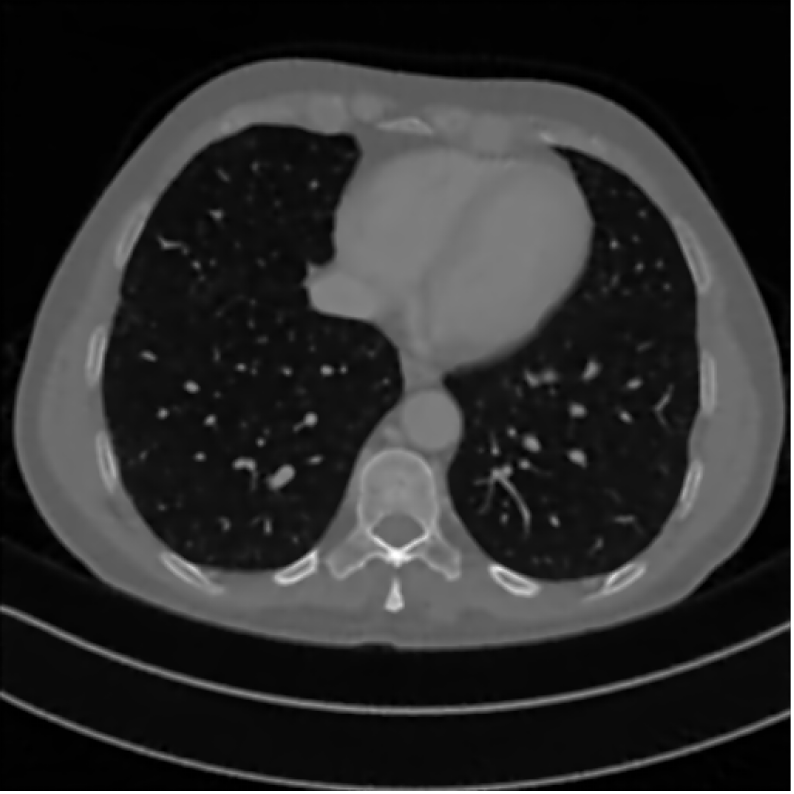}};

\node at (231.5,3) {\color{white} $37.0$};
 \end{axis}
 
\end{tikzpicture} 
\caption{
The parallel-beam CT reconstructed images with wavelet, TV, and HS$_1$ based regularization
at iterations $10$, $20$, $40$, and $60$.
Columns $1$, $3$, $5$, and $7$ (respectively, $2$, $4$, $6$, and $8$)
show the reconstructions without (respectively, with) \RNP.
The associated PSNR values are listed at the right corner of each image.
}
\label{fig:CTParallel:RecoImages}
\end{figure*}

\begin{figure}
    \centering
    \subfigure[]{\includegraphics[width=0.485\linewidth]{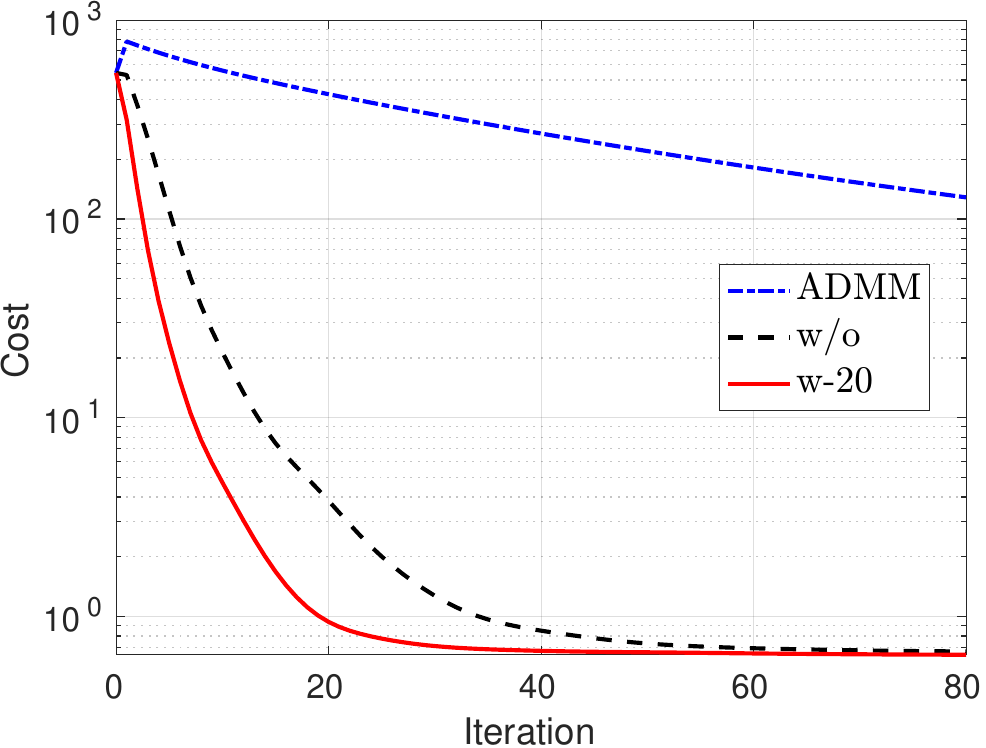}\label{fig:CTParallel:Wav:lossiter}}
    \subfigure[]{\includegraphics[width=0.485\linewidth]{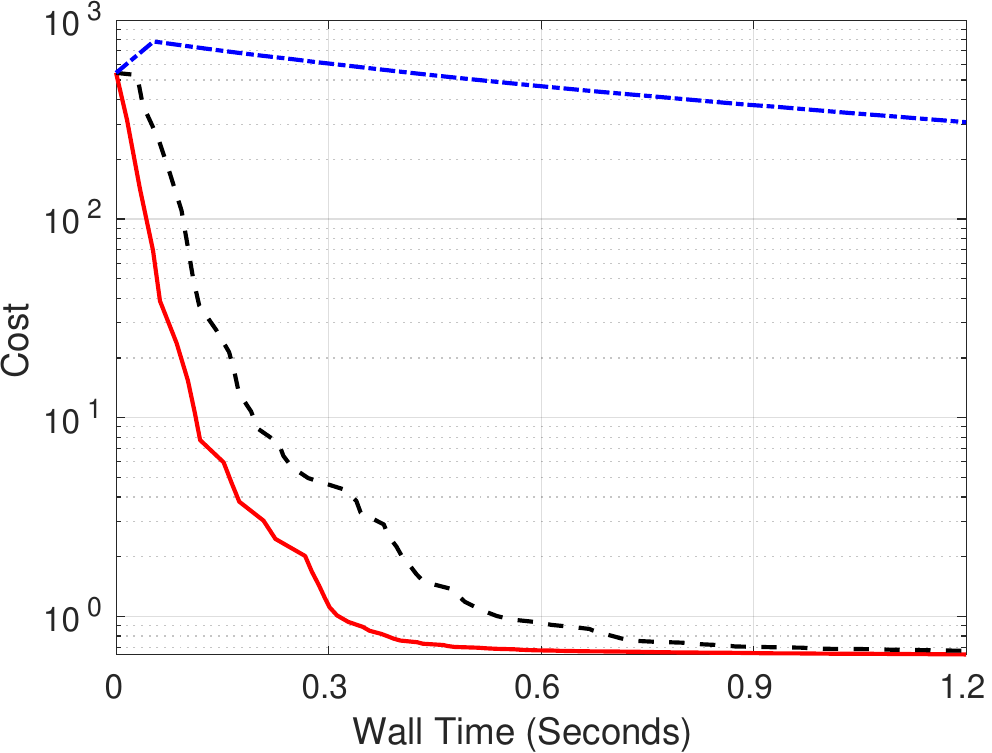}\label{fig:CTParallel:Wav:losstime}}
    
    \subfigure[]{\includegraphics[width=0.48\linewidth]{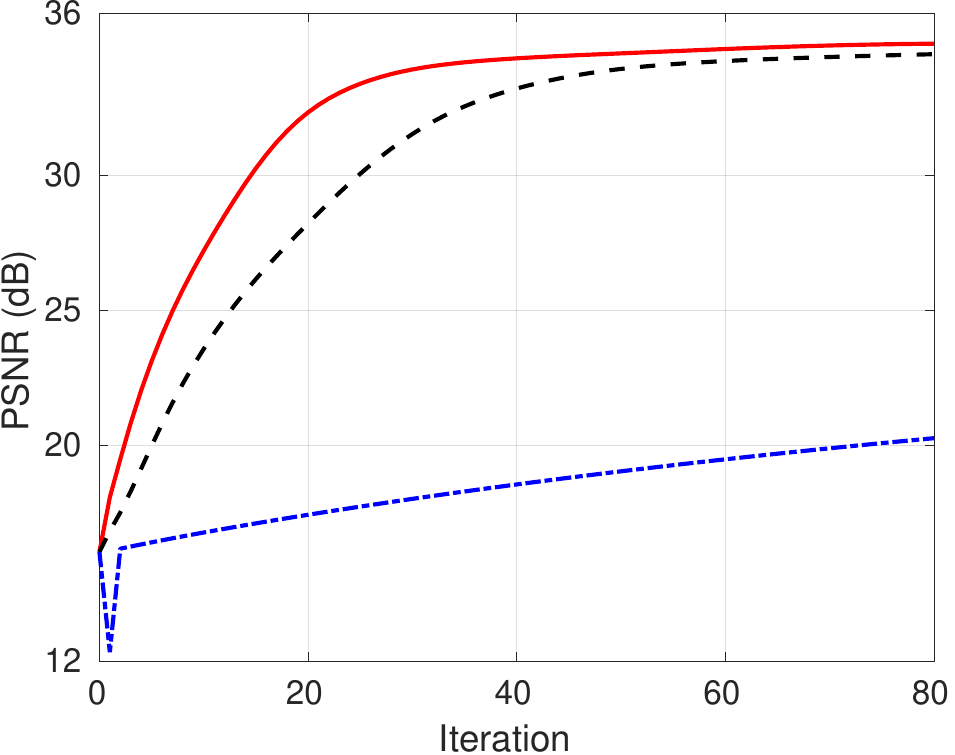}\label{fig:CTParallel:Wav:psnriter}}
    \subfigure[]{\includegraphics[width=0.49\linewidth]{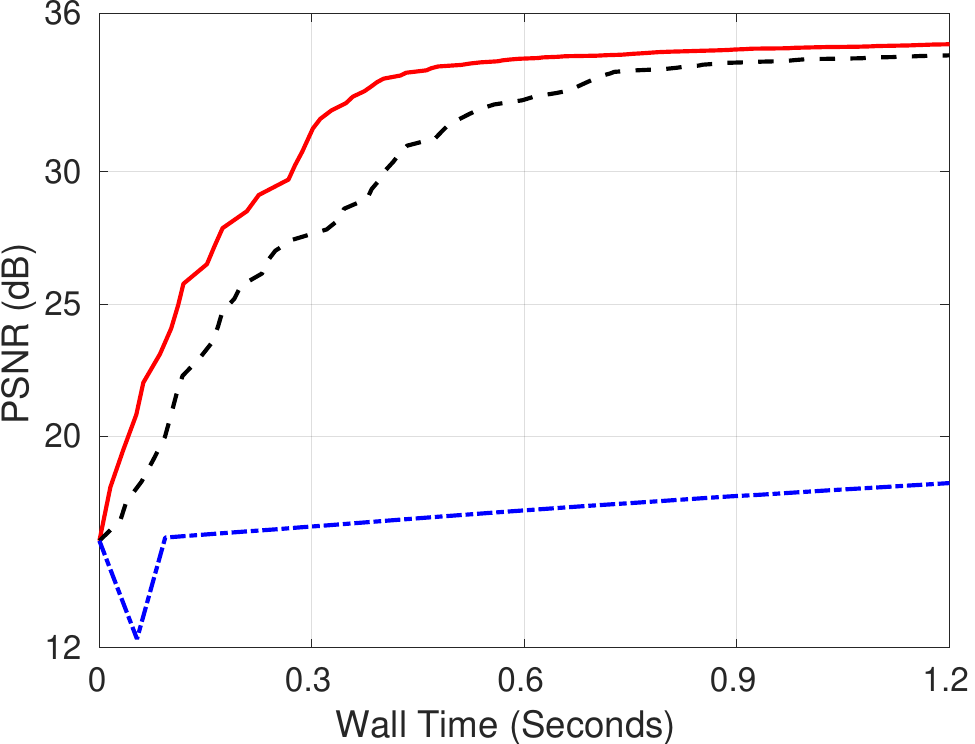}\label{fig:CTParallel:Wav:psnrtime}}
    \caption{Comparison of using \RNP ~\MRcb{and ADMM} for wavelet based CT reconstruction. w/o denotes the one without  \RNP.}
    \label{fig:CTParallel:Wav}
\end{figure}

\begin{figure}
    \centering
    \subfigure[]{\includegraphics[width=0.485\linewidth]{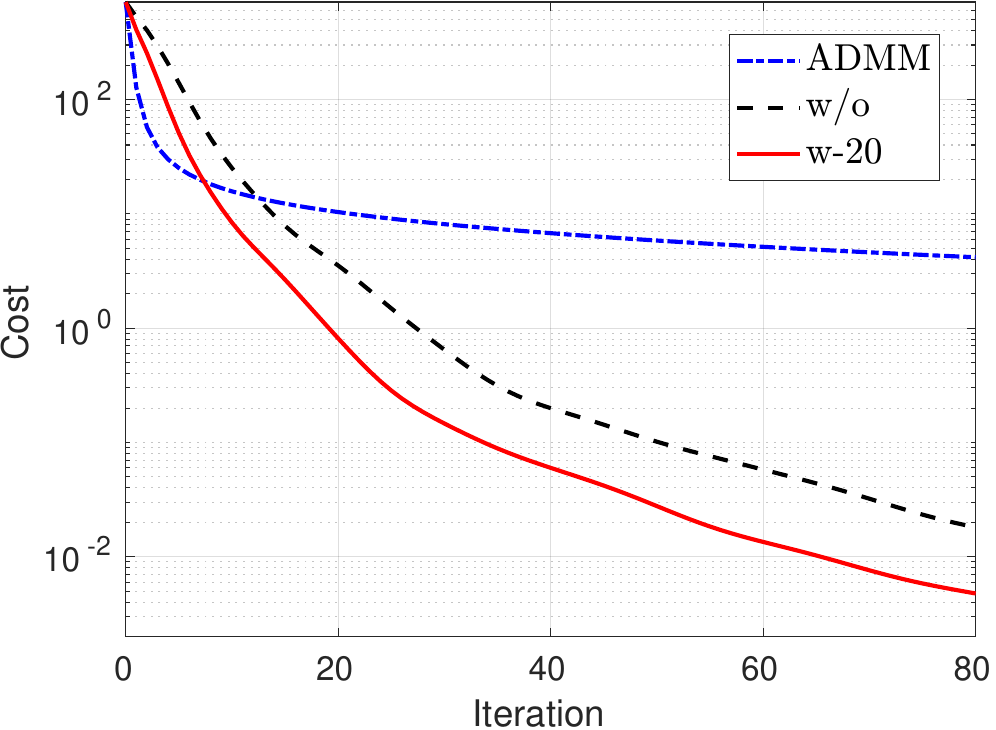}}
    \subfigure[]{\includegraphics[width=0.485\linewidth]{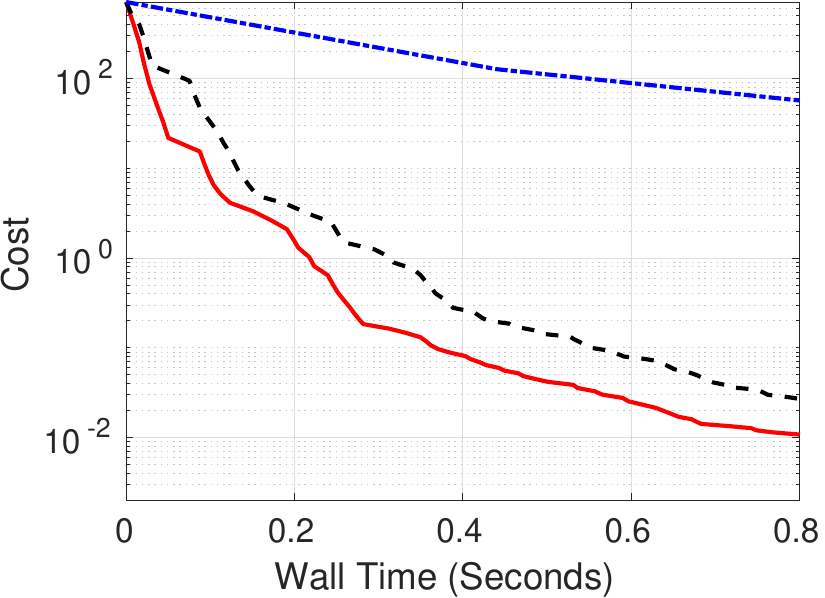}}
    
    \subfigure[]{\includegraphics[width=0.48\linewidth]{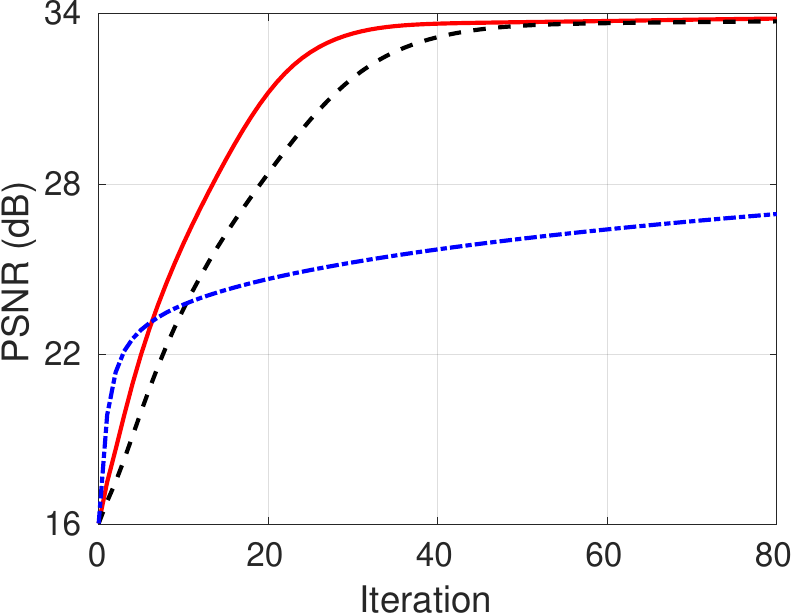}}
    \subfigure[]{\includegraphics[width=0.49\linewidth]{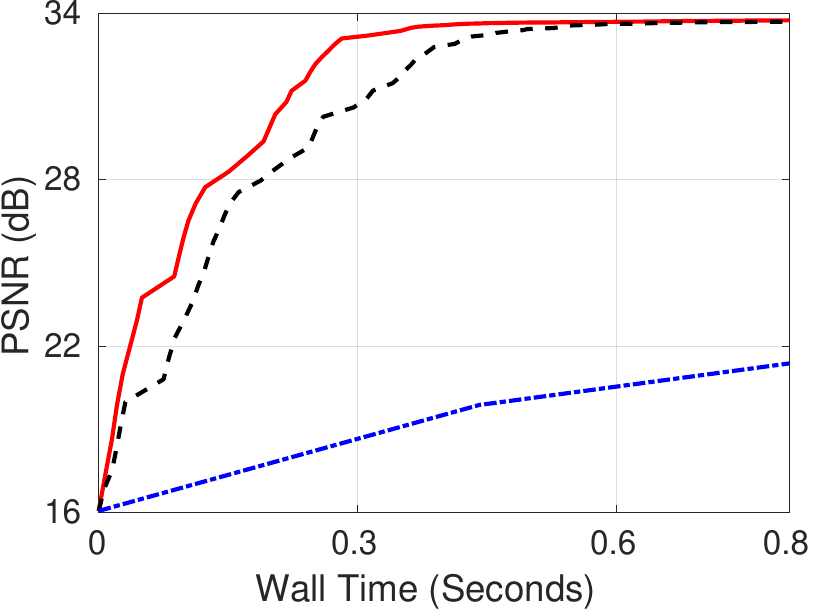}}

    \caption{Comparison of using \RNP ~\MRcb{and ADMM} for TV based CT reconstruction. w/o denotes the one without  \RNP.}
    \label{fig:CTParallel:TV}
\end{figure}

\Cref{fig:CTParallel:Wav} presents the comparison of using \RNP for wavelet based CT reconstruction.
\Cref{fig:CTParallel:Wav:lossiter,fig:CTParallel:Wav:psnriter}
shows that w-$20$ converged faster than w/o in terms of iterations,
illustrating the effectiveness of using \RNP.
Moreover,
\Cref{fig:CTParallel:Wav:losstime,fig:CTParallel:Wav:psnrtime}
shows that w-$20$ converged faster than w/o in terms of wall time,
demonstrating the wall time
for solving the related WPM with \Cref{them:structuredWPM:evaluation} is negligible.
\Cref{fig:CTParallel:TV} presents the results of TV-based reconstruction
that shows similar trends as in wavelet-based regularization. \MRcb{Obviously, ADMM is the slowest algorithm in these comparisons.}
\Cref{fig:CTParallel:RecoImages} illustrates
the reconstructed images at iterations $10$, $20$, $40$, and $60$,
for both w-$K$ and w/o.


\begin{figure}[t]
    \centering
    \subfigure[]{\includegraphics[width=0.485\linewidth]{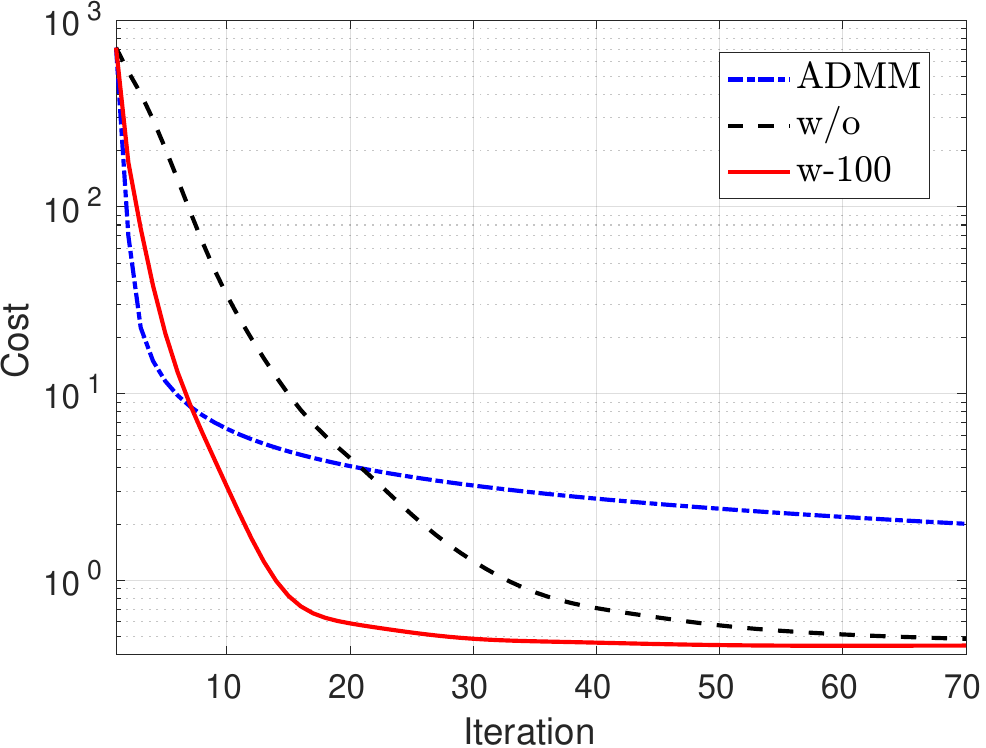}}
    \subfigure[]{\includegraphics[width=0.485\linewidth]{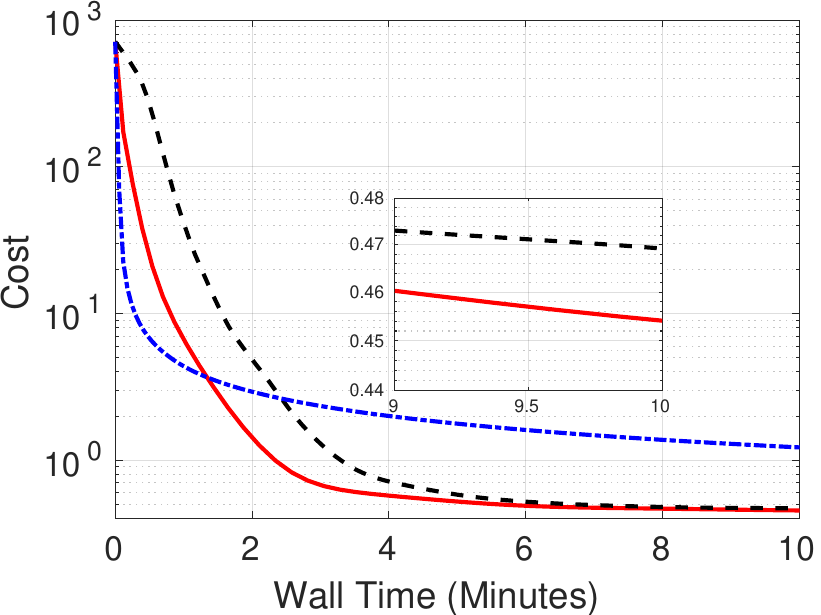}}
    
    \subfigure[]{\includegraphics[width=0.485\linewidth]{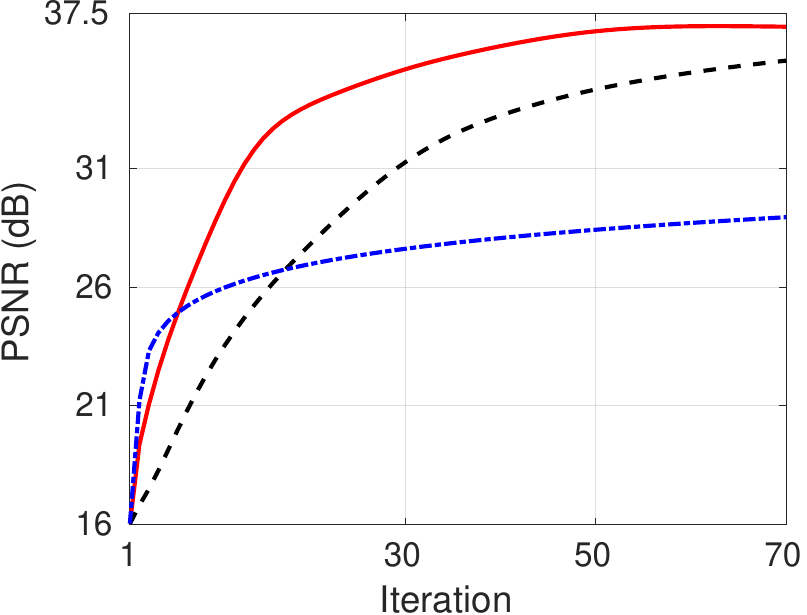}}
    \subfigure[]{\includegraphics[width=0.49\linewidth]{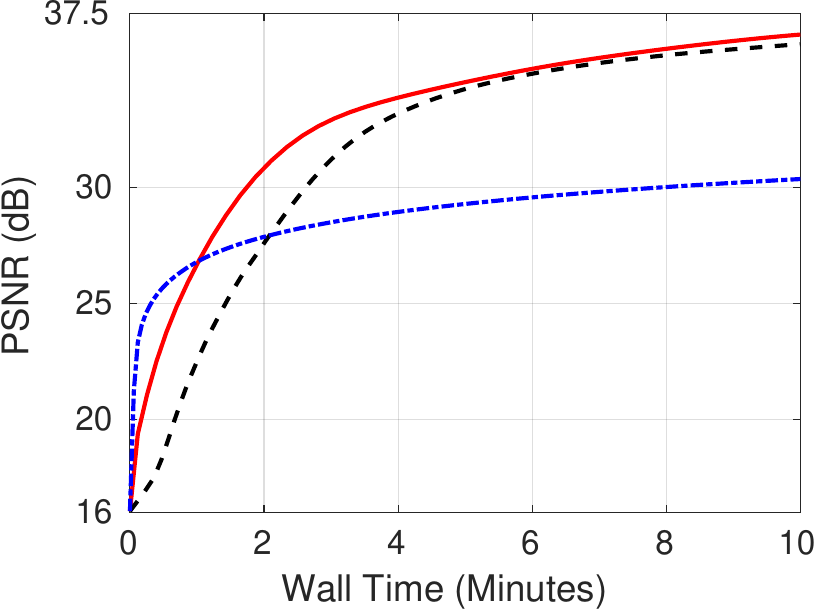}}

    \caption{Comparison of using \RNP ~\MRcb{and ADMM} for HS$_1$ based CT reconstruction. w/o denotes reconstruction  without \RNP.}
    \label{fig:CTParallel:HS1}
\end{figure}

Since TV based reconstruction will introduce blocking artifacts \cite{lefkimmiatis2013hessian},
we included the study of HS norm based reconstruction.
\Cref{fig:CTParallel:HS1} shows the results of HS$_1$ based reconstruction,
where we observed that using \RNP accelerated the convergence compared to not using \RNP.
Moreover, we observed that HS$_1$ yielded a higher PSNR than wavelet and TV regularizers.
The supplementary material reports the performance of HS$_2$ and HS$_\infty$ based reconstruction
that trended similar to HS$_1$.




\subsubsection{\MRcb{Comparison of IRM and GKS}}
\label{sec:numericalExps:sub:CT:sub:CompIRM_GKS}
\MRcb{We compared the performance of WAPG using \RNP with \IRM and GKS for wavelet-based parallel-beam CT reconstruction. \Cref{fig:CT:Wav:IRM:GKS} presents the cost values versus the number of iterations and wall time for different methods.
For \IRM and GKS, the $\ell_1$ norm is smoothed, but the cost values presented here are still computed with the $\ell_1$ norm. We observed that \IRM is the fastest algorithm in terms of the number of iterations. However, since \IRM requires higher computation per iteration, we found that w-$20$ is the fastest algorithm in terms of wall time.  Compared to \IRM, GKS significantly reduces the computational cost per iteration, which is also validated in our experiment. However, we observed that w-$20$ is still faster than GKS in terms of wall time in our setting.

\begin{figure}[t]
	\centering
	\subfigure[]{
\includegraphics[width=0.485\linewidth]{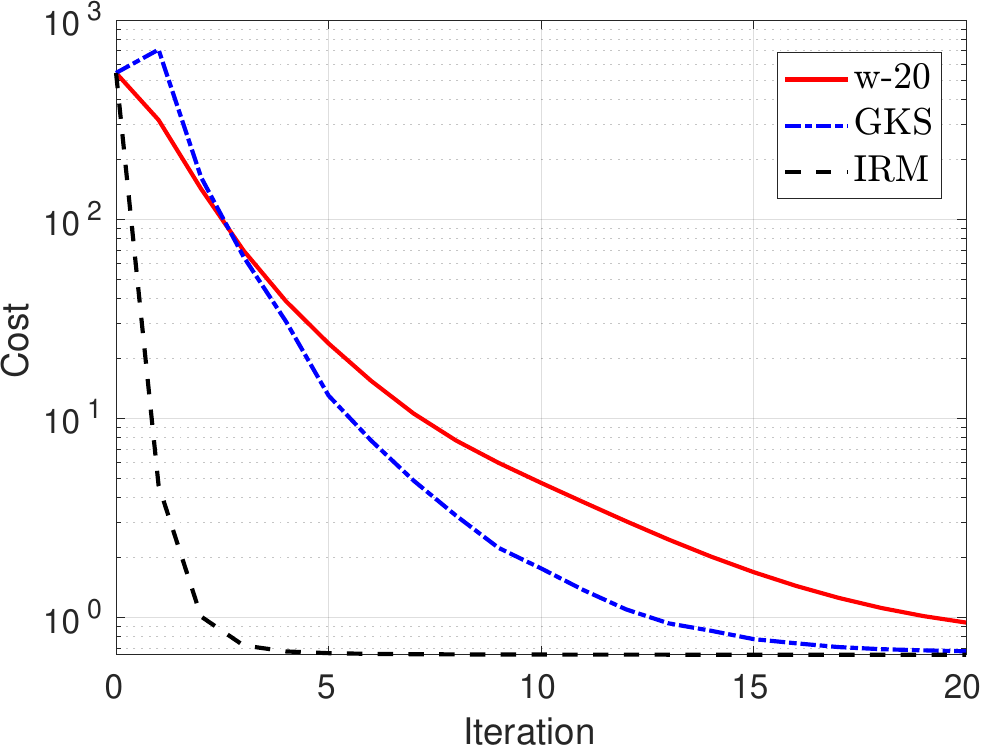}}
\subfigure[]{\includegraphics[width=0.485\linewidth]{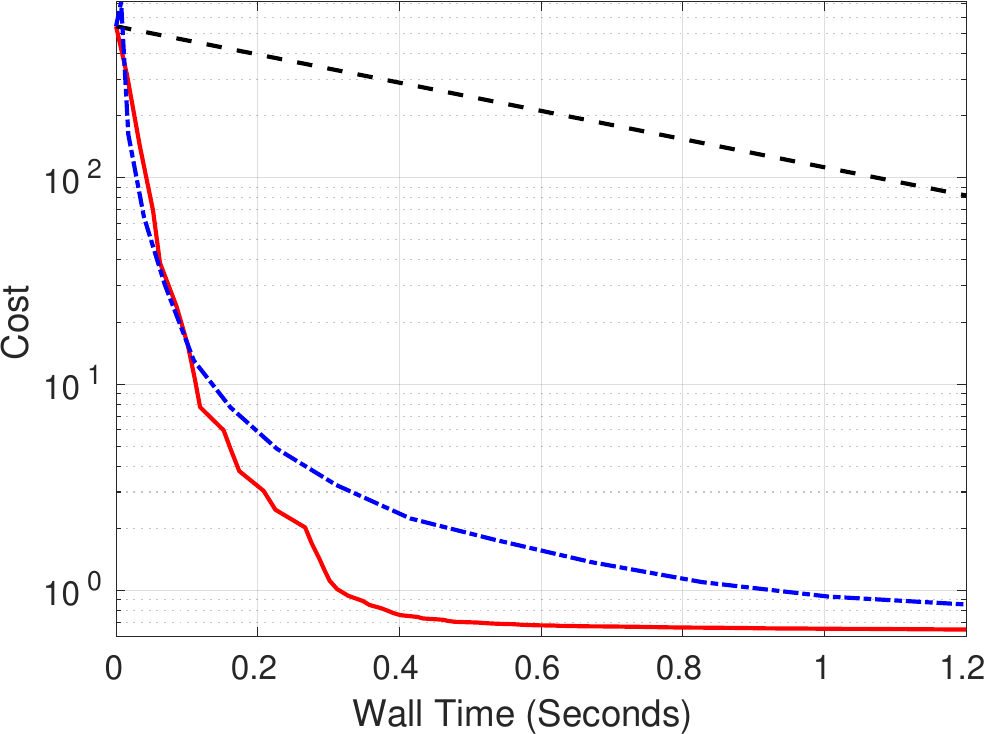}}
\caption{\MRcb{Comparison of using \RNP , IRM, and GKS for wavelet based CT reconstruction.}}\label{fig:CT:Wav:IRM:GKS}
\end{figure}

}

\subsection{\MRcbT{Computed Tomography Reconstruction with Real Data}}
\label{sec:numericalExp:sub:RealCTReco}
\MRcbT{We used the publicly available walnut CT dataset
provided by the Finnish Inverse Problems Society,
which was acquired with a fan-beam geometry \cite{hamalainen2015tomographic}.
The dataset includes a given matrix $\umA$
and the corresponding $120$ projection measurements $\uvy$.
We evaluated the effectiveness of \RNP for image reconstruction
by solving \eqref{eq:lpl1qMixedNorm} with TV-based regularization,
with the sketch size $K=100$.
In this experiment, we subsampled the data and used only $40$ projections.
\Cref{fig:realCostIterTime} shows the cost values
versus the number of iterations and wall time,
demonstrating our approach converged faster than others.
\Cref{fig:CTReal:RecoImages} presents the reconstructed images.
Using \RNP recovered sharper details than APM without \RNP at the same number of iterations,
illustrating the effectiveness of \RNP.
Because walnut data did not include the ground truth image,
PSNR values are not reported. 

\begin{figure}
	\centering
	\subfigure[]{\includegraphics[width=0.485\linewidth]{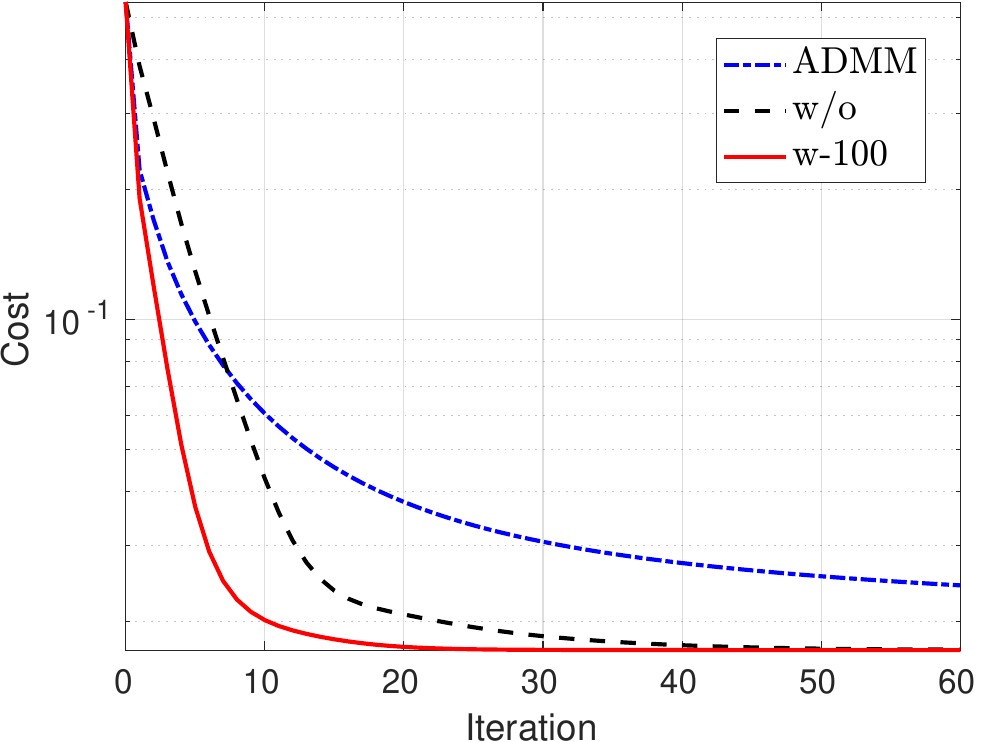}}
	\subfigure[]{\includegraphics[width=0.485\linewidth]{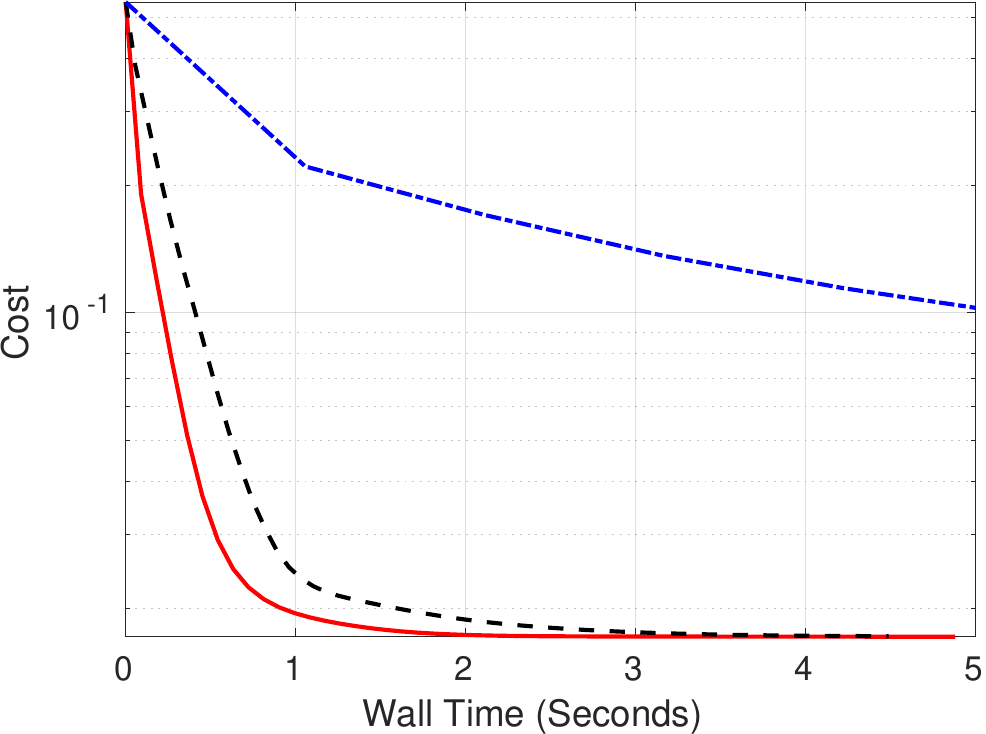}}
	\caption{\MRcbT{Comparison of using \RNP and ADMM for TV based reconstruction on the walnut data. w/o denotes reconstruction without \RNP.}}
	\label{fig:realCostIterTime}
\end{figure}

\begin{figure*}[ht]
\hspace{0.6cm}
\begin{tikzpicture}
    \begin{axis}[at={(0,0)},anchor = north west,
    xmin = 0,xmax = 250,ymin = 0,ymax = 70,width=0.95\textwidth,
        scale only axis,
        enlargelimits=false,
       axis line style={draw=none},
       tick style={draw=none},
        axis equal image,
        xticklabels={,,},yticklabels={,,},
        ylabel style={yshift=-0.3cm,xshift=-1.4cm},
       ]

    \node[inner sep=0.5pt, anchor = south west] (p1_1) at (0,0) {\includegraphics[width=0.11\textwidth]{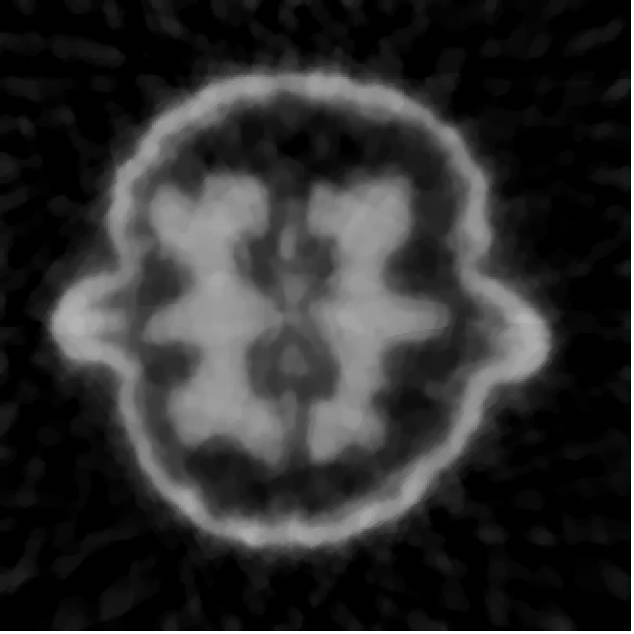}};
    
    \node at (26,26.5) {\color{white} $10$};
    
    \node at (5,3) {\color{white} w/o};

    \node[inner sep=0.5pt, anchor = west] (p1_2) at (p1_1.east) {\includegraphics[ width=0.11\textwidth]{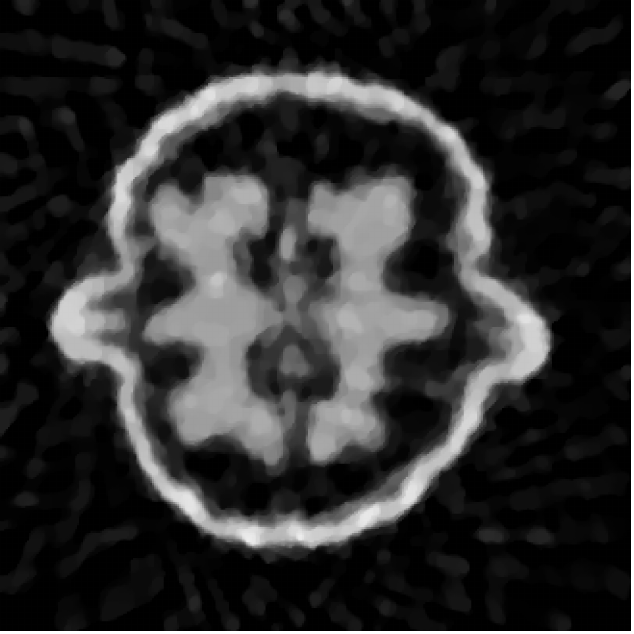}};

    \node at (33,3) {\color{white} w};
    
    \node[inner sep=0.5pt, anchor = west] (p1_3) at (p1_2.east) {\includegraphics[ width=0.11\textwidth]{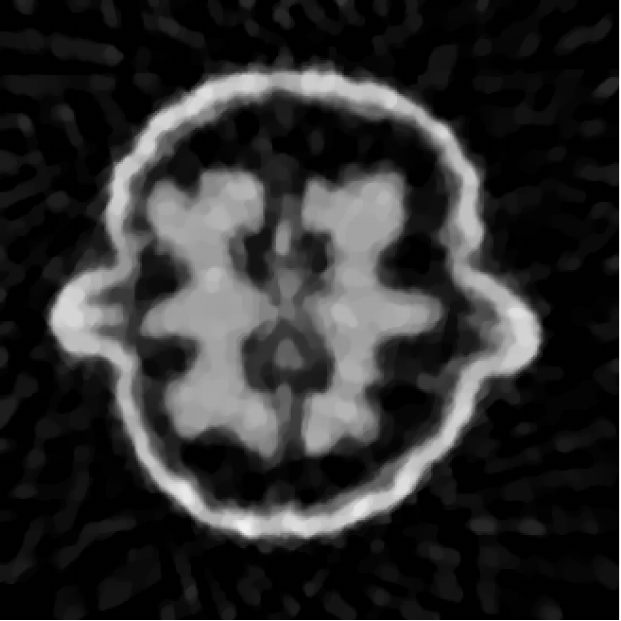}};

    \node at (85,26.5) {\color{white} $20$};
    
    \node at (64,3) {\color{white} w/o};

 \node[inner sep=0.5pt, anchor = west] (p1_4) at (p1_3.east) {\includegraphics[ width=0.11\textwidth]{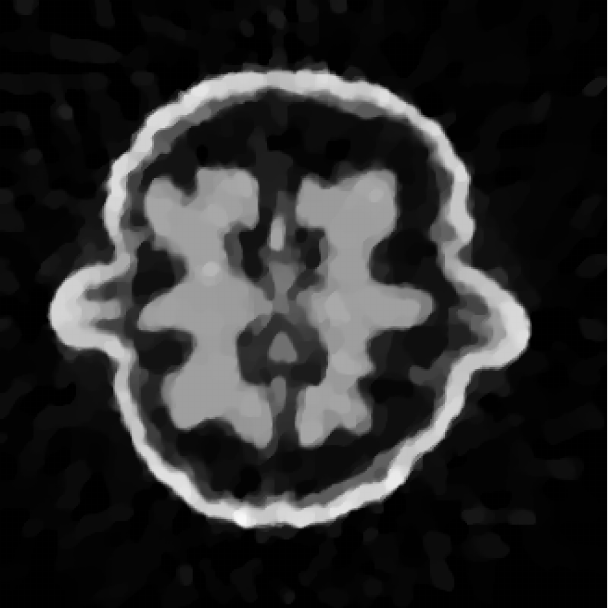}};

  \node at (92,3) {\color{white} w};

 \node[inner sep=0.5pt, anchor = west] (p1_5) at (p1_4.east) {\includegraphics[ width=0.11\textwidth]{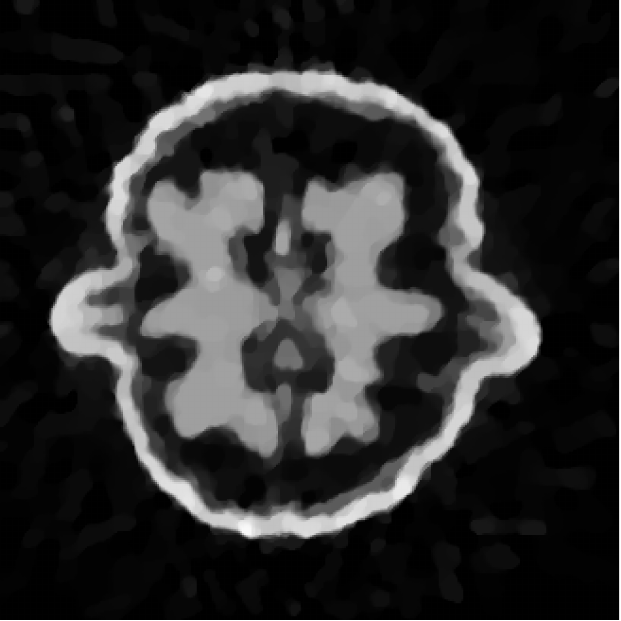}};

    \node at (145,26.5) {\color{white} $40$};
    
    \node at (123,3) {\color{white} w/o};

;
 \node[inner sep=0.5pt, anchor = west] (p1_6) at (p1_5.east) {\includegraphics[ width=0.11\textwidth]{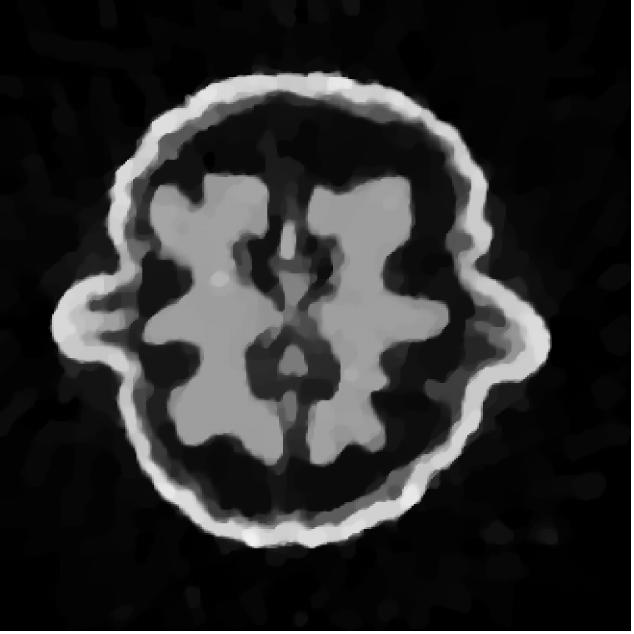}};

    \node at (151,3) {\color{white} w};

\node[inner sep=0.5pt, anchor = west] (p1_7) at (p1_6.east) {\includegraphics[ width=0.11\textwidth]{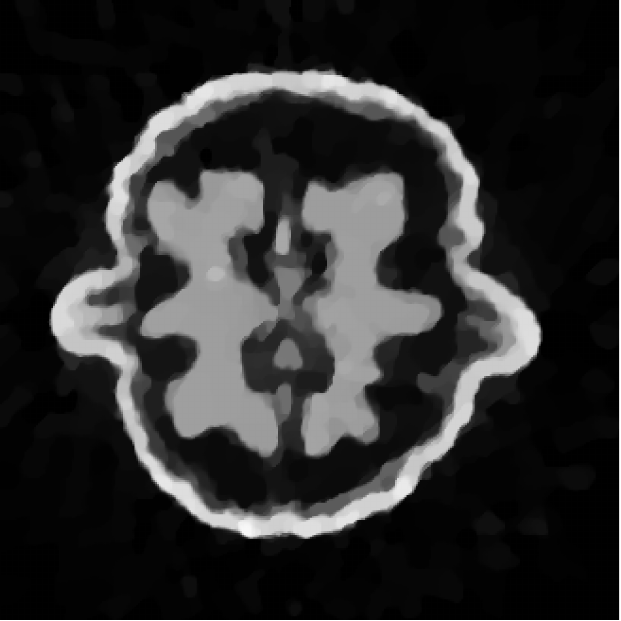}};

     \node at (204,26.5) {\color{white} $60$};
    
    \node at (183,3) {\color{white} w/o};

\node[inner sep=0.5pt, anchor = west] (p1_8) at (p1_7.east) {\includegraphics[ width=0.11\textwidth]{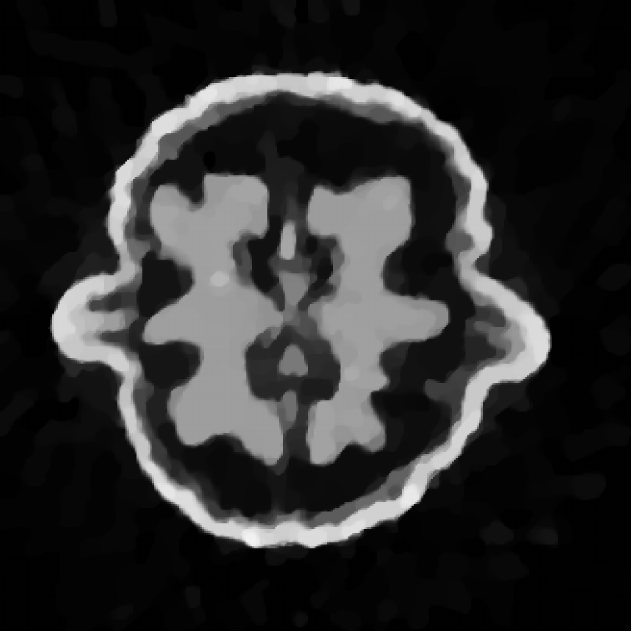}};

\node at (210,3) {\color{white} w};

\end{axis}

\end{tikzpicture}
\caption{
\MRcbT{CT reconstructions of the walnut data using TV based regularization are shown at iterations $10$, $20$, $40$, and $60$.
Columns $1$, $3$, $5$, and $7$ (respectively, $2$, $4$, $6$, and $8$)
show the reconstructions without (respectively, with) \RNP.}}
\label{fig:CTReal:RecoImages}
\end{figure*}

}
\section{Conclusion and Future Work}
\label{sec:Conclusion}
In this paper, we introduced \RNP to accelerate variational image reconstruction.
Additionally, we showed how to efficiently adapt \RNP
to solve reconstruction problems that involve nonsmooth priors.
By leveraging modern GPU computational platforms and PyTorch,
we achieved an on-the-fly implementation,
making \RNP suitable for applications that require real-time processing.
We extensively evaluated the performance of \RNP on image deblurring, super-resolution tasks,
and computed tomography with various regularizers,
demonstrating its effectiveness and efficiency in accelerating variational image reconstruction. 
 
Extending \RNP to magnetic resonance imaging (MRI) reconstruction
presents an interesting future direction.
Modern MRI scanners use multi-coils to acquire k-space data,
which causes the forward model to change for each scan,
requiring the preconditioner to be computed on-the-fly.
Additionally, in functional MRI applications,
the blood oxygenation level-dependent (BOLD) contrast,
which is sensitive to magnetic field inhomogeneity, is primarily used.
In some cases,
images are jointly reconstructed with the associated field maps \cite{olafsson2008fast},
causing the forward model to change at each iteration.
Given \RNP's on-the-fly implementation,
we believe it holds great potential as a preconditioner for acceleration.
 
In this paper, we considered impulsive noise,
and set the data-fidelity term to be the $\ell_p$ norm with $p \leq 1$.
Another interesting direction for future work is to incorporate deep learning to address this problem. Developing a novel PnP/RED framework for impulsive noise would be an exciting research direction.
 
 
\appendices
\crefalias{section}{appendix}
\section{Examples of \texorpdfstring{$\|\umL\uvx\|_{1,\phi}$}{}}
\label{app:sec:examplesMixedNorm}
Let $\umX\in\mathbb R^{N_1\times N_2}$ be the matrix form of $\uvx$,
where $\uvx$ is the column stacking of $\umX$.
So if $\umL$ represents the first-order differential operator, we have $G=N$ and
$$
\uvv_l = \bmat \umX_{i,j}-\umX_{i-1,j} \\ \umX_{i,j}-\umX_{i,j-1} \emat,
$$
with $l = j \, N_1+i$ and $\umX_{i,j}$
denotes the component of $\umX$ at the $i$th row and the $j$th column.
So for $\phi=1$ (respectively, $\phi=2$), we have
$\|\umL\uvx\|_{1,1}=\sum_l |\umX_{i,j}-\umX_{i-1,j}|+|\umX_{i,j}-\umX_{i,j-1}|$
(respectively, \MRcb{$\|\umL\uvx\|_{1,2}=\sum_l
\sqrt{(\umX_{i,j}-\umX_{i-1,j})^2+(\umX_{i,j}-\umX_{i,j-1})^2}$}),
which represents the anisotropic (respectively, isotropic) total variation.

If $\umL$ represents the second-order finite-difference operator, we have 
$$
\uvv_l = \bmat  v_l^{11} & v_l^{12} \\[5pt] v_l^{21} & v_l^{22} \emat,
$$
with $l=j \, N_1+i$ and
$v_l^{11}=\umX_{i-1,j}-2\umX_{i,j}+\umX_{i+1,j},\,
v_l^{22}=\umX_{i,j-1}-2\umX_{i,j}+\umX_{i,j+1},\,
v_l^{12}=v_l^{21}=\frac{1}{4}\big(\umX_{i+1,j+1}-\umX_{i+1,j-1}-\umX_{i-1,j+1}+\umX_{i-1,j-1}\big)$.
Thus $\|\umL\uvx\|_{1,\phi}=\sum_l \|\uvv_l\|_\phi$.



\bibliographystyle{IEEEtran}
\bibliography{IEEEabrv,Refs}

\end{document}


%

\title{Supplementary Material: On Adapting Randomized Nystr\"{o}m Preconditioners to Accelerate Variational Image Reconstruction}

 \author{ 
 Tao Hong, \IEEEmembership{Member, IEEE}, Zhaoyi Xu, Jason Hu, \IEEEmembership{Student Member, IEEE}
 and Jeffrey A. Fessler, \IEEEmembership{Fellow, IEEE}
 \thanks{T. Hong is with the Department of Radiology,
 University of Michigan, Ann Arbor, MI 48109,  
 (Email: \texttt{tahong@umich.edu}). TH was partly supported by National Institutes of Health grant R01NS112233.
 }
 \thanks{Z. Xu is with the Department of Mechanical Engineering, University of Michigan, Ann Arbor, MI 48109,  
 (Email: \texttt{zhaoyix@umich.edu}).}
 \thanks{J. Hu and J. Fessler
 are with the Department of Electrical and Computer Engineering,
 University of Michigan, Ann Arbor, MI 48109, USA 
 (Email: \texttt{jashu,fessler@umich.edu}).}
}

%

%
%

\markboth{}
{Shell \MakeLowercase{\mrmit{et al.}}: Bare Demo of IEEEtran.cls for IEEE Journals}
%



\maketitle



%
\IEEEpeerreviewmaketitle
\section{Image Deblurring and Super-Resolution}
\begin{figure}[ht]
    \centering
    \subfigure[$1$\% noise level.]{\includegraphics[width=0.42\linewidth]{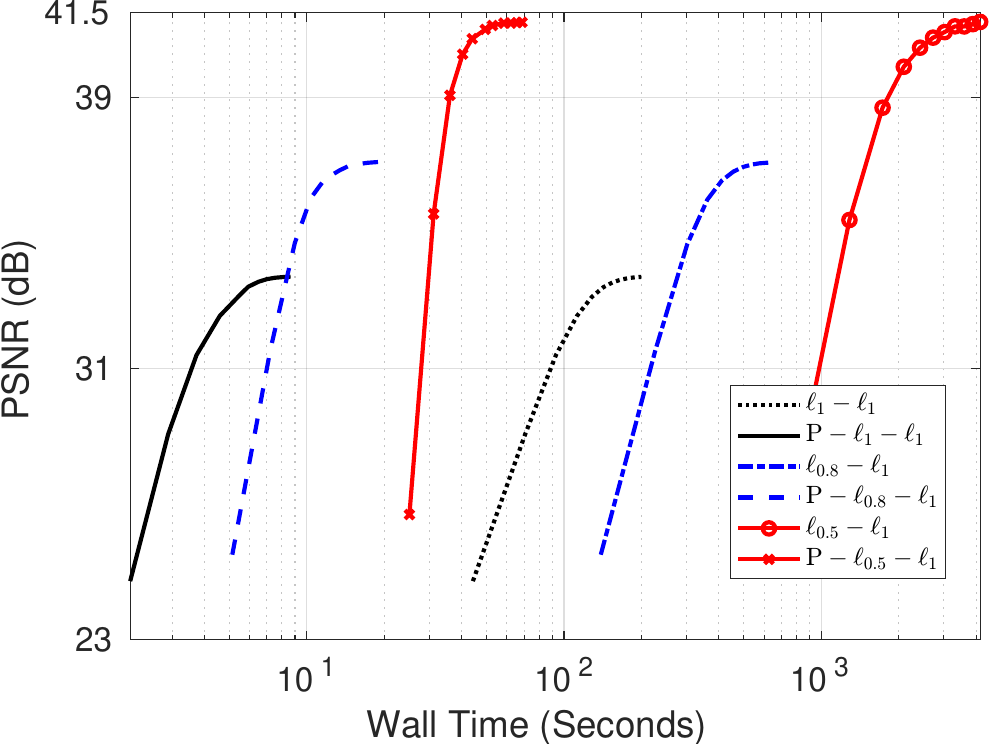}}
    \subfigure[$15$\% noise level.]{\includegraphics[width=0.42\linewidth]{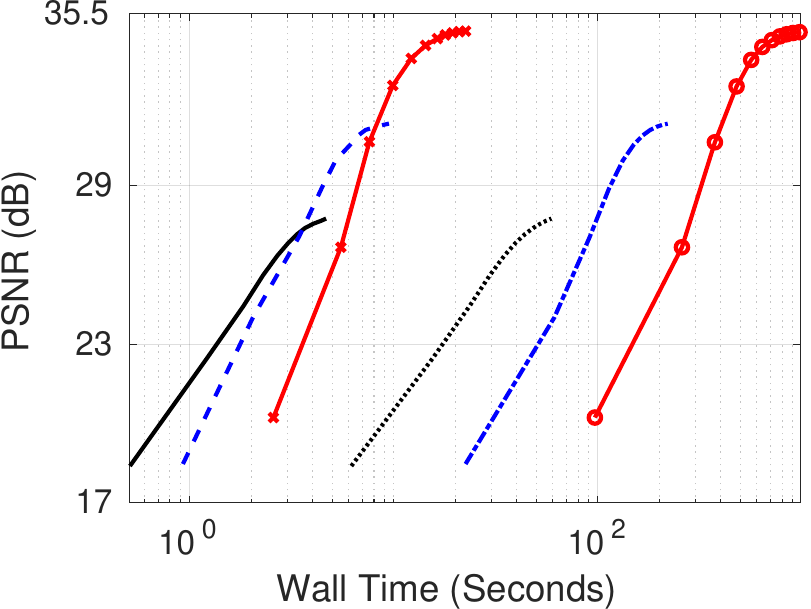}}
    \caption{PSNR values versus wall time in the image deblurring task with a uniform kernel at different noise levels for the starfish image.}
    \label{fig:supp:imagedeblurUniform}
\end{figure}

\begin{figure}[ht]
    \centering
    \subfigure[$1$\% noise level.]{\includegraphics[width=0.42\linewidth]{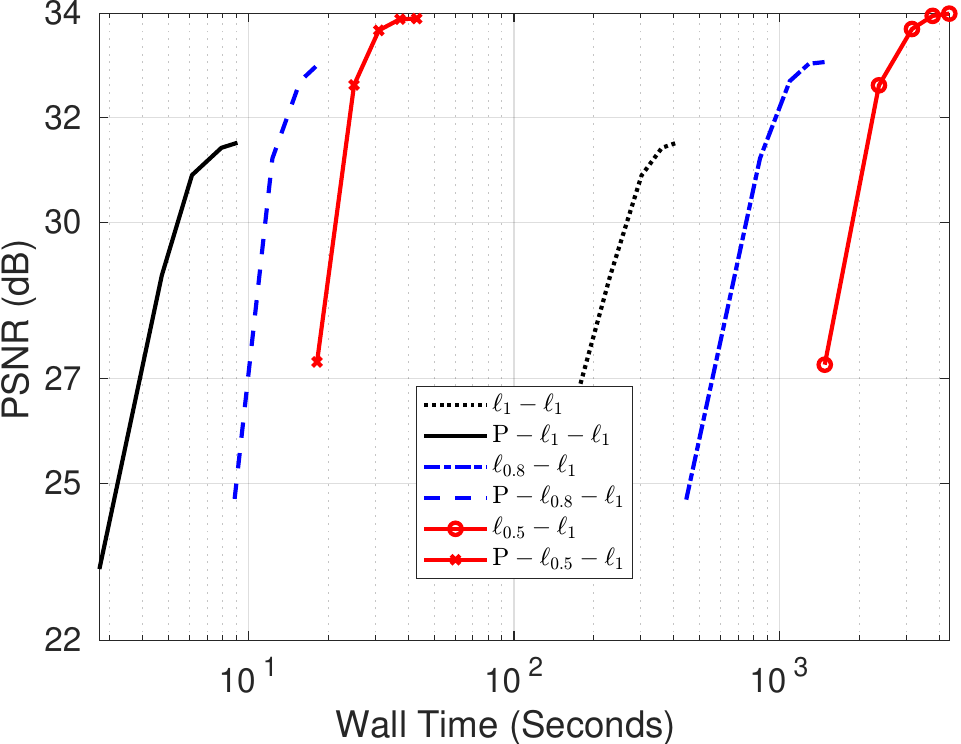}}
\subfigure[$15$\% noise level.]{\includegraphics[width=0.42\linewidth]{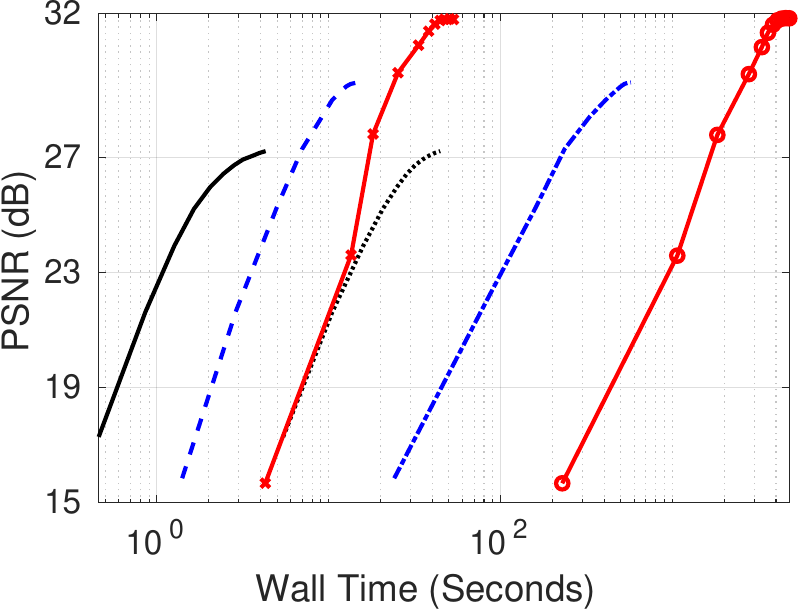}}
    \caption{PSNR values versus wall time in the image deblurring task with a Gaussian kernel at different noise levels for the leaves image.}
    \label{fig:supp:imagedeblurGaussian}
\end{figure}

\begin{figure}[ht]
    \centering
    \subfigure[$1$\% noise level.]{\includegraphics[width=0.42\linewidth]{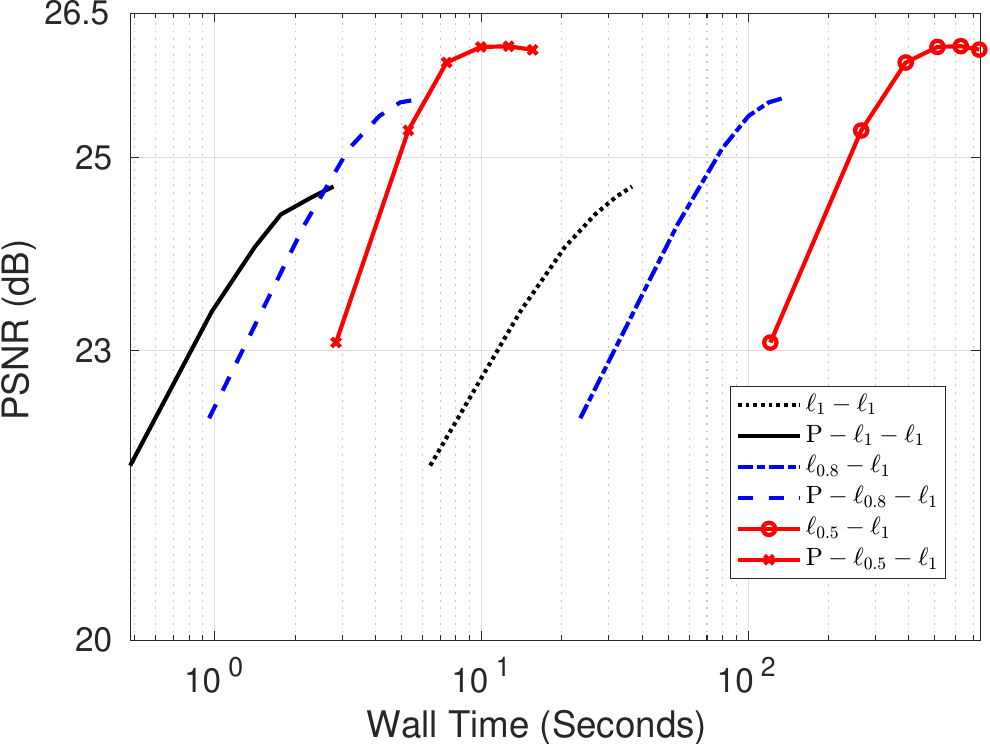}}
   \subfigure[$15$\% noise level.]{ \includegraphics[width=0.42\linewidth]{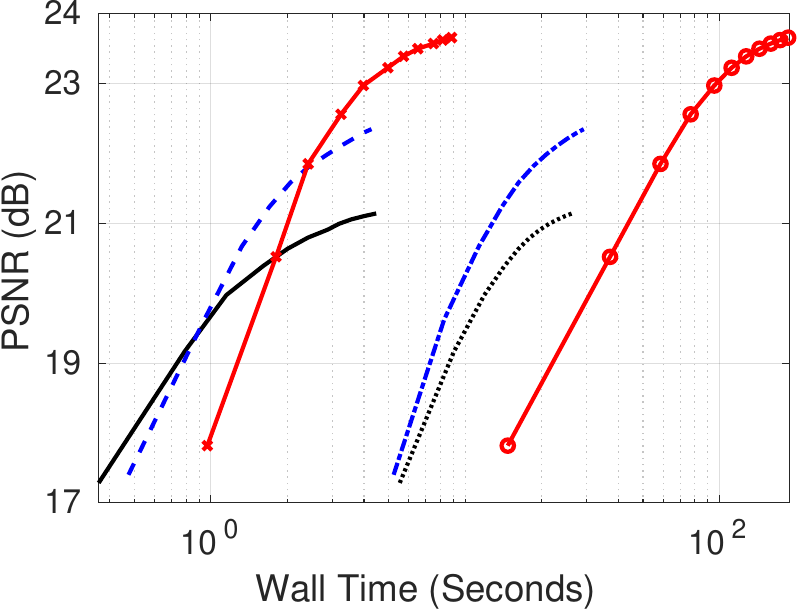}}
    \caption{PSNR values versus wall time in the image super-resolution task with a downsample scale $2$ at different noise levels for the bike image.}
    \label{fig:supp:imageSR}
\end{figure}

We present additional results for image deblurring and super-resolution tasks with two different noise levels, $1$\% and $15$\%.  \Cref{fig:supp:imagedeblurUniform,fig:supp:imagedeblurGaussian,fig:supp:imageSR} show the PSNR values versus wall time. Clearly,  we saw that using \RNP significantly accelerated the convergence of \IRM in terms of wall time. Moreover, a smaller $p$ yielded a higher PSNR. These trends align with the observations in the main paper.

\section{Computed Tomography Reconstruction}
\begin{figure}
    \centering
    \subfigure[]{\includegraphics[width=0.35\linewidth]{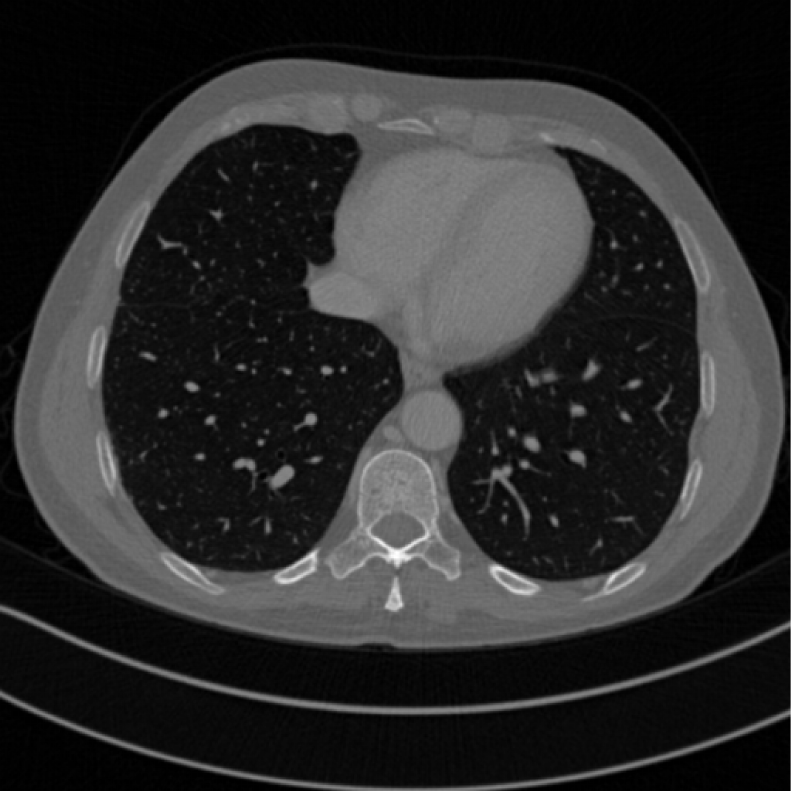}\label{fig:CT:GT:a}}
    \subfigure[]{\includegraphics[width=0.35\linewidth]{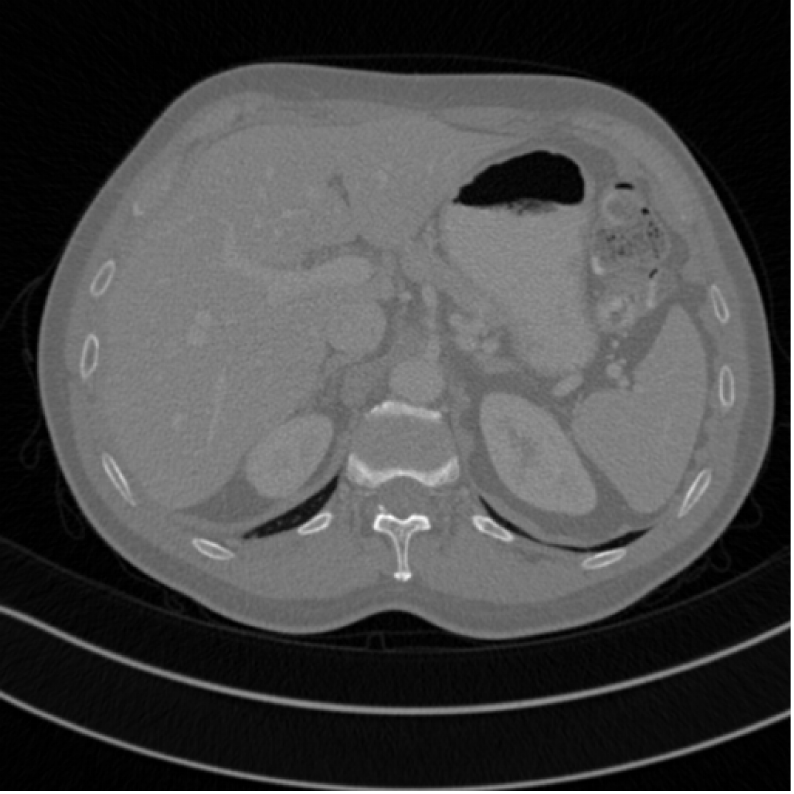}\label{fig:CT:GT:b}}
    \caption{Ground truth CT images.}
    \label{fig:CT:GT}
\end{figure}

We propose additional results for using \RNP for Fig. S.\ref{fig:CT:GT:b}, the fan-beam acquisition geometry, and the HS$_2$ and H$_\infty$ based reconstruction. 

\begin{figure*}[ht]
\vspace{-2.5cm}
	\centering
\begin{tikzpicture}
    \begin{axis}[at={(0,0)},anchor = north west,
    xmin = 0,xmax = 250,ymin = 0,ymax = 70,ylabel = Wavelet,width=0.95\textwidth,
        scale only axis,
        enlargelimits=false,
       axis line style={draw=none},
       tick style={draw=none},
        axis equal image,
        xticklabels={,,},yticklabels={,,},
        ylabel style={yshift=-0.3cm,xshift=-1.4cm},
       ]

    \node[inner sep=0.5pt, anchor = south west] (p1_1) at (0,0) {\includegraphics[ width=0.11\textwidth]{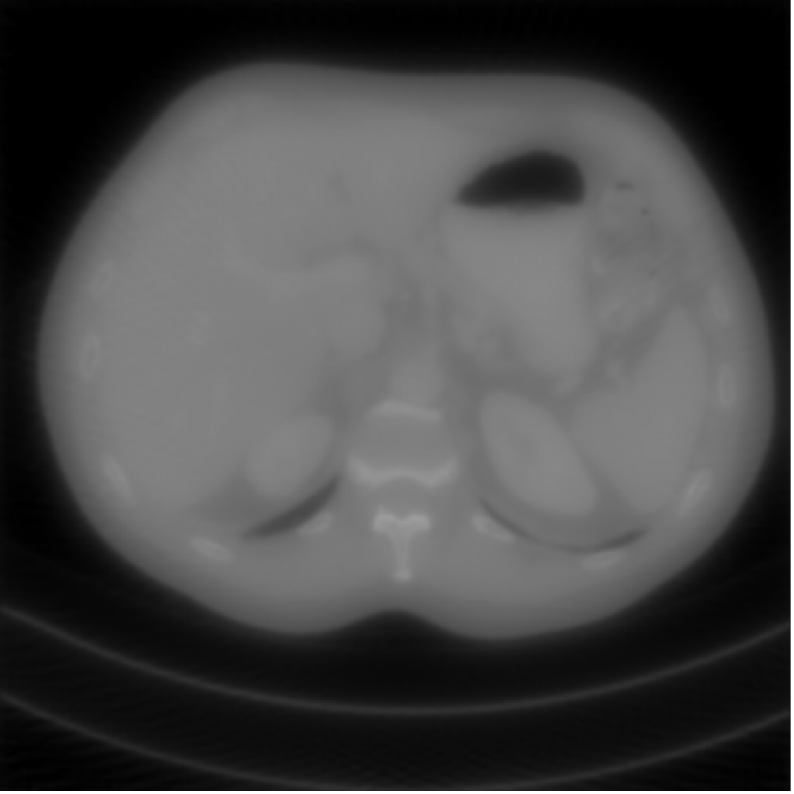}};
    
    \node at (26,26.5) {\color{white} $10$};
    
    \node at (5,3) {\color{white} w/o};
    \node at (24,3) {\color{white} $24.5$};
    
    \node[inner sep=0.5pt, anchor = west] (p1_2) at (p1_1.east) {\includegraphics[ width=0.11\textwidth]{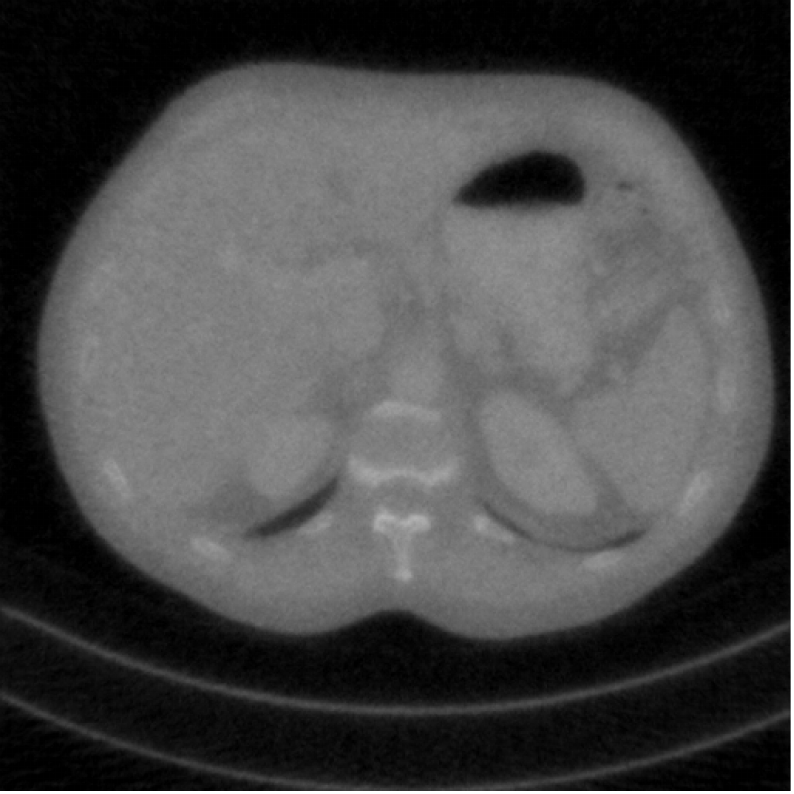}};

    
    
     \node at (54,3) {\color{white} $26.7$};
    \node at (33,3) {\color{white} w};
    
    \node[inner sep=0.5pt, anchor = west] (p1_3) at (p1_2.east) {\includegraphics[ width=0.11\textwidth]{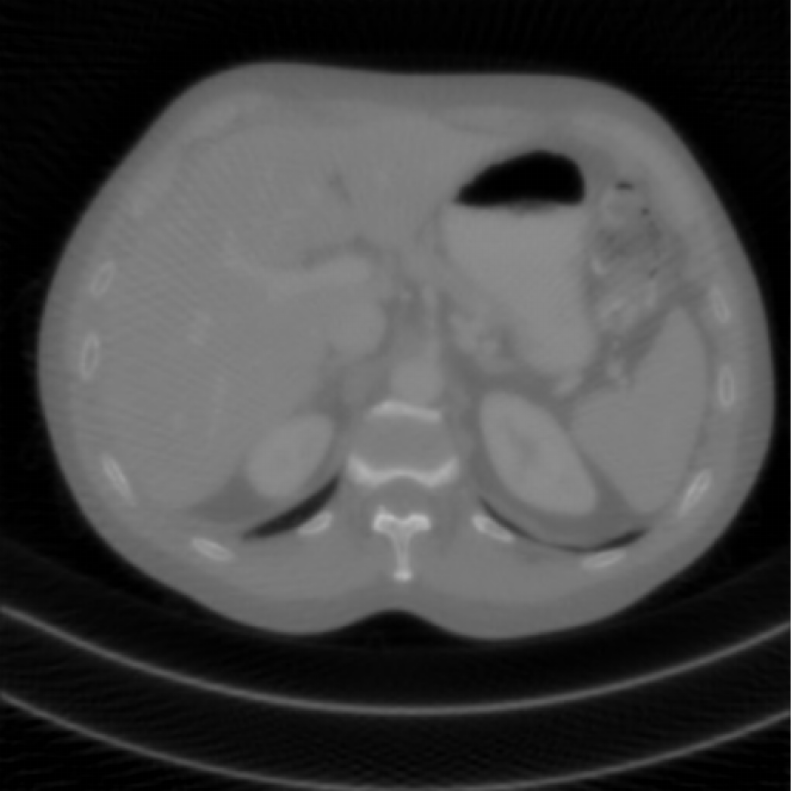}};

    \node at (85,26.5) {\color{white} $20$};
    
    \node at (64,3) {\color{white} w/o};
    
      \node at (83.5,3) {\color{white} $28.8$};
   
 \node[inner sep=0.5pt, anchor = west] (p1_4) at (p1_3.east) {\includegraphics[ width=0.11\textwidth]{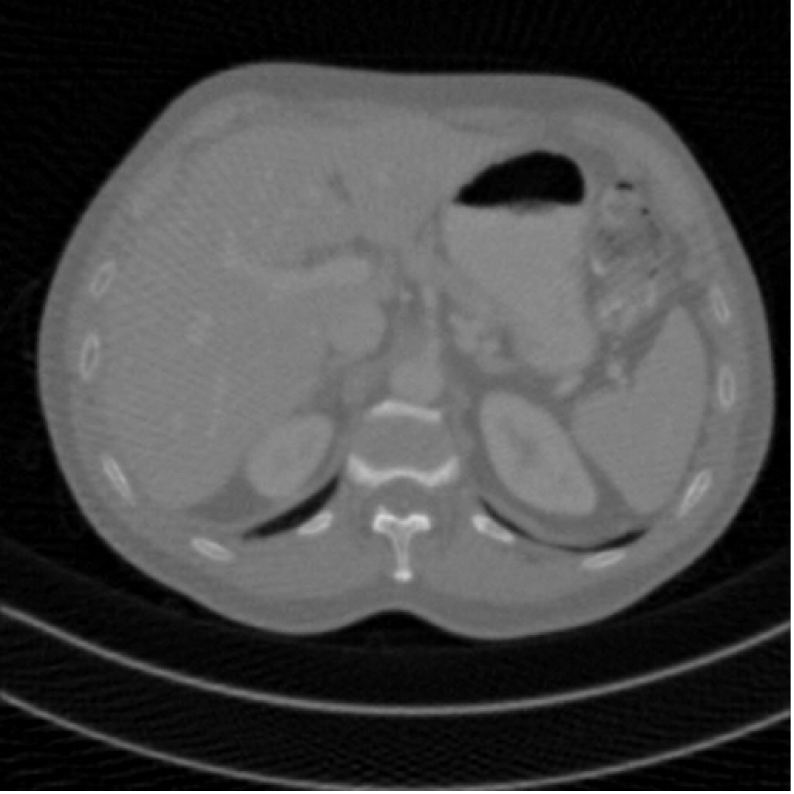}};

    
    \node at (113,3) {\color{white} $32.0$};
    
  \node at (92,3) {\color{white} w};

 \node[inner sep=0.5pt, anchor = west] (p1_5) at (p1_4.east) {\includegraphics[ width=0.11\textwidth]{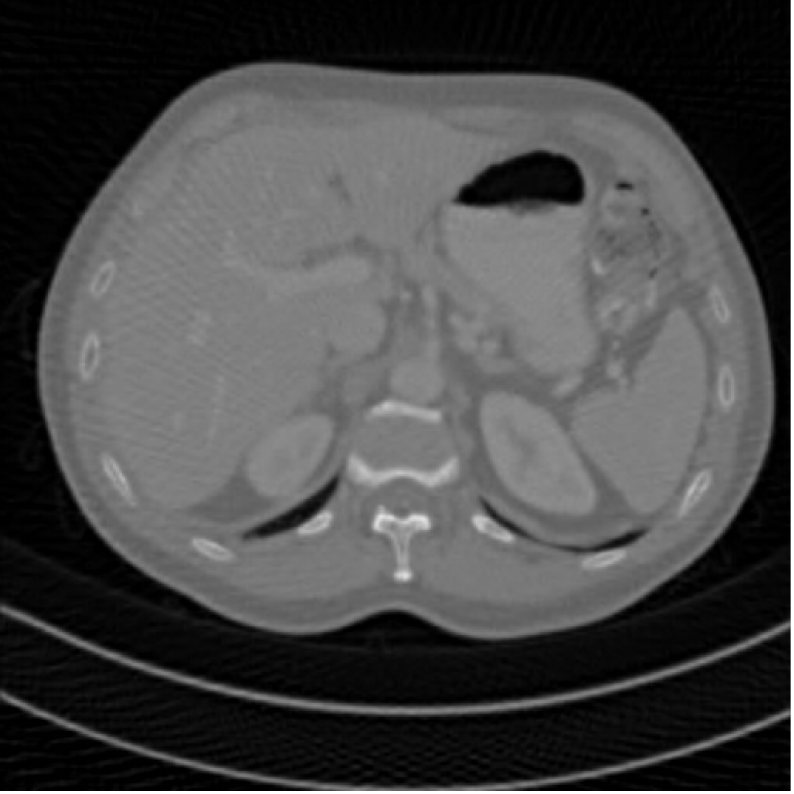}};

    \node at (145,26.5) {\color{white} $40$};
    
    \node at (123,3) {\color{white} w/o};
    
     \node at (143,3) {\color{white} $34.0$};
    
 \node[inner sep=0.5pt, anchor = west] (p1_6) at (p1_5.east) {\includegraphics[ width=0.11\textwidth]{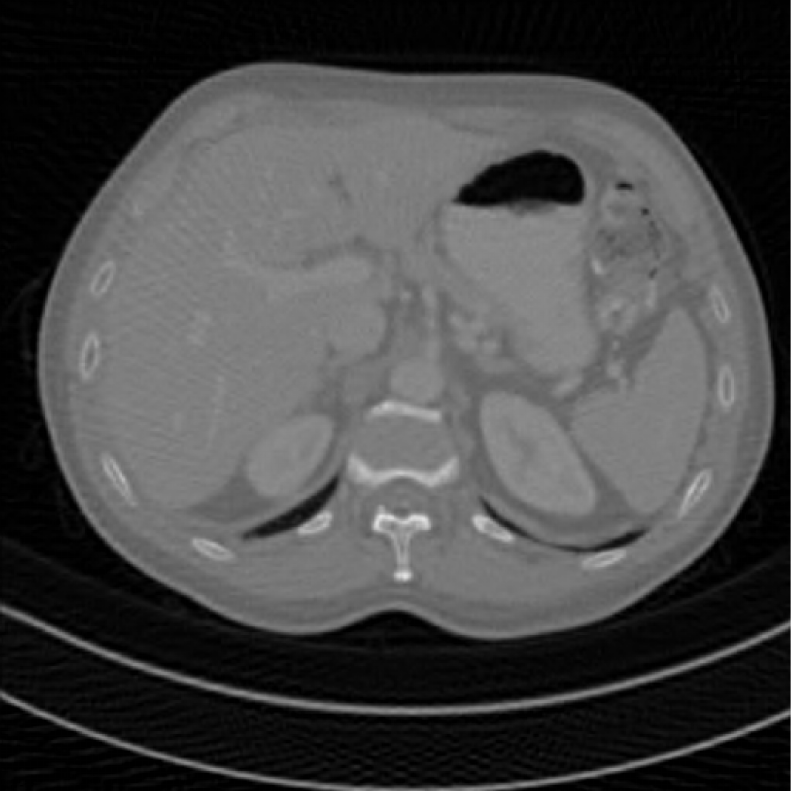}};

    \node at (151,3) {\color{white} w};
    
    \node at (172.5,3) {\color{white} $34.9$};
    
\node[inner sep=0.5pt, anchor = west] (p1_7) at (p1_6.east) {\includegraphics[ width=0.11\textwidth]{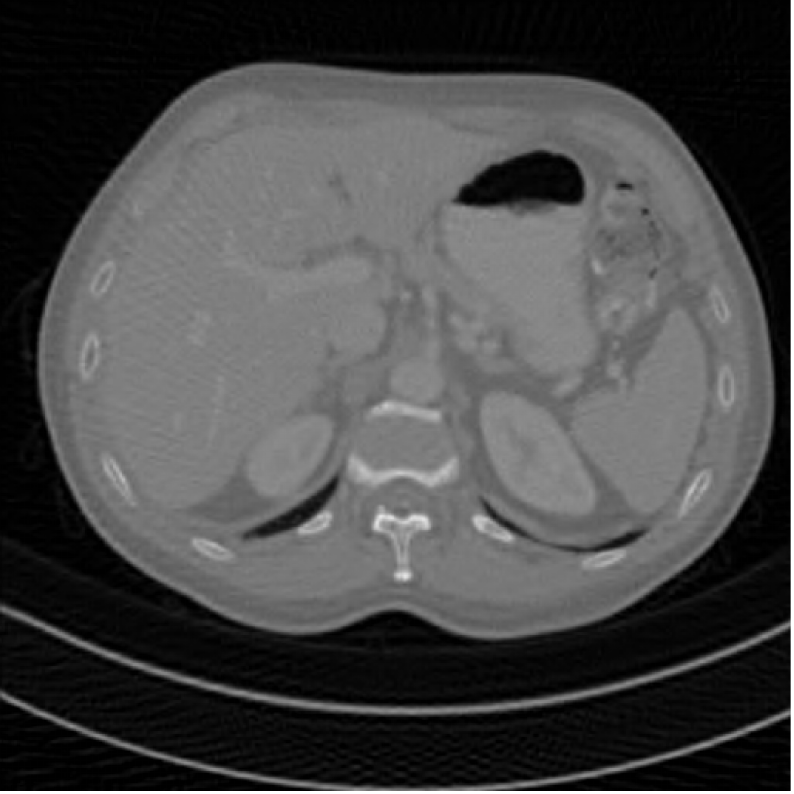}};

    \node at (204,26.5) {\color{white} $60$};
    
    \node at (183,3) {\color{white} w/o};
    
    \node at (201.5,3) {\color{white} $35.0$};

\node[inner sep=0.5pt, anchor = west] (p1_8) at (p1_7.east) {\includegraphics[ width=0.11\textwidth]{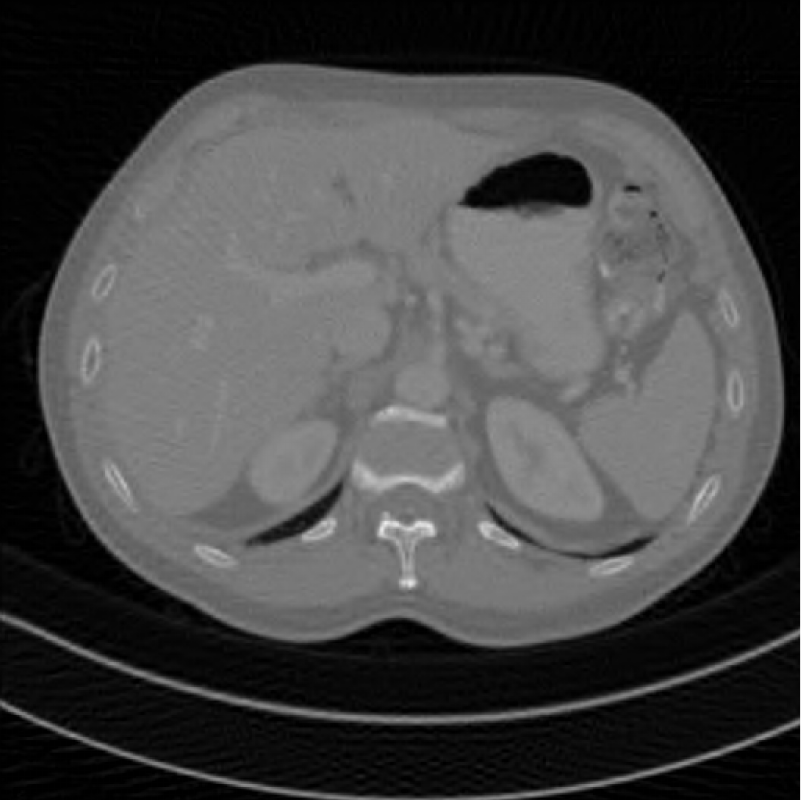}};

\node at (210,3) {\color{white} w};
    

\node at (231.5,3) {\color{white} $35.2$};
\end{axis}

 \begin{axis}[at={(p1_1.south west)},anchor = north west,
     xmin = 0,xmax = 250,ymin = 0,ymax = 70, width=0.95\textwidth,ylabel = TV,
         scale only axis,
         enlargelimits=false,
         yshift=2.8cm,
        axis line style={draw=none},
        tick style={draw=none},
         axis equal image,
         xticklabels={,,},yticklabels={,,},
         ylabel style={yshift=-0.3cm,xshift=-1.4cm},
        ]
        
   \node[inner sep=0.5pt, anchor = south west] (TVp1_1) at (0,0) {\includegraphics[width=0.11\textwidth]{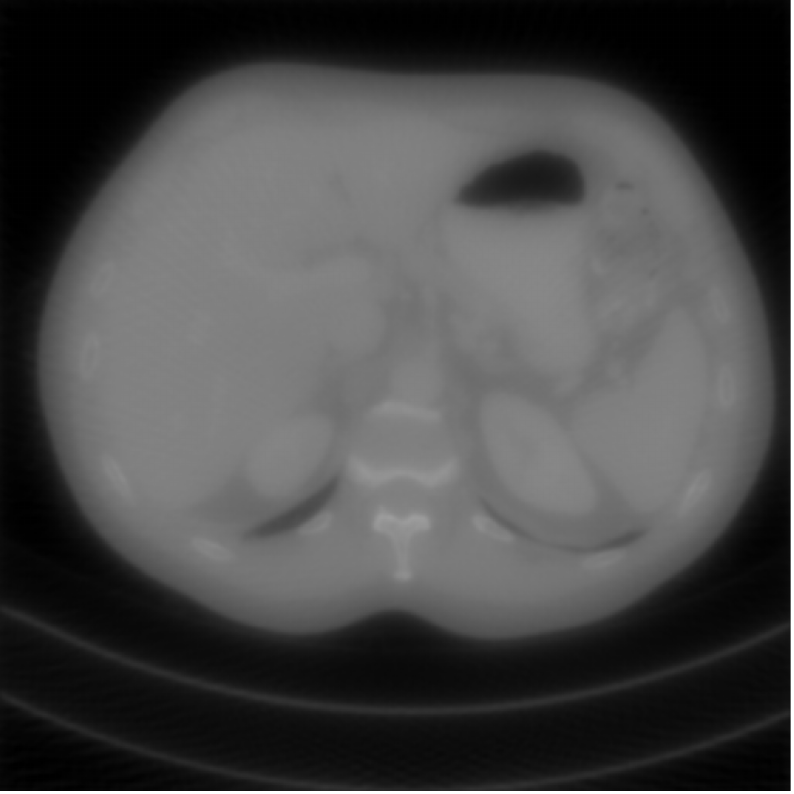}};
   
    \node at (24,3) {\color{white} $24.5$};
    
    \node[inner sep=0.5pt, anchor = west] (TVp1_2) at (TVp1_1.east) {\includegraphics[ width=0.11\textwidth]{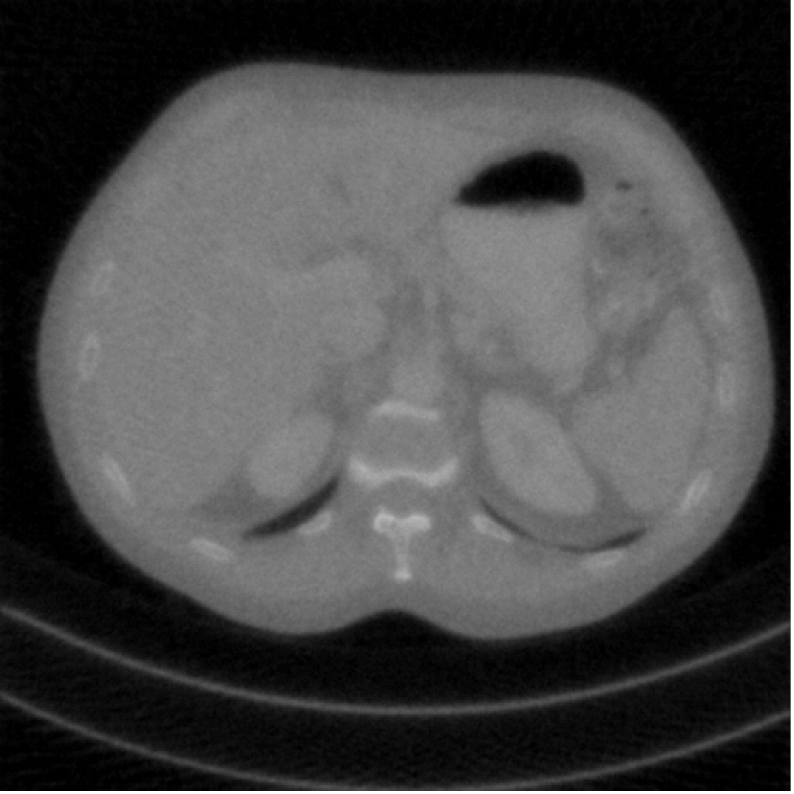}};
 
     \node at (53,3) {\color{white} $27.7$};
  
    \node[inner sep=0.5pt, anchor = west] (TVp1_3) at (TVp1_2.east) {\includegraphics[ width=0.11\textwidth]{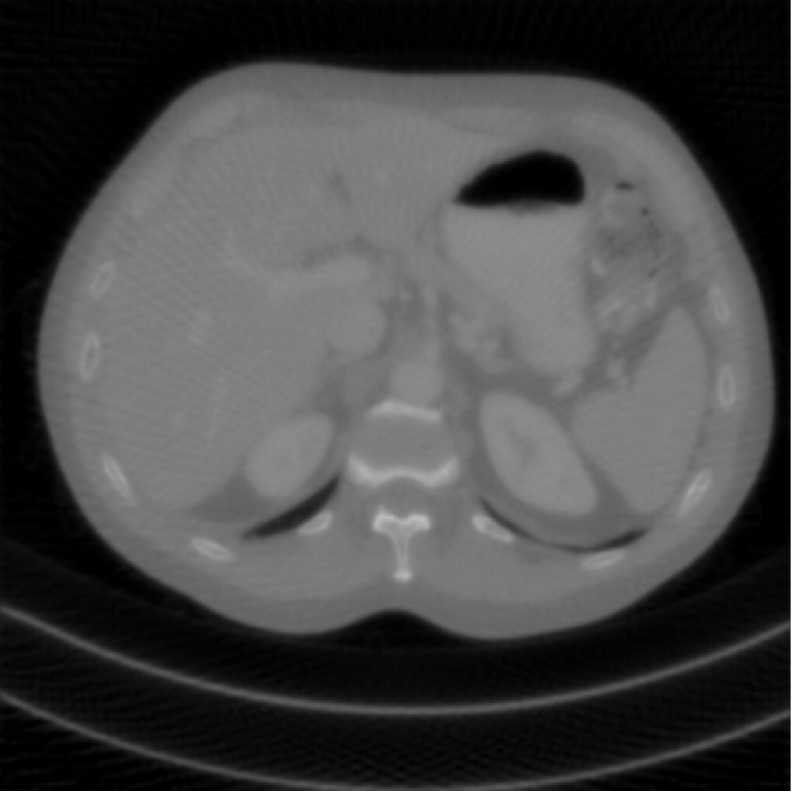}};

    
    \node at (83.5,3) {\color{white} $29.0$};
   
 \node[inner sep=0.5pt, anchor = west] (TVp1_4) at (TVp1_3.east) {\includegraphics[ width=0.11\textwidth]{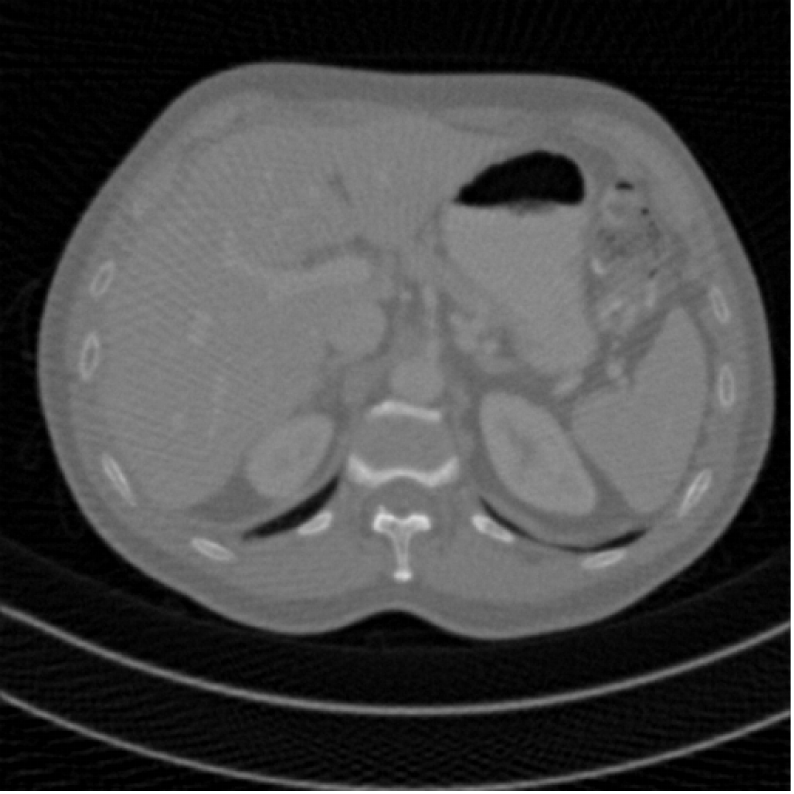}};

    
\node at (113,3) {\color{white} $33.0$};
    
 \node[inner sep=0.5pt, anchor = west] (TVp1_5) at (TVp1_4.east) {\includegraphics[ width=0.11\textwidth]{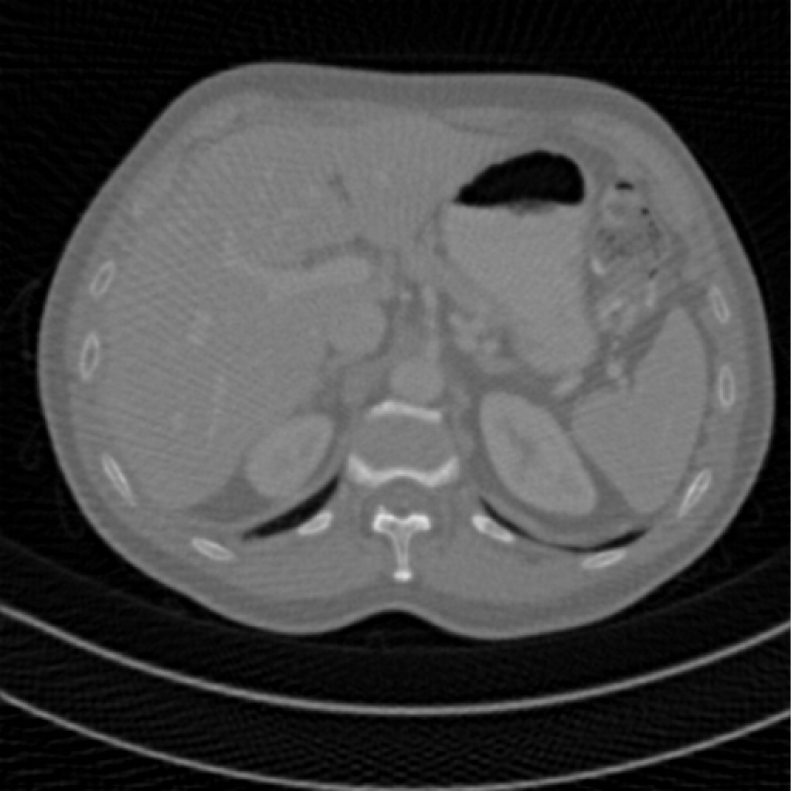}};

    
       \node at (143,3) {\color{white} $34.0$};
    
 \node[inner sep=0.5pt, anchor = west] (TVp1_6) at (TVp1_5.east) {\includegraphics[ width=0.11\textwidth]{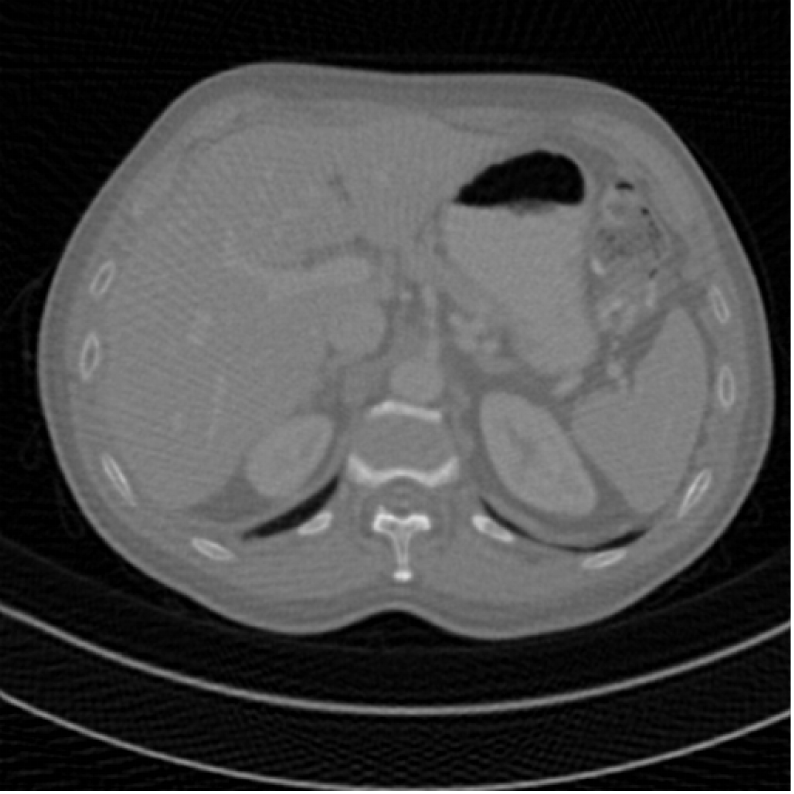}};

    \node at (172.5,3) {\color{white} $34.5$};
    
\node[inner sep=0.5pt, anchor = west] (TVp1_7) at (TVp1_6.east) {\includegraphics[ width=0.11\textwidth]{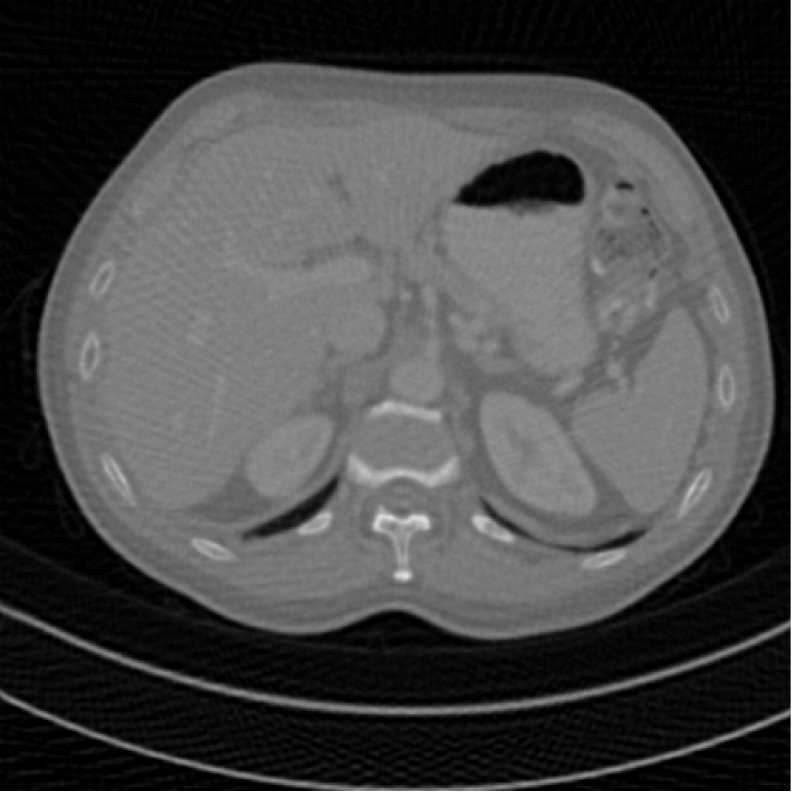}};

    
\node at (201.5,3) {\color{white} $33.7$};

\node[inner sep=0.5pt, anchor = west] (TVp1_8) at (TVp1_7.east) {\includegraphics[ width=0.11\textwidth]{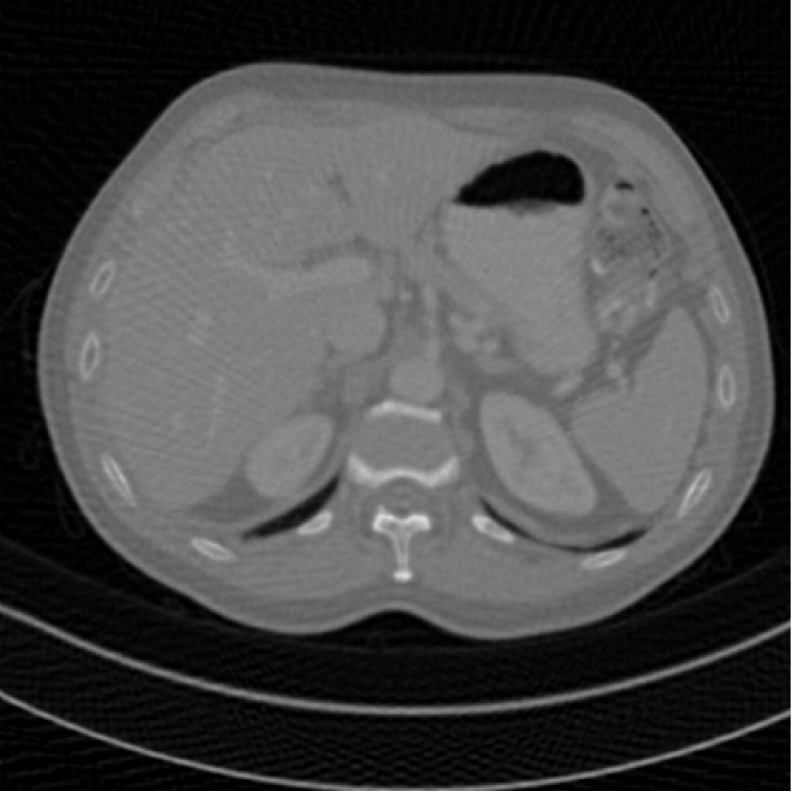}};


\node at (231.5,3) {\color{white} $34.7$};
 \end{axis}

 \begin{axis}[at={(TVp1_1.south west)},anchor = north west,
     xmin = 0,xmax = 250,ymin = 0,ymax = 70, width=0.95\textwidth,ylabel = HS$_1$,
         scale only axis,
         enlargelimits=false,
         yshift=2.8cm,
        axis line style={draw=none},
        tick style={draw=none},
         axis equal image,
         xticklabels={,,},yticklabels={,,},
         ylabel style={yshift=-0.3cm,xshift=-1.4cm},
        ]
        
   \node[inner sep=0.5pt, anchor = south west] (HSp1_1) at (0,0) {\includegraphics[width=0.11\textwidth]{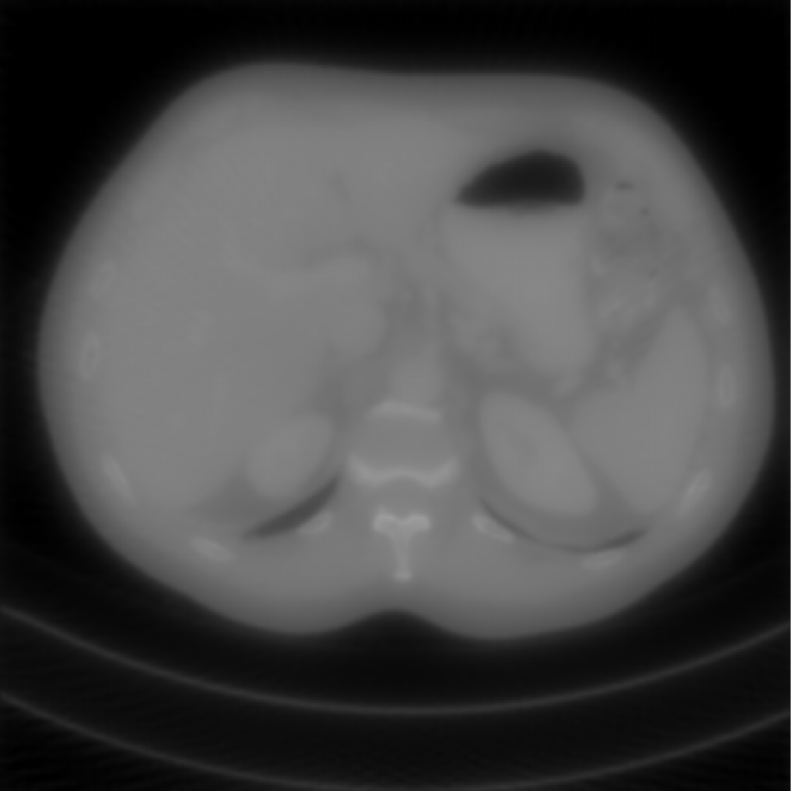}};

       \node at (24,3) {\color{white} $24.5$};
    
    \node[inner sep=0.5pt, anchor = west] (HSp1_2) at (HSp1_1.east) {\includegraphics[ width=0.11\textwidth]{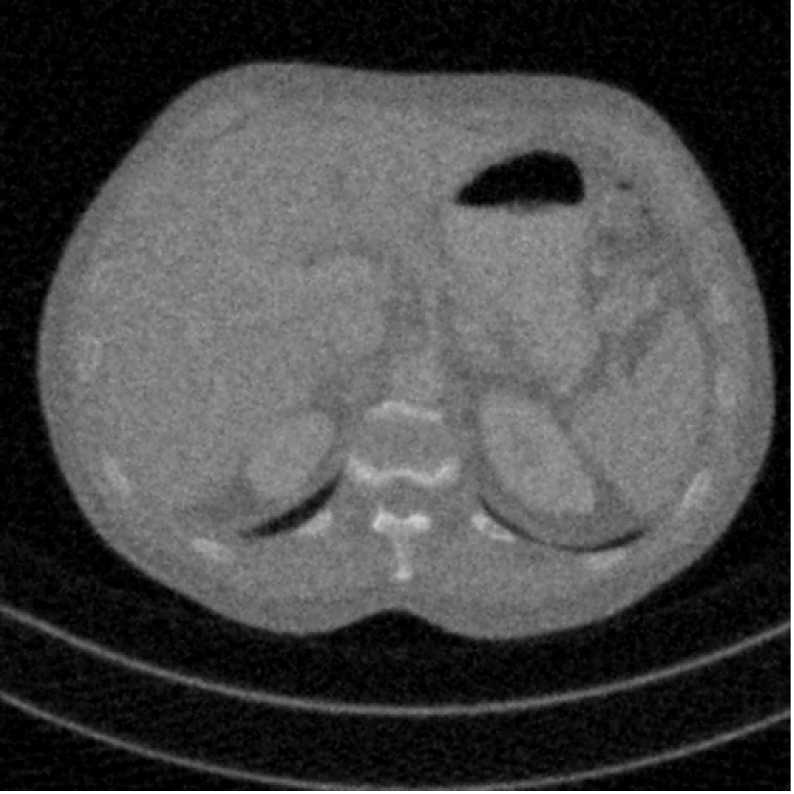}};
 
    
  \node at (54,3) {\color{white} $28.4$};
    \node[inner sep=0.5pt, anchor = west] (HSp1_3) at (HSp1_2.east) {\includegraphics[ width=0.11\textwidth]{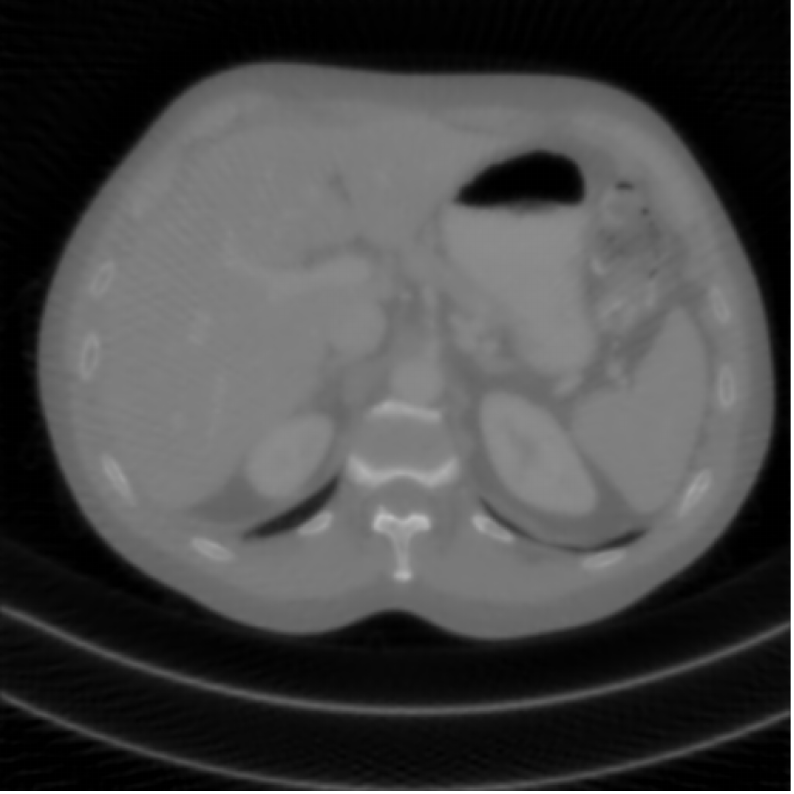}};

     \node at (83.5,3) {\color{white} $28.9$};
   
 \node[inner sep=0.5pt, anchor = west] (HSp1_4) at (HSp1_3.east) {\includegraphics[ width=0.11\textwidth]{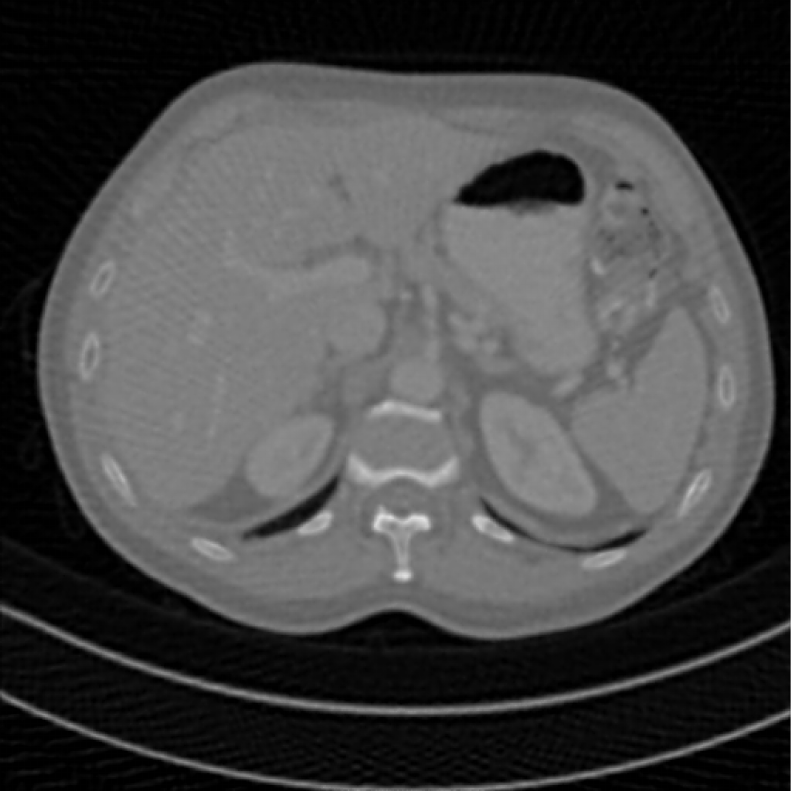}};

    
\node at (113,3) {\color{white} $34.3$};
    
 \node[inner sep=0.5pt, anchor = west] (HSp1_5) at (HSp1_4.east) {\includegraphics[ width=0.11\textwidth]{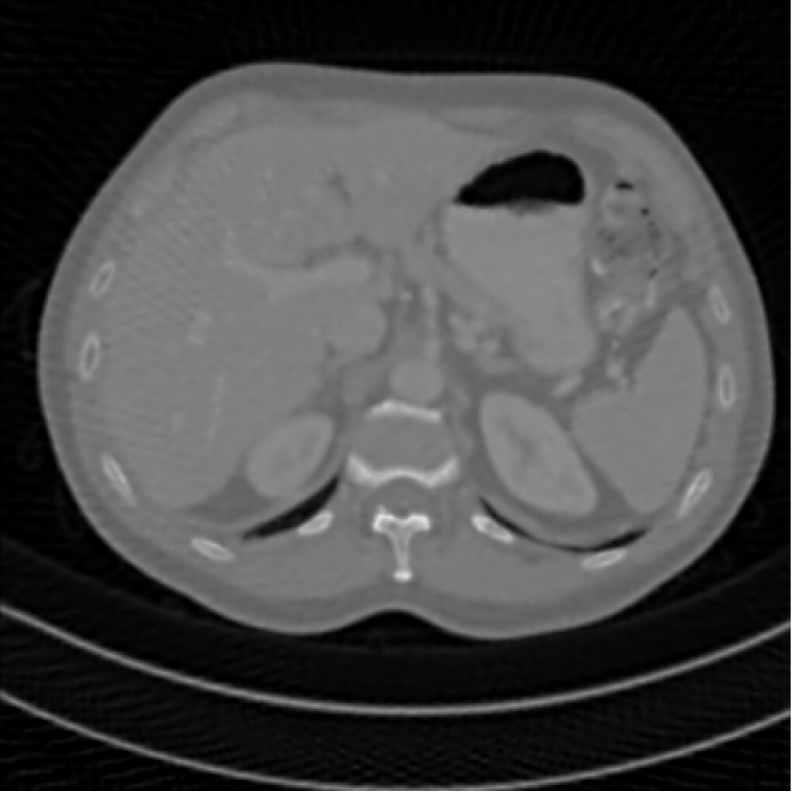}};

    \node at (143,3) {\color{white} $34.1$};
    
 \node[inner sep=0.5pt, anchor = west] (HSp1_6) at (HSp1_5.east) {\includegraphics[ width=0.11\textwidth]{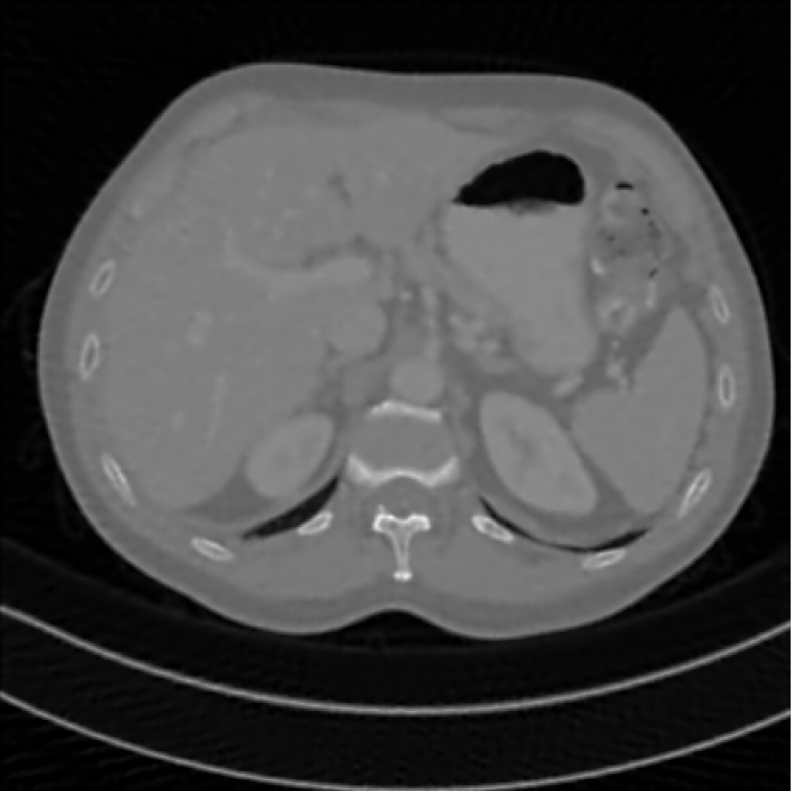}};

    \node at (172.5,3) {\color{white} $36.3$};
\node[inner sep=0.5pt, anchor = west] (HSp1_7) at (HSp1_6.east) {\includegraphics[ width=0.11\textwidth]{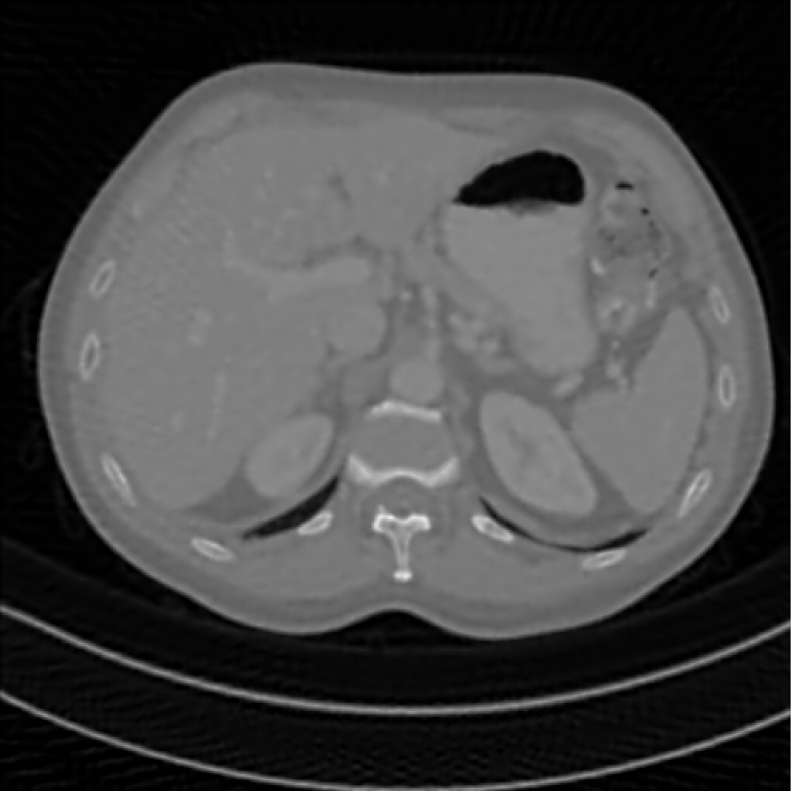}};

 \node at (201.5,3) {\color{white} $35.6$};
 
\node[inner sep=0.5pt, anchor = west] (HSp1_8) at (HSp1_7.east) {\includegraphics[ width=0.11\textwidth]{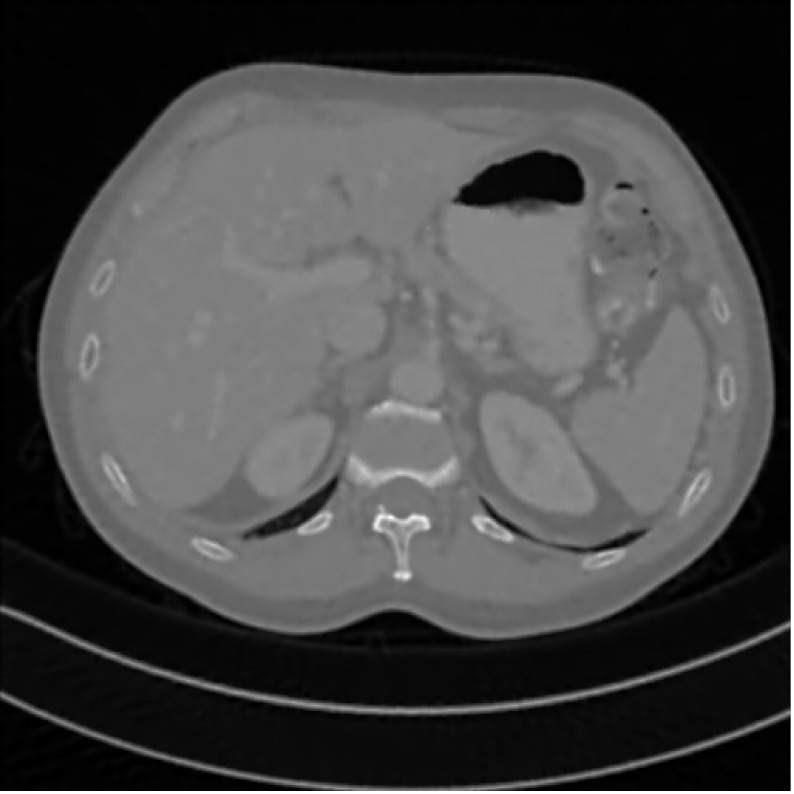}};


\node at (231.5,3) {\color{white} $37.5$};
 \end{axis}
 
\end{tikzpicture} 
\caption{The parallel-beam CT reconstructed images with wavelet, TV, and HS$_1$ based regularization
at iterations $10$, $20$, $40$, and $60$.
Columns $1$, $3$, $5$, and $7$ (respectively, $2$, $4$, $6$, and $8$)
show the reconstructions without (respectively, with) \RNP.
The associated PSNR values are listed at the right corner of each image.
}
\label{fig:CTParallel:RecoImages:b}
\end{figure*}

\subsection{Additional Results for Parallel-Beam Acquisition Geometry}
\begin{figure}
    \centering
    \subfigure[]{
\includegraphics[width=0.44\linewidth]{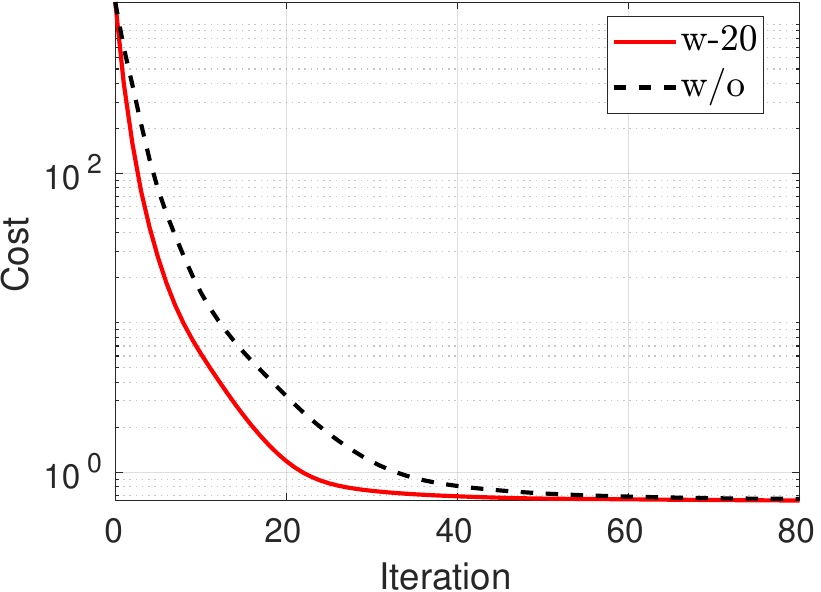}}
    \subfigure[]{
 \includegraphics[width=0.43\linewidth]{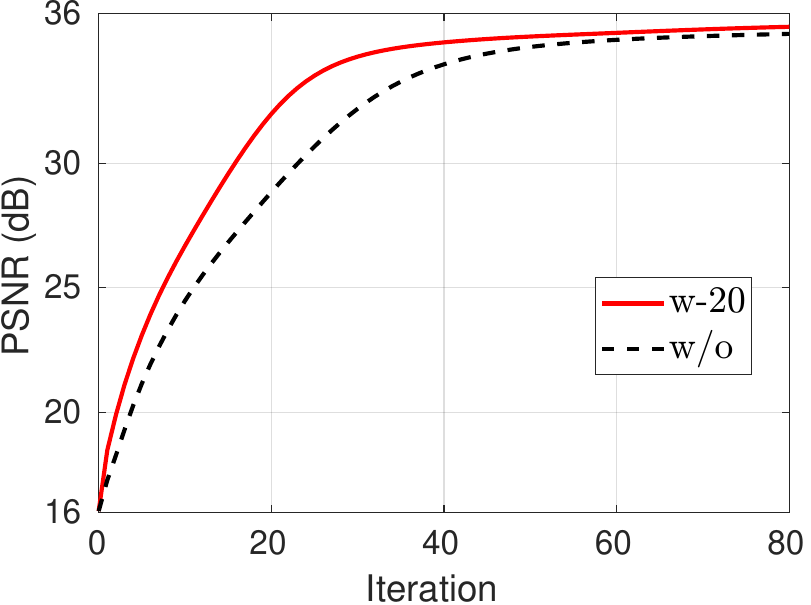}} 
    \caption{Comparison of using \RNP for wavelet based parallel-beam CT reconstruction for Fig. S.\ref{fig:CT:GT:b}. w/o denotes the one without using \RNP.}
    \label{fig:supp:parallel:im:b:wav}
\end{figure}
\begin{figure}
    \centering

    \subfigure[]{
\includegraphics[width=0.43\linewidth]{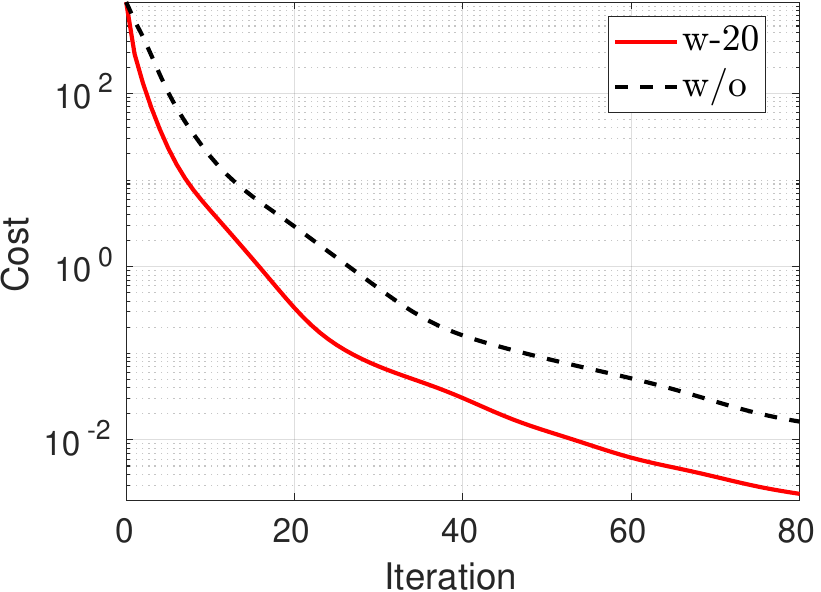}}    
    \subfigure[]{
 \includegraphics[width=0.42\linewidth]{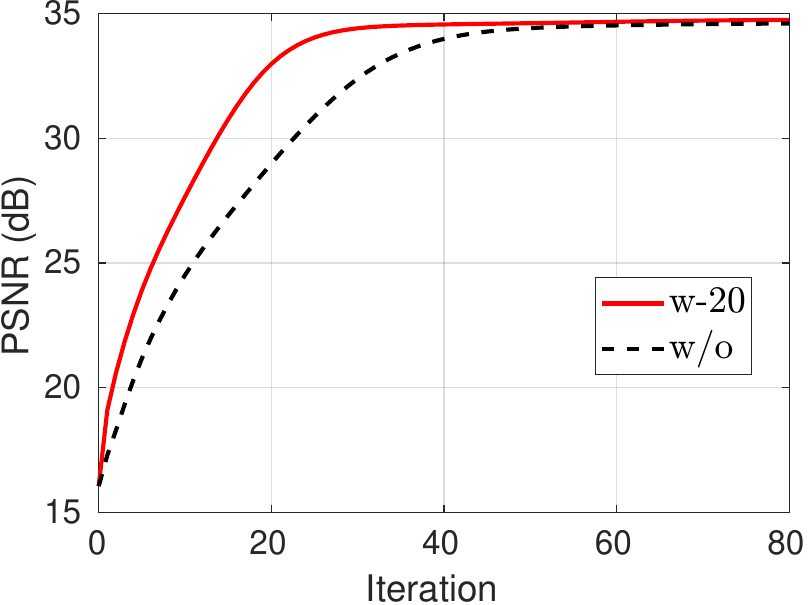}}

    \caption{Comparison of using \RNP for total variation based parallel-beam CT reconstruction for Fig. S.\ref{fig:CT:GT:b}. w/o denotes the one without using \RNP.}
    \label{fig:supp:parallel:im:b:TV}
\end{figure}

\begin{figure}
    \centering
    \subfigure[]{
\includegraphics[width=0.42\linewidth]{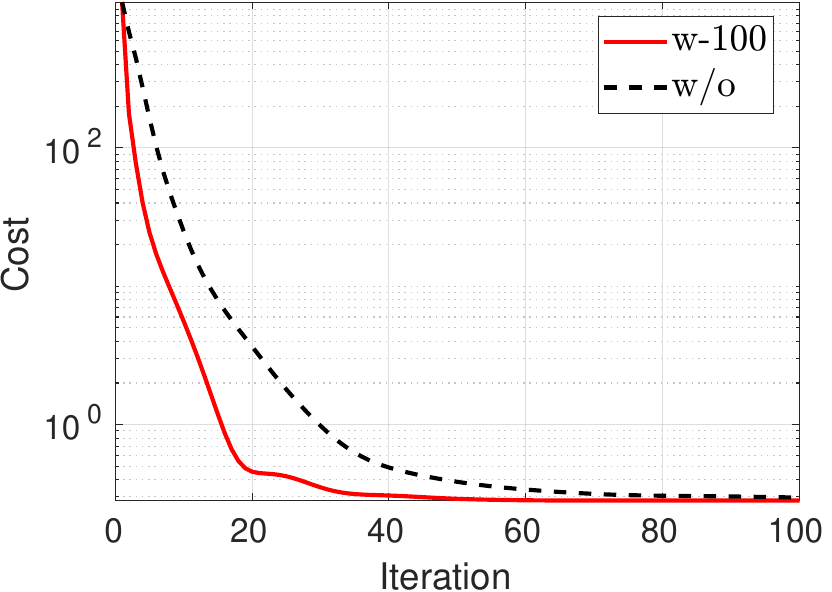}}
    \subfigure[]{
 \includegraphics[width=0.42\linewidth]{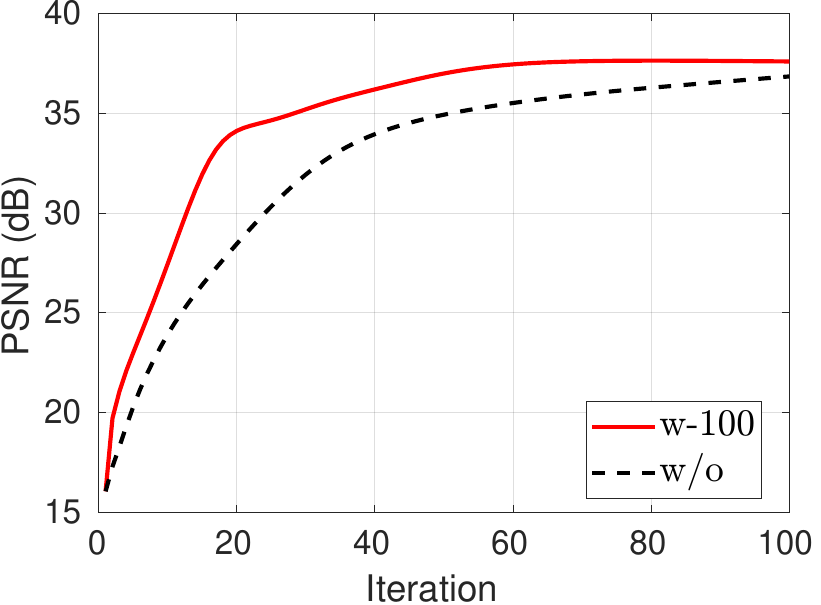}}

    \caption{Comparison of using \RNP for HS$_1$  based parallel-beam CT reconstruction for Fig. S.\ref{fig:CT:GT:b}. w/o denotes the one without using \RNP.}
    \label{fig:supp:parallel:im:b:HS1}
\end{figure}

\Cref{fig:supp:parallel:im:b:wav,fig:supp:parallel:im:b:TV,fig:supp:parallel:im:b:HS1} present the cost and PSNR values of wavelet, total variation, HS$_1$ norm based parallel-beam CT reconstruction for Fig. S.\ref{fig:CT:GT:b}. Clearly, we observed that using \RNP significantly accelerated APG. \Cref{fig:CTParallel:RecoImages:b} shows the reconstructed images at different iterations, where we observed that the method with \RNP yielded clearer images at the same number of iterations illustrating the effectiveness of using \RNP.

\subsection{Results for Fan-Beam Acquisition Geometry}
\begin{figure*}[ht]
\vspace{-2.5cm}
	\centering
\begin{tikzpicture}
    \begin{axis}[at={(0,0)},anchor = north west,
    xmin = 0,xmax = 250,ymin = 0,ymax = 70,ylabel = Wavelet,width=0.95\textwidth,
        scale only axis,
        enlargelimits=false,
       axis line style={draw=none},
       tick style={draw=none},
        axis equal image,
        xticklabels={,,},yticklabels={,,},
        ylabel style={yshift=-0.3cm,xshift=-1.4cm},
       ]

    \node[inner sep=0.5pt, anchor = south west] (p1_1) at (0,0) {\includegraphics[ width=0.11\textwidth]{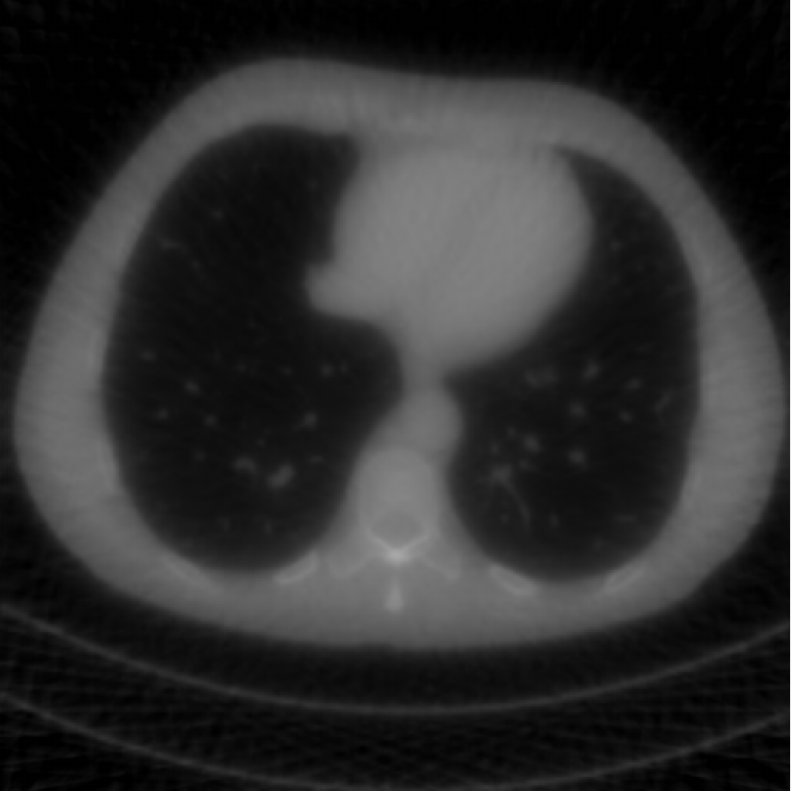}};
    
    \node at (26,26.5) {\color{white} $10$};
    
    \node at (5,3) {\color{white} w/o};
        \node at (24,3) {\color{white} $23.8$};
    
    \node[inner sep=0.5pt, anchor = west] (p1_2) at (p1_1.east) {\includegraphics[ width=0.11\textwidth]{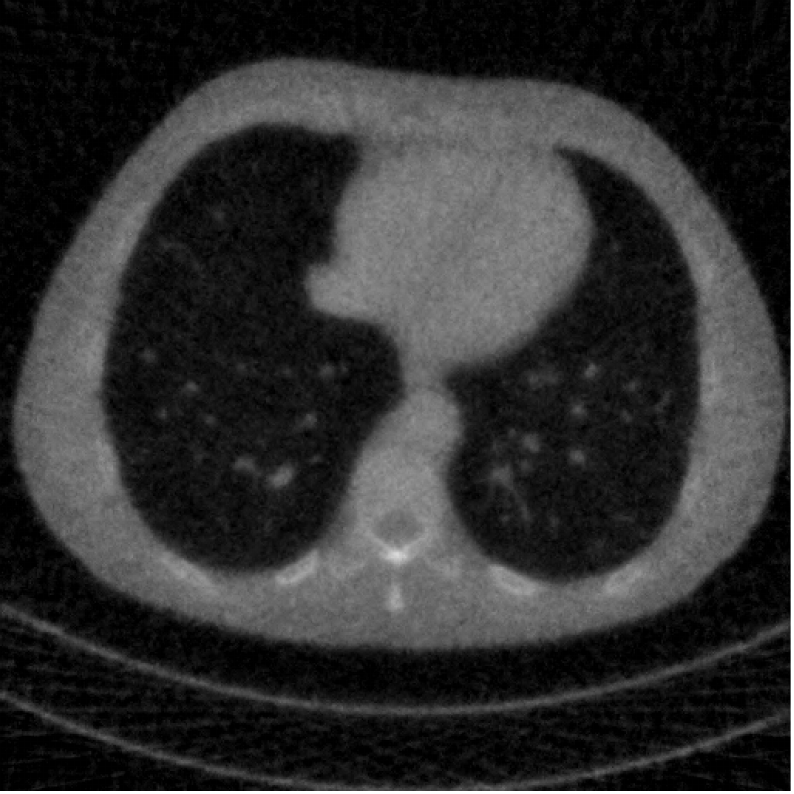}};

    
     \node at (54,3) {\color{white} $26.1$};
    \node at (33,3) {\color{white} w};
    
    \node[inner sep=0.5pt, anchor = west] (p1_3) at (p1_2.east) {\includegraphics[ width=0.11\textwidth]{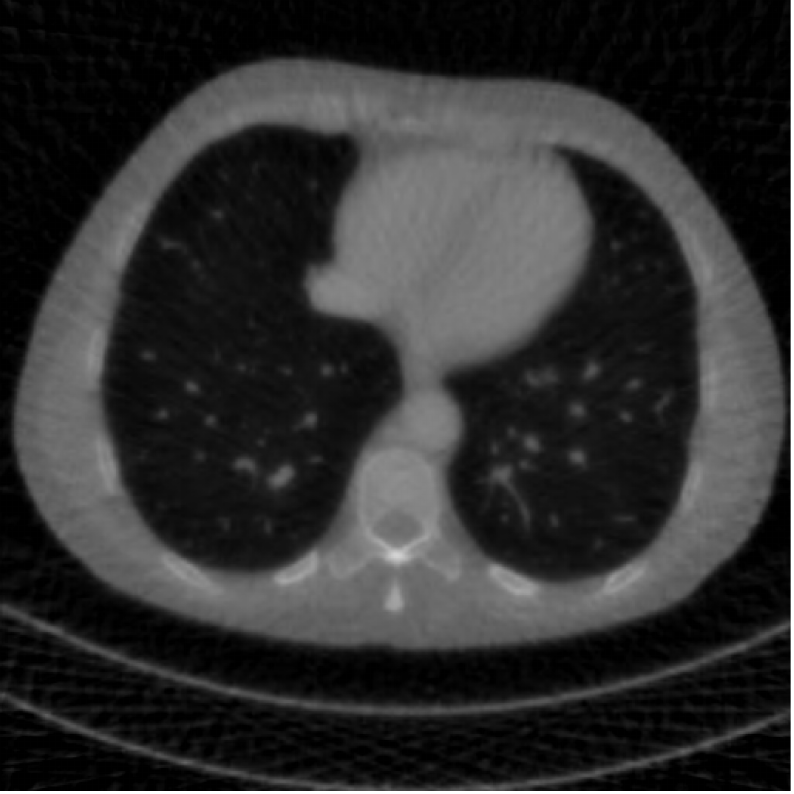}};

    \node at (85,26.5) {\color{white} $20$};
    
    \node at (64,3) {\color{white} w/o};
    
     \node at (83.5,3) {\color{white} $27.6$};
   
 \node[inner sep=0.5pt, anchor = west] (p1_4) at (p1_3.east) {\includegraphics[ width=0.11\textwidth]{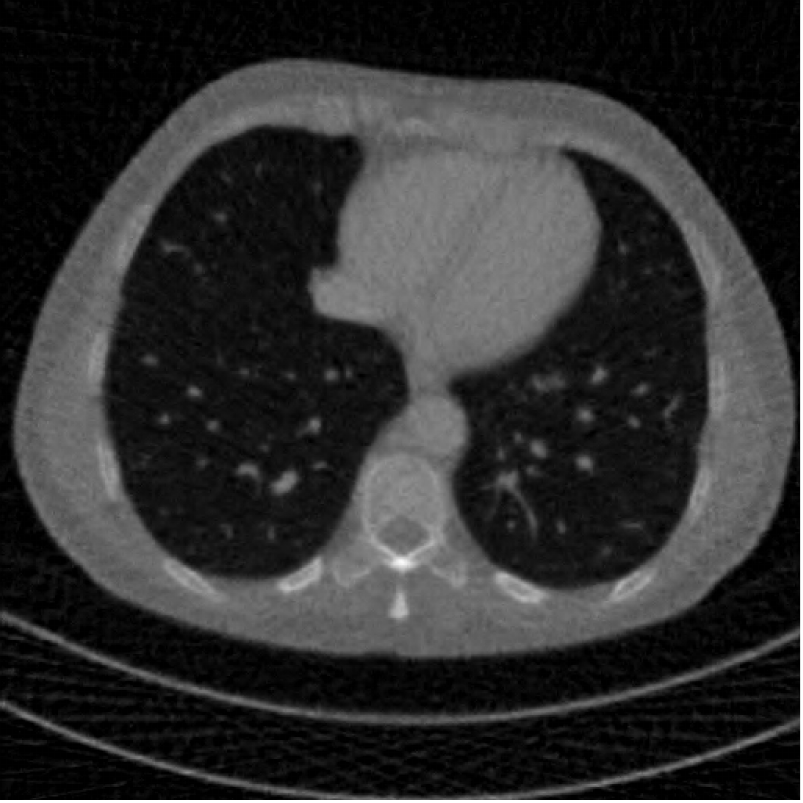}};

  \node at (113,3) {\color{white} $29.7$};
    
  \node at (92,3) {\color{white} w};

 \node[inner sep=0.5pt, anchor = west] (p1_5) at (p1_4.east) {\includegraphics[ width=0.11\textwidth]{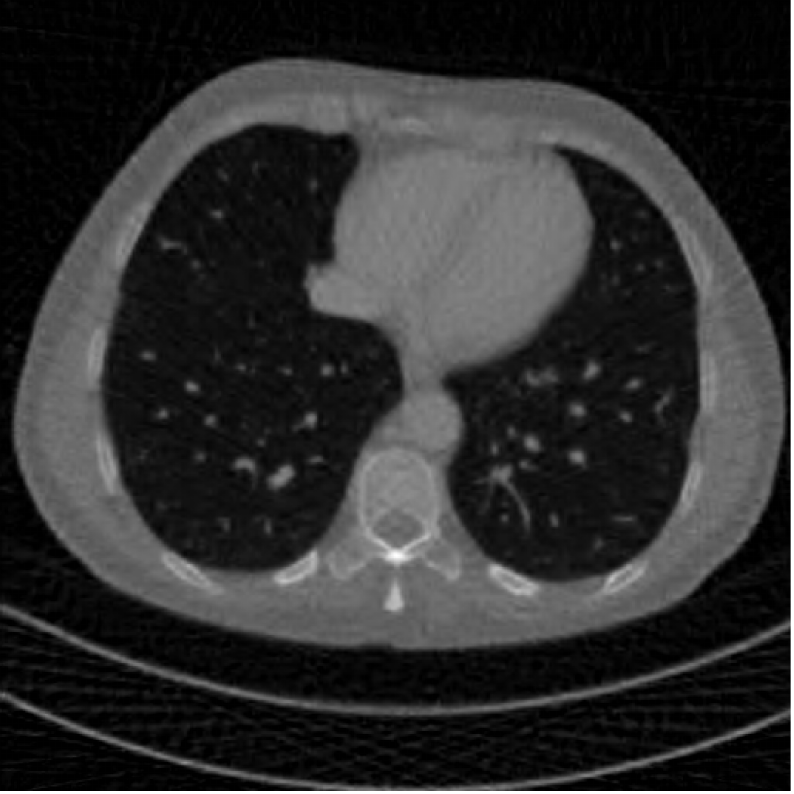}};

    \node at (145,26.5) {\color{white} $40$};
    
    \node at (123,3) {\color{white} w/o};
    
    \node at (143,3) {\color{white} $30.8$};
    
 \node[inner sep=0.5pt, anchor = west] (p1_6) at (p1_5.east) {\includegraphics[ width=0.11\textwidth]{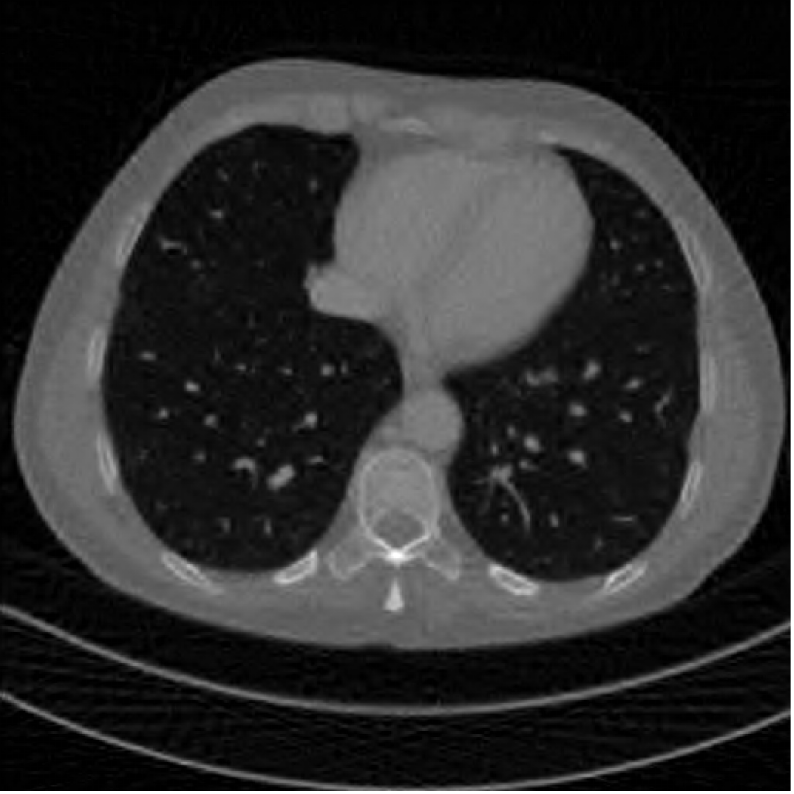}};

    \node at (151,3) {\color{white} w};
    
    \node at (172.5,3) {\color{white} $32.1$};
    
\node[inner sep=0.5pt, anchor = west] (p1_7) at (p1_6.east) {\includegraphics[ width=0.11\textwidth]{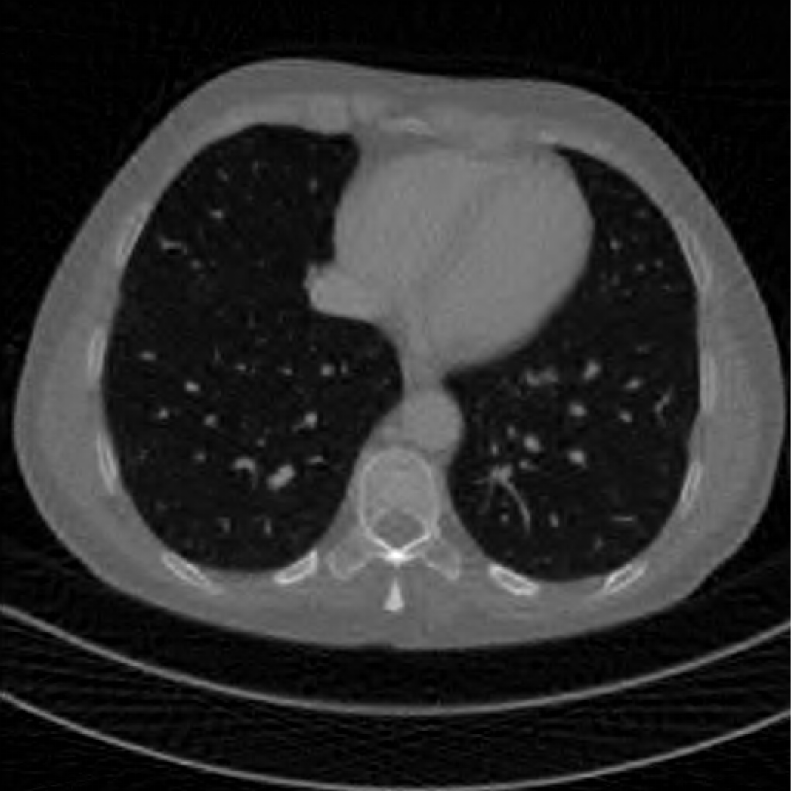}};

    \node at (204,26.5) {\color{white} $60$};
    
    \node at (183,3) {\color{white} w/o};
    
    \node at (201.5,3) {\color{white} $32.0$};

\node[inner sep=0.5pt, anchor = west] (p1_8) at (p1_7.east) {\includegraphics[ width=0.11\textwidth]{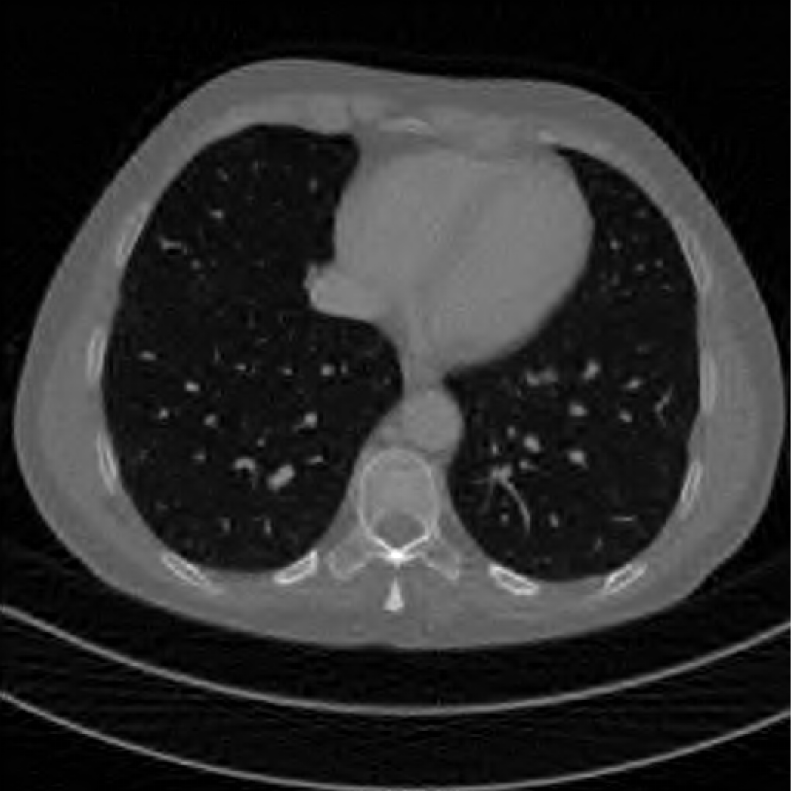}};

\node at (210,3) {\color{white} w};
    
\node at (231.5,3) {\color{white} $33.0$};

\end{axis}

 \begin{axis}[at={(p1_1.south west)},anchor = north west,
     xmin = 0,xmax = 250,ymin = 0,ymax = 70, width=0.95\textwidth,ylabel = TV,
         scale only axis,
         enlargelimits=false,
         yshift=2.8cm,
        axis line style={draw=none},
        tick style={draw=none},
         axis equal image,
         xticklabels={,,},yticklabels={,,},
         ylabel style={yshift=-0.3cm,xshift=-1.4cm},
        ]
        
   \node[inner sep=0.5pt, anchor = south west] (TVp1_1) at (0,0) {\includegraphics[width=0.11\textwidth]{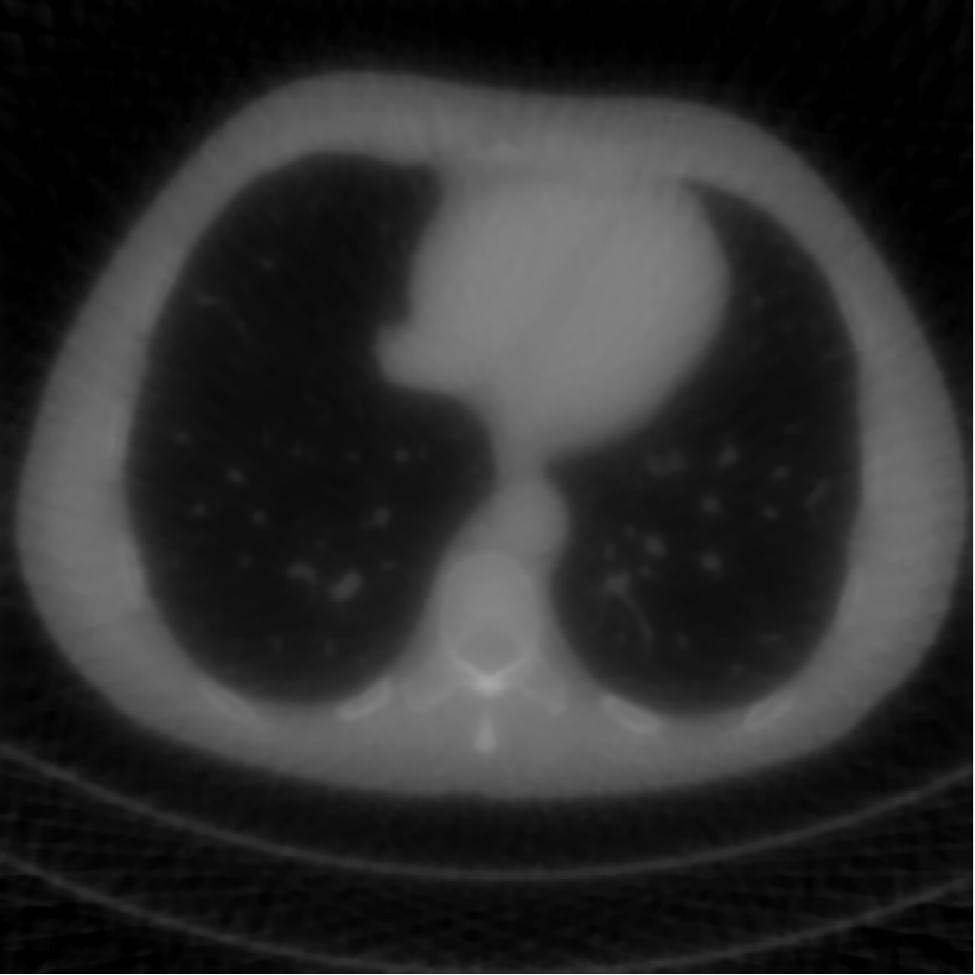}};
   
    \node at (24,3) {\color{white} $23.8$};
    
    \node[inner sep=0.5pt, anchor = west] (TVp1_2) at (TVp1_1.east) {\includegraphics[ width=0.11\textwidth]{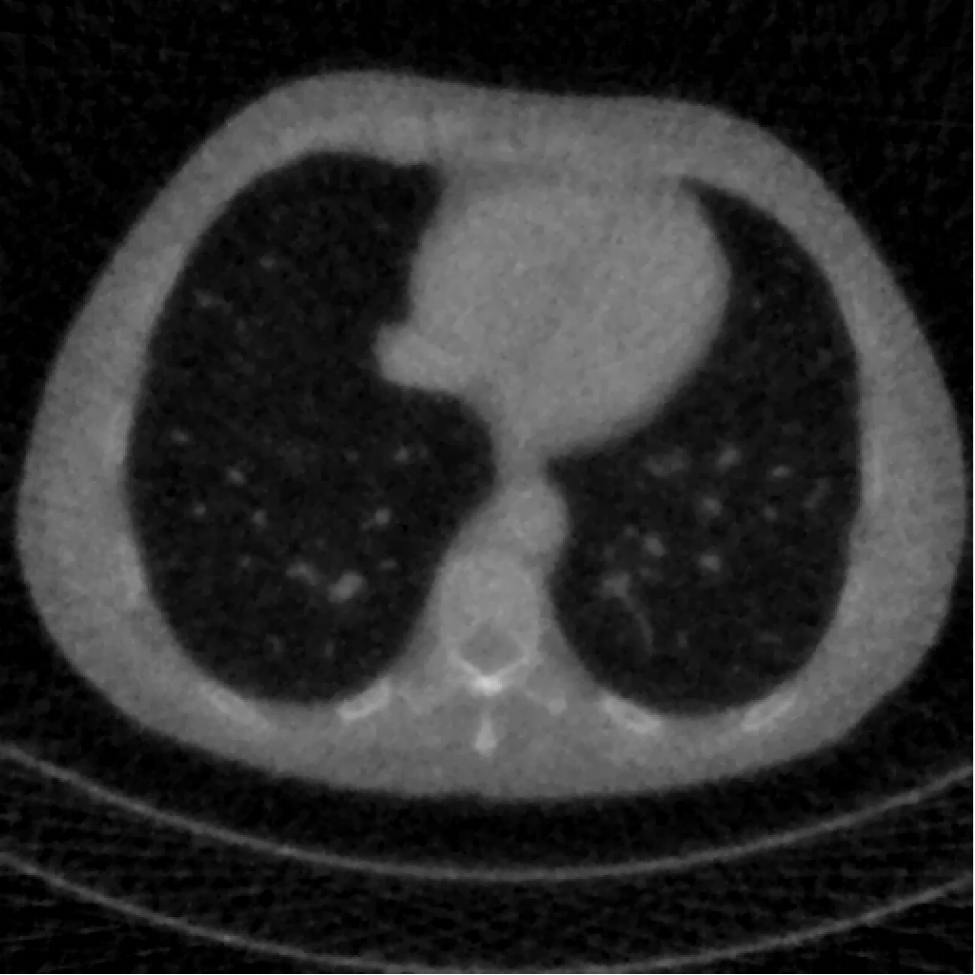}};
 
      \node at (54,3) {\color{white} $26.5$};
  
    \node[inner sep=0.5pt, anchor = west] (TVp1_3) at (TVp1_2.east) {\includegraphics[ width=0.11\textwidth]{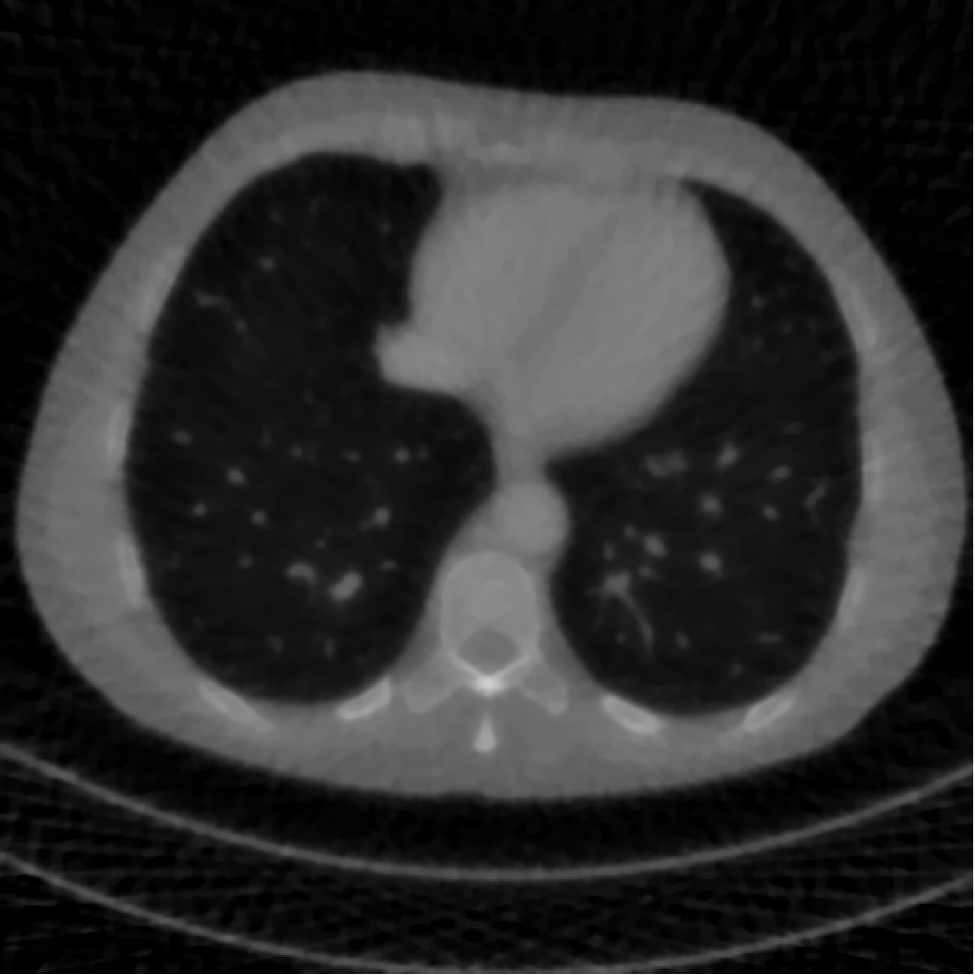}};

        \node at (83.5,3) {\color{white} $27.8$};
   
 \node[inner sep=0.5pt, anchor = west] (TVp1_4) at (TVp1_3.east) {\includegraphics[ width=0.11\textwidth]{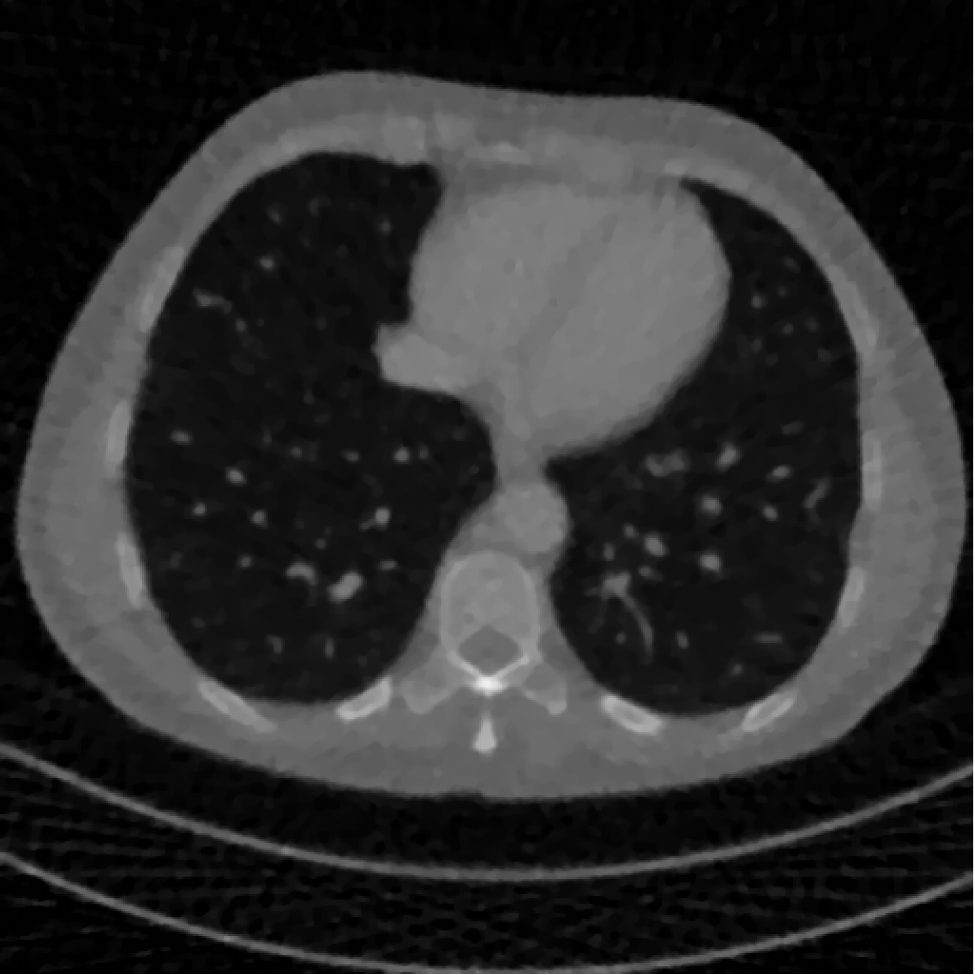}};

  \node at (113,3) {\color{white} $30.6$};
    
 \node[inner sep=0.5pt, anchor = west] (TVp1_5) at (TVp1_4.east) {\includegraphics[ width=0.11\textwidth]{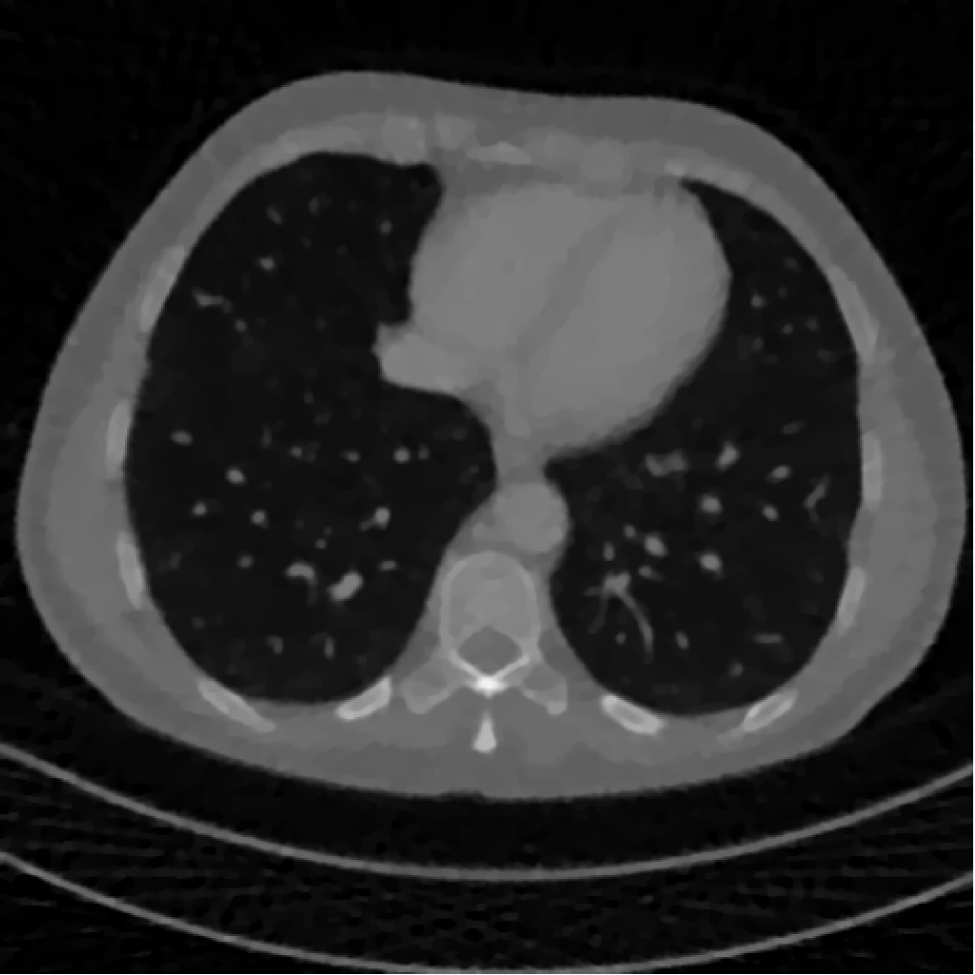}};

    \node at (143,3) {\color{white} $32.1$};
    
 \node[inner sep=0.5pt, anchor = west] (TVp1_6) at (TVp1_5.east) {\includegraphics[ width=0.11\textwidth]{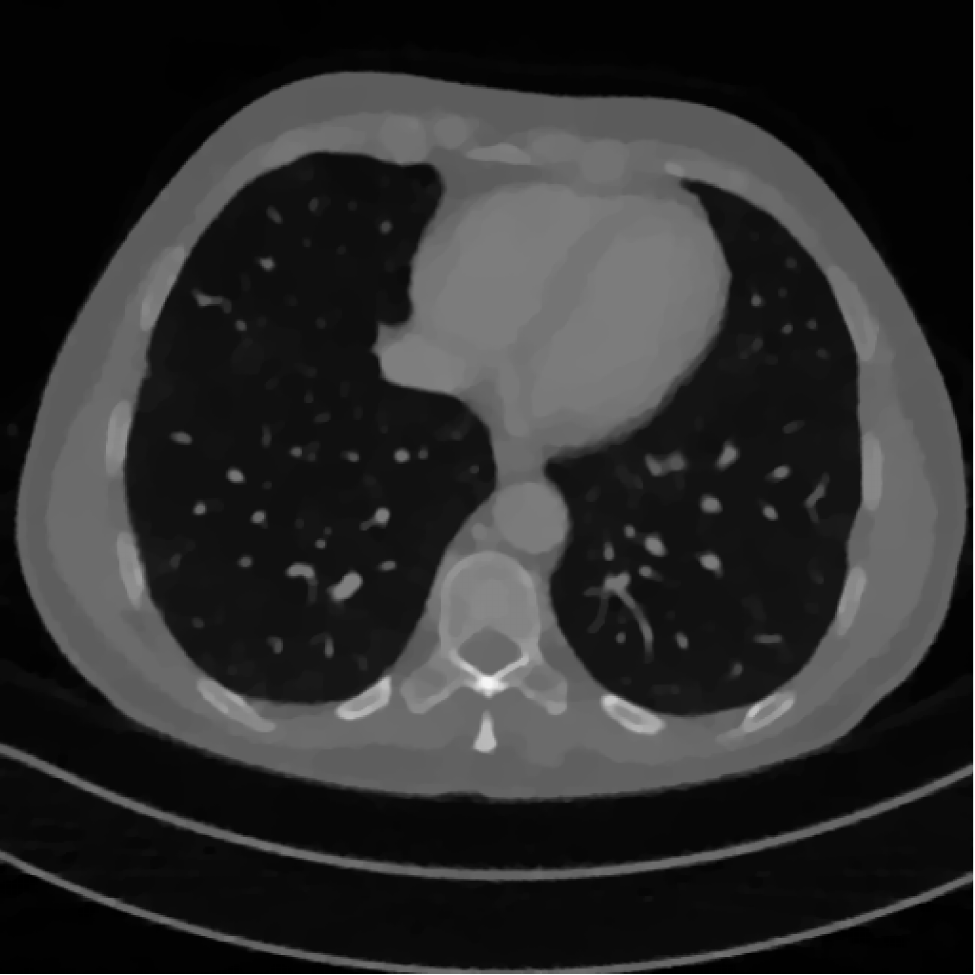}};

    \node at (172.5,3) {\color{white} $34.3$};
    
\node[inner sep=0.5pt, anchor = west] (TVp1_7) at (TVp1_6.east) {\includegraphics[ width=0.11\textwidth]{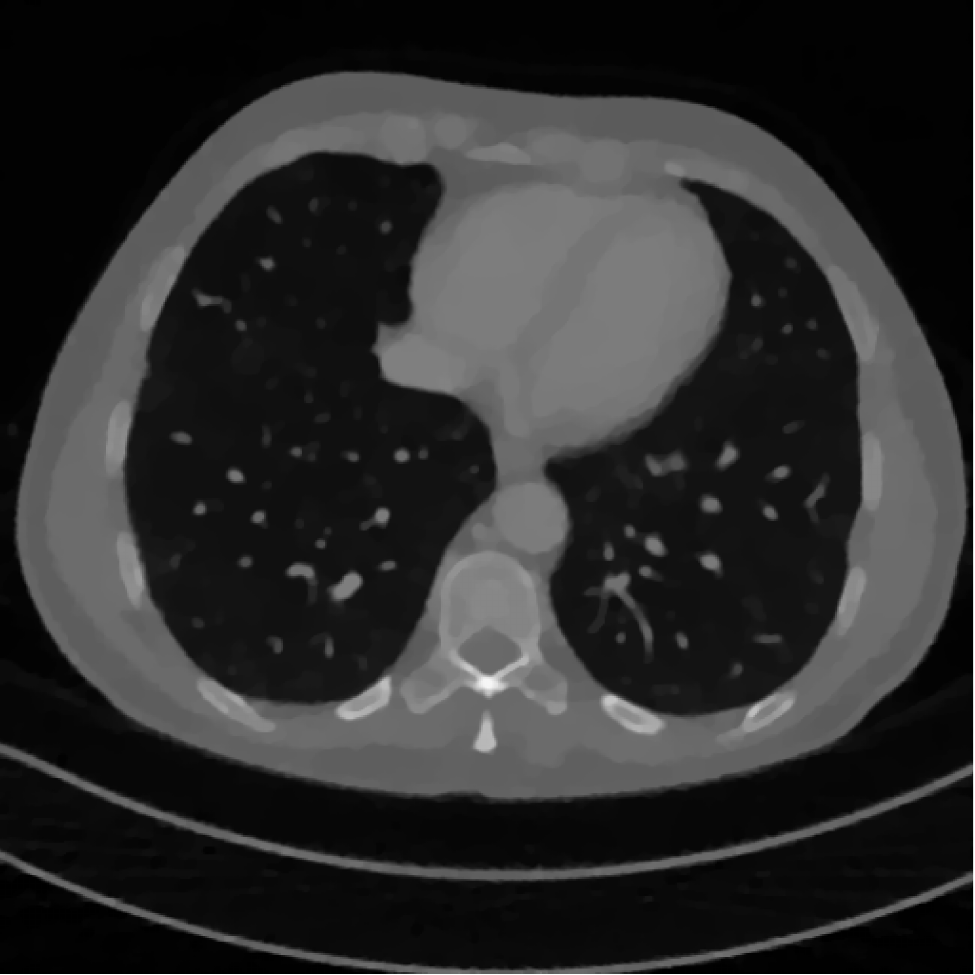}};

    \node at (200.5,3) {\color{white} $34.1$};

\node[inner sep=0.5pt, anchor = west] (TVp1_8) at (TVp1_7.east) {\includegraphics[ width=0.11\textwidth]{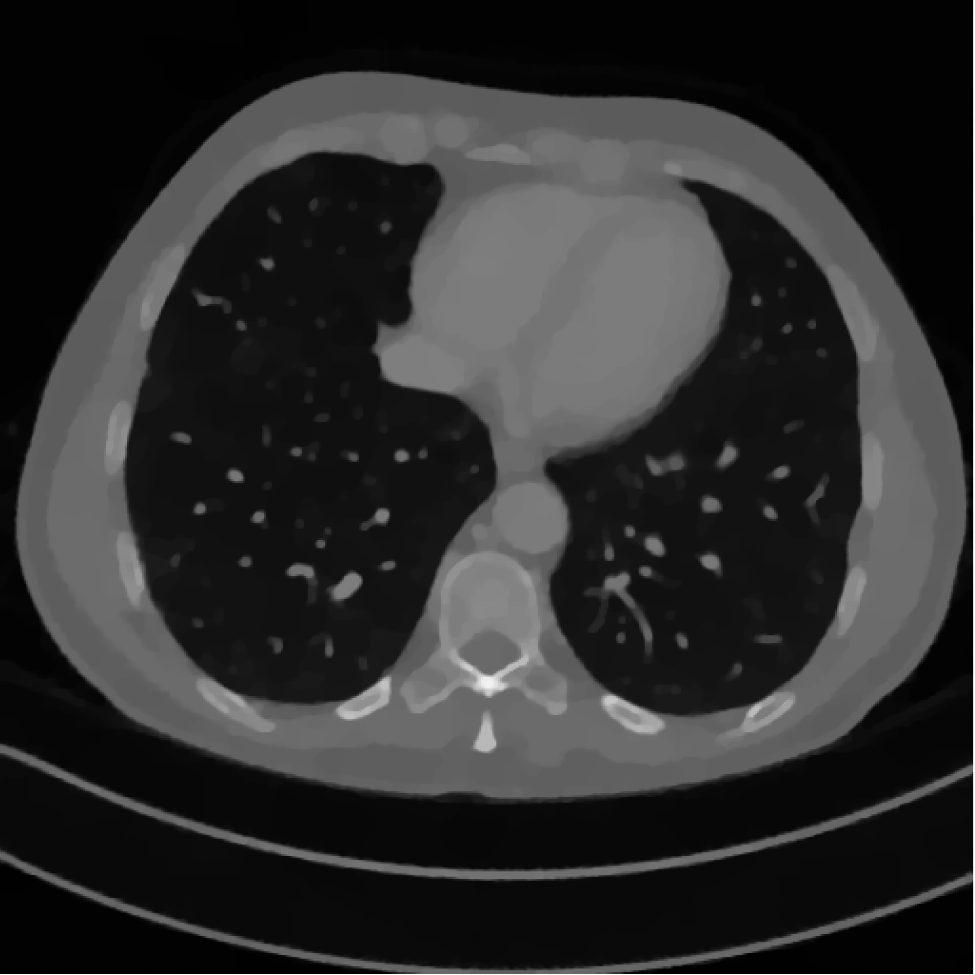}};

\node at (231.5,3) {\color{white} $34.6$};
 \end{axis}

 \begin{axis}[at={(TVp1_1.south west)},anchor = north west,
     xmin = 0,xmax = 250,ymin = 0,ymax = 70, width=0.95\textwidth,ylabel = HS$_1$,
         scale only axis,
         enlargelimits=false,
         yshift=2.8cm,
        axis line style={draw=none},
        tick style={draw=none},
         axis equal image,
         xticklabels={,,},yticklabels={,,},
         ylabel style={yshift=-0.3cm,xshift=-1.4cm},
        ]
        
   \node[inner sep=0.5pt, anchor = south west] (HSp1_1) at (0,0) {\includegraphics[width=0.11\textwidth]{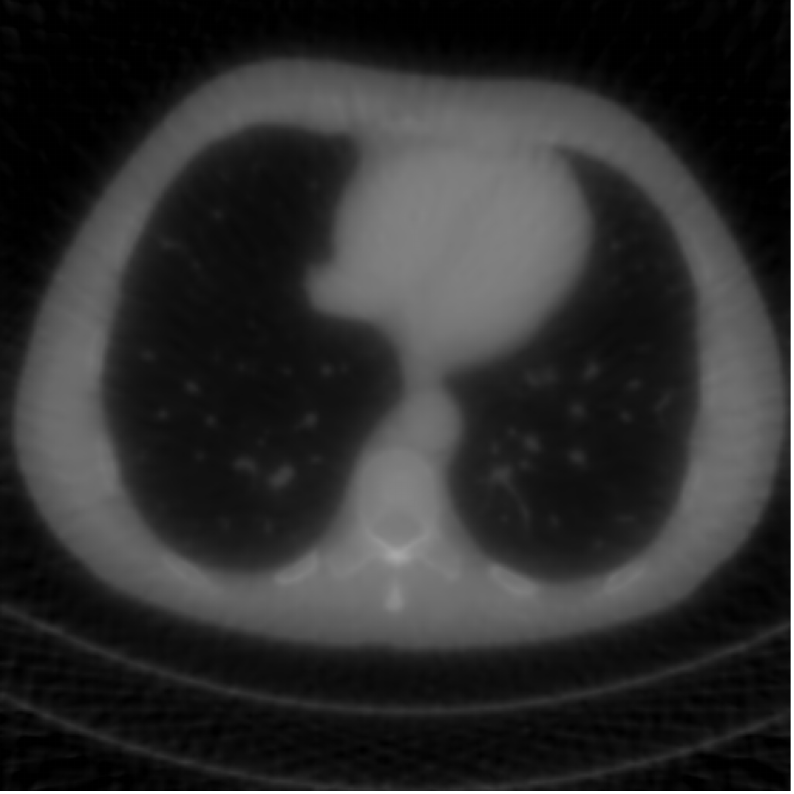}};

    \node at (23,3) {\color{white} $23.8$};
    
    \node[inner sep=0.5pt, anchor = west] (HSp1_2) at (HSp1_1.east) {\includegraphics[ width=0.11\textwidth]{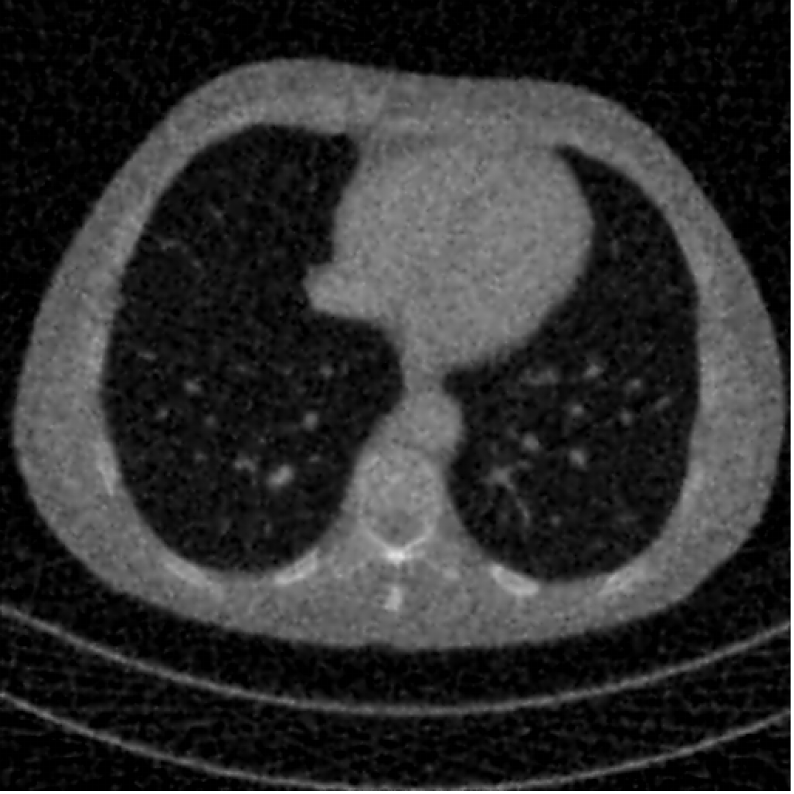}};
 
    \node at (53,3) {\color{white} $27.0$};
    
    \node[inner sep=0.5pt, anchor = west] (HSp1_3) at (HSp1_2.east) {\includegraphics[ width=0.11\textwidth]{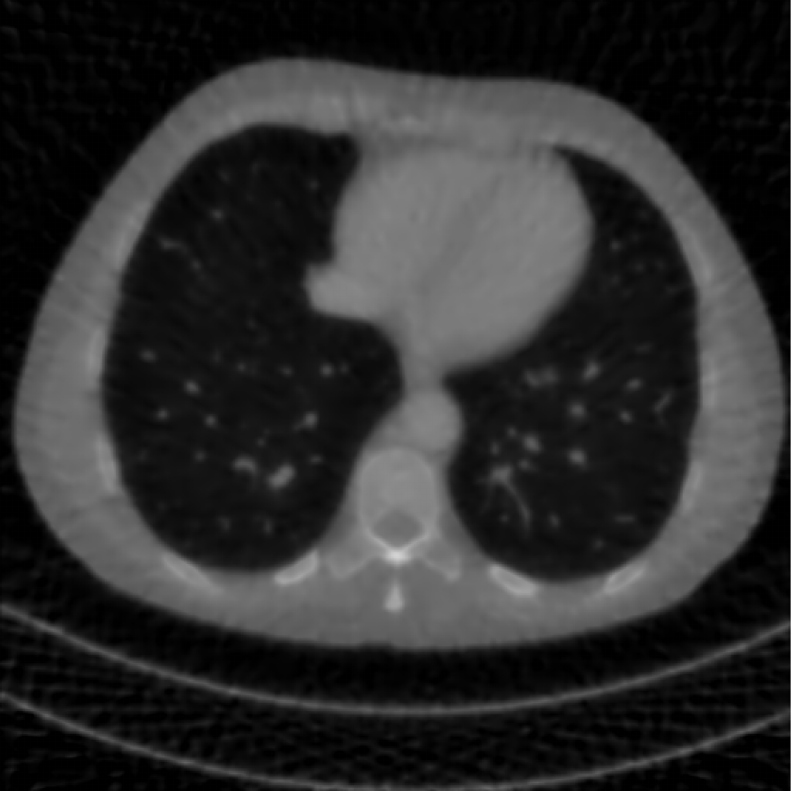}};

    \node at (82.5,3) {\color{white} $27.6$};
 \node[inner sep=0.5pt, anchor = west] (HSp1_4) at (HSp1_3.east) {\includegraphics[ width=0.11\textwidth]{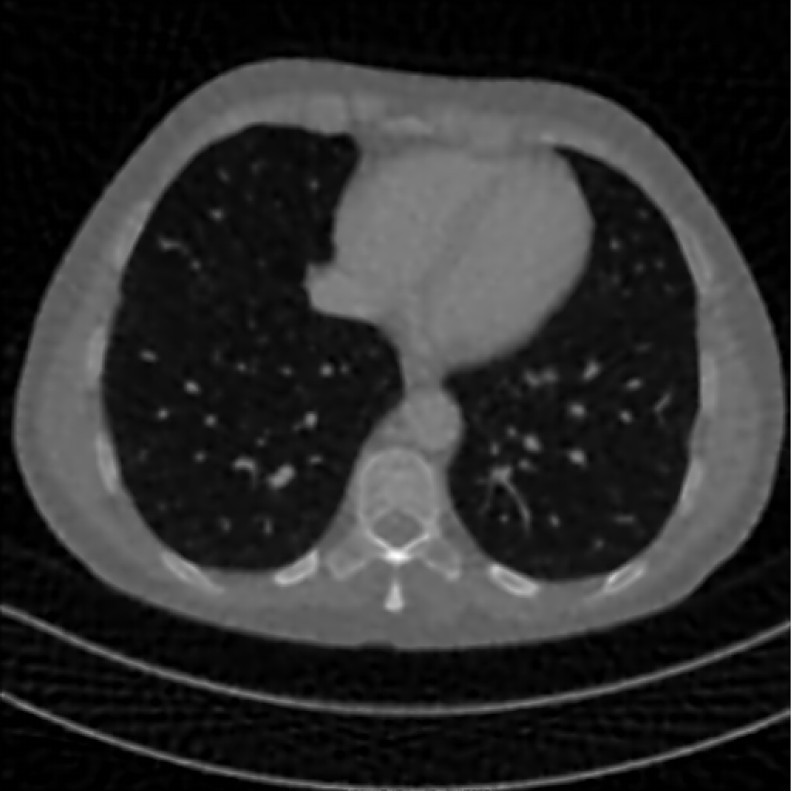}};

  \node at (113,3) {\color{white} $31.5$};
    
    
 \node[inner sep=0.5pt, anchor = west] (HSp1_5) at (HSp1_4.east) {\includegraphics[ width=0.11\textwidth]{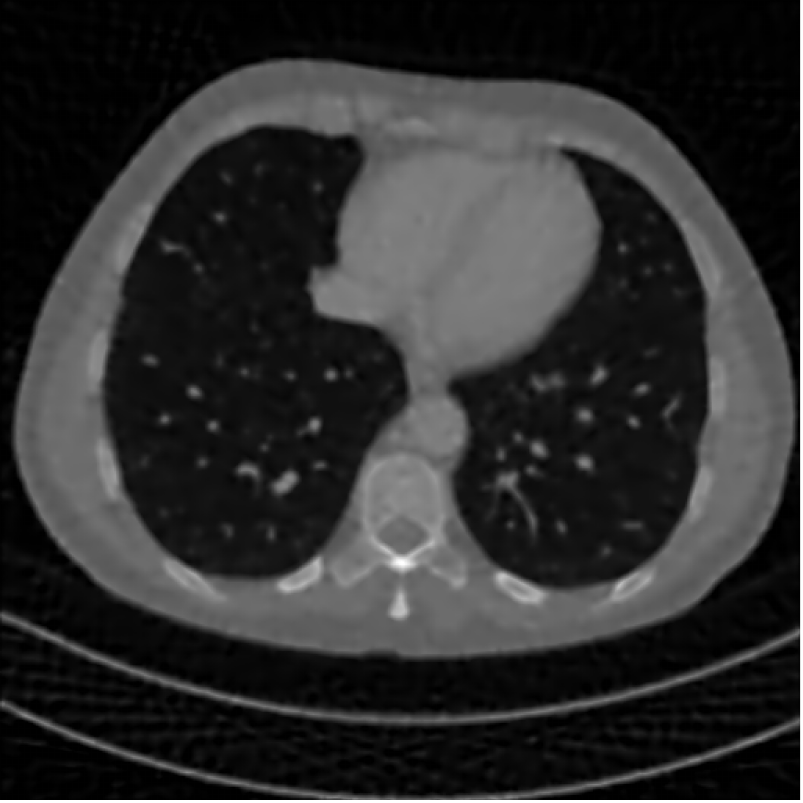}};

    \node at (143,3) {\color{white} $31.4$};
    
 \node[inner sep=0.5pt, anchor = west] (HSp1_6) at (HSp1_5.east) {\includegraphics[ width=0.11\textwidth]{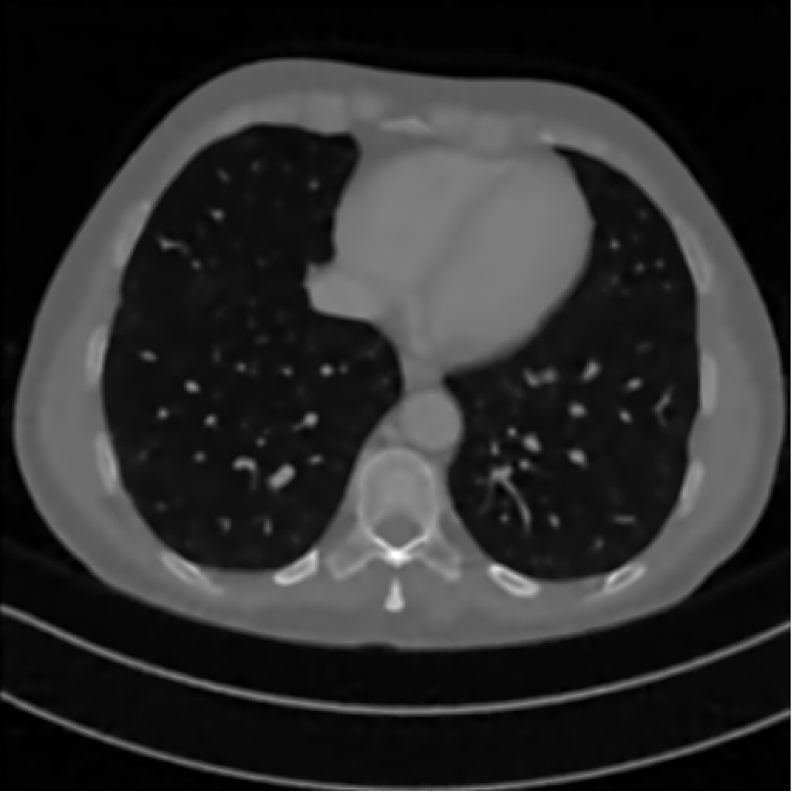}};

    \node at (172.5,3) {\color{white} $33.8$};
    
\node[inner sep=0.5pt, anchor = west] (HSp1_7) at (HSp1_6.east) {\includegraphics[ width=0.11\textwidth]{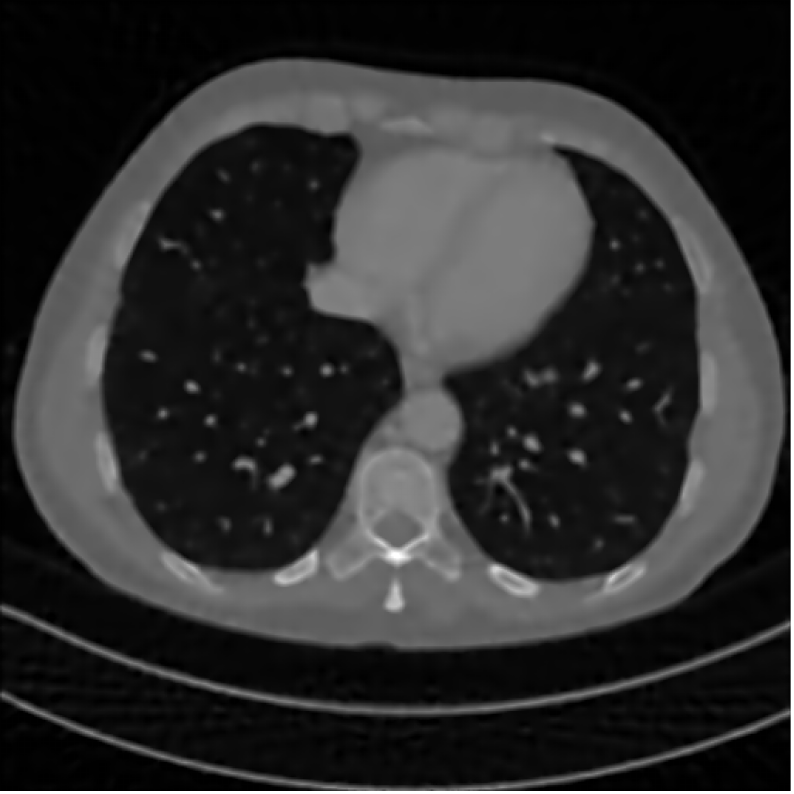}};

\node at (201.5,3) {\color{white} $33.1$};

\node[inner sep=0.5pt, anchor = west] (HSp1_8) at (HSp1_7.east) {\includegraphics[width=0.11\textwidth]{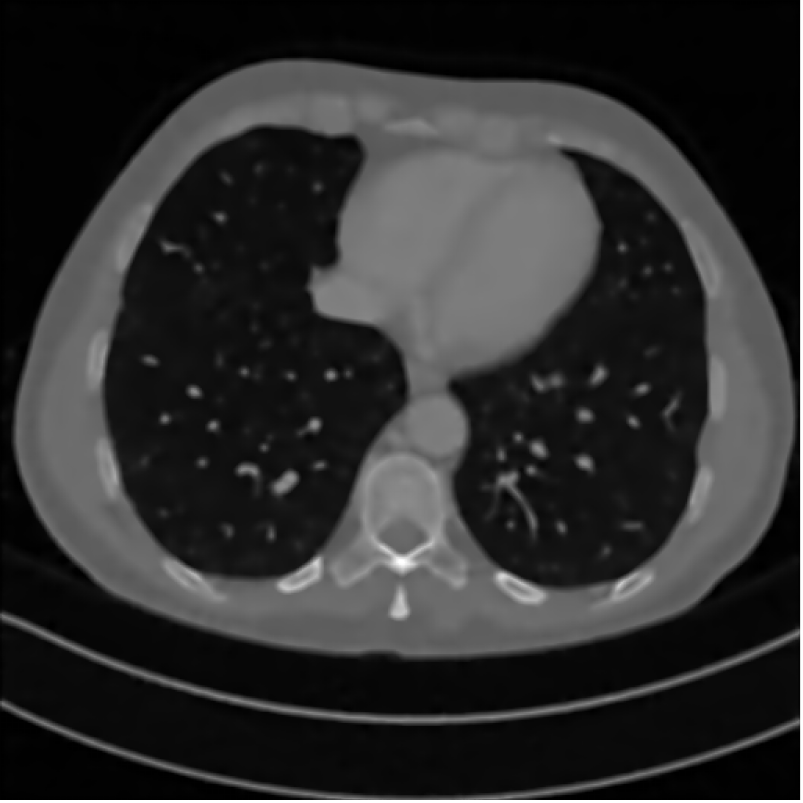}};

\node at (231.5,3) {\color{white} $33.9$};

 \end{axis}
 
\end{tikzpicture} 
\caption{The fan-beam CT reconstructed images with wavelet, TV, and HS$_1$ based regularization
at iterations $10$, $20$, $40$, and $60$.
Columns $1$, $3$, $5$, and $7$ (respectively, $2$, $4$, $6$, and $8$)
show the reconstructions without (respectively, with) \RNP.
The associated PSNR values are listed at the right corner of each image.
}
\label{fig:CTFan:RecoImages:a}
\end{figure*}

\begin{figure}
    \centering
    \subfigure[]{
\includegraphics[width=0.42\linewidth]{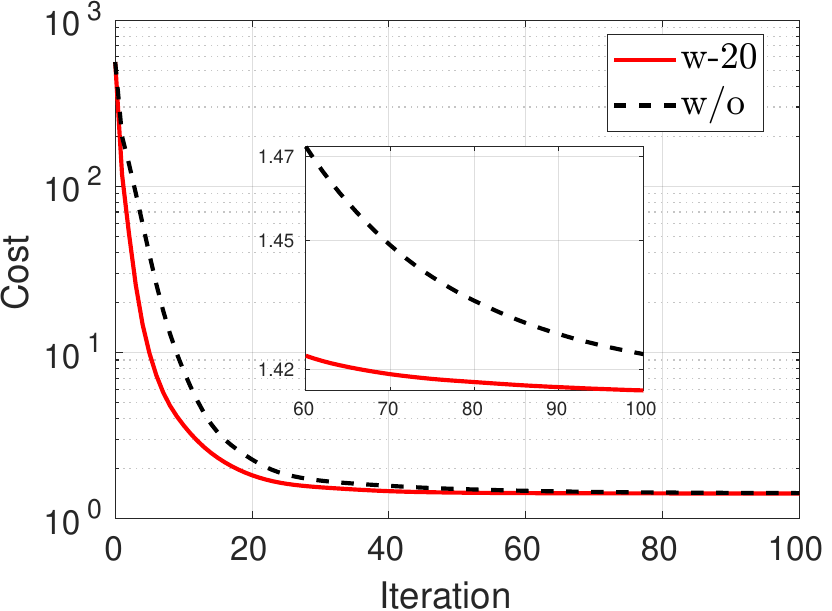}}
\subfigure[]{
 \includegraphics[width=0.42\linewidth]{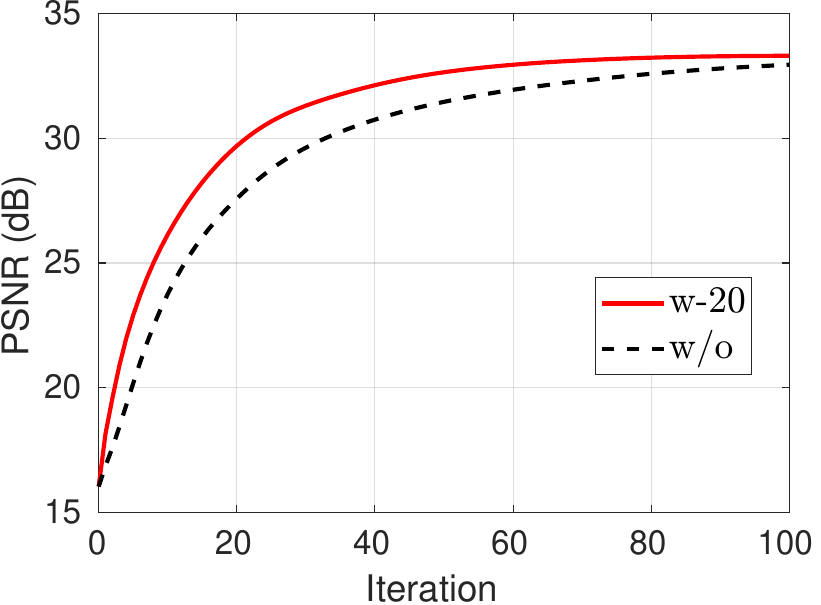}}    
%
    \caption{Comparison of using \RNP for wavelet based parallel-beam CT reconstruction for Fig. S.\ref{fig:CT:GT:a}. w/o denotes the one without using \RNP.}
    \label{fig:supp:fan:im:a:wav}
\end{figure}

\begin{figure}
    \centering

    \subfigure[]{
\includegraphics[width=0.43\linewidth]{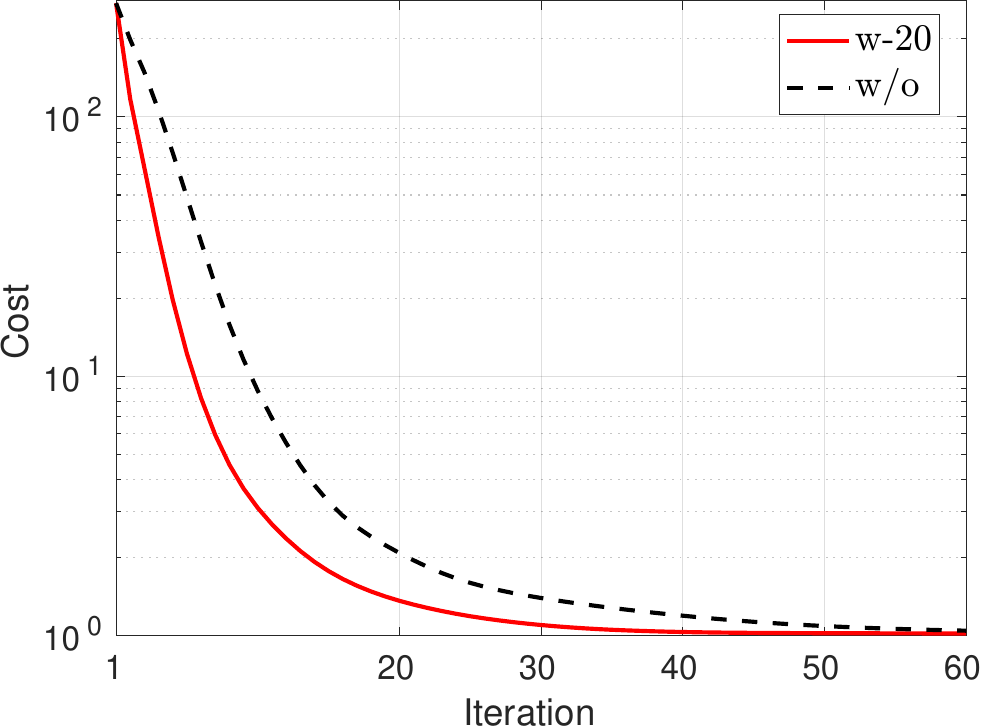}}
    \subfigure[]{
 \includegraphics[width=0.42\linewidth]{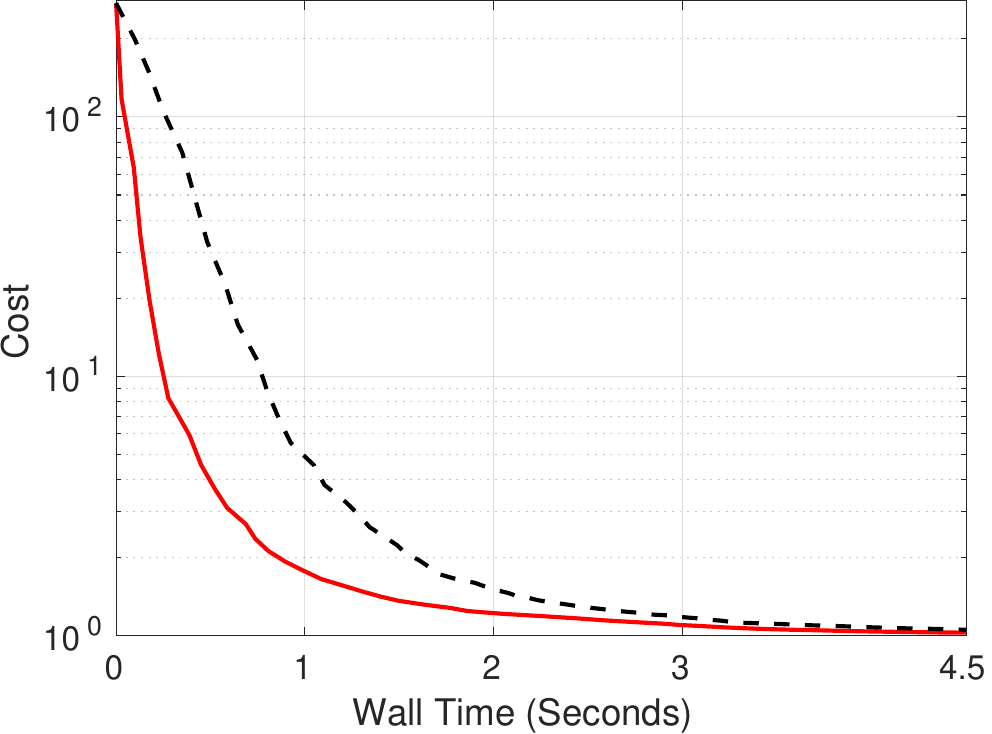}}

    \subfigure[]{
 \includegraphics[width=0.42\linewidth]{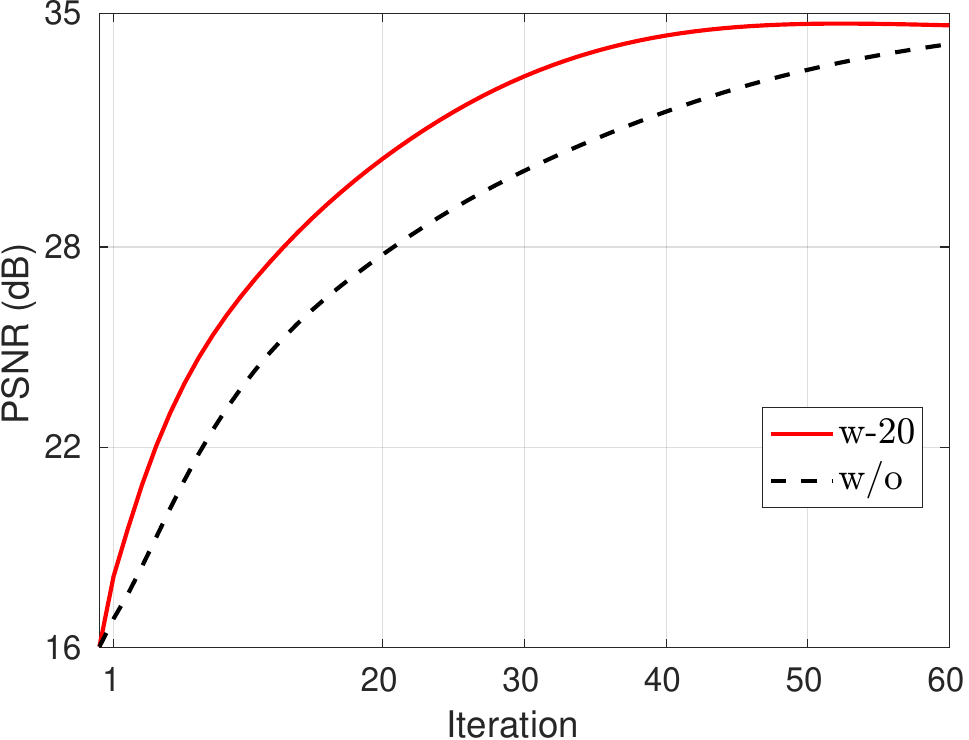}}
    \subfigure[]{
 \includegraphics[width=0.41\linewidth]{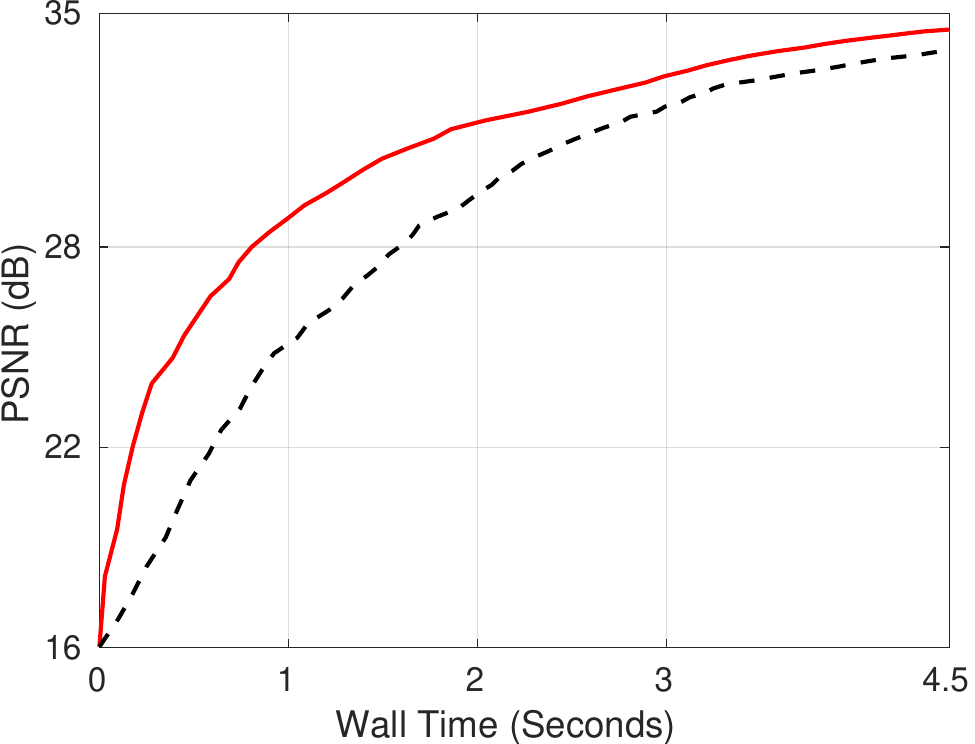}}
    \caption{Comparison of using \RNP for total variation based parallel-beam CT reconstruction for Fig. S.\ref{fig:CT:GT:a}. w/o denotes the one without using \RNP.}
    \label{fig:supp:fan:im:a:TV}
\end{figure}

\begin{figure}
    \centering
    
    \subfigure[]{
\includegraphics[width=0.42\linewidth]{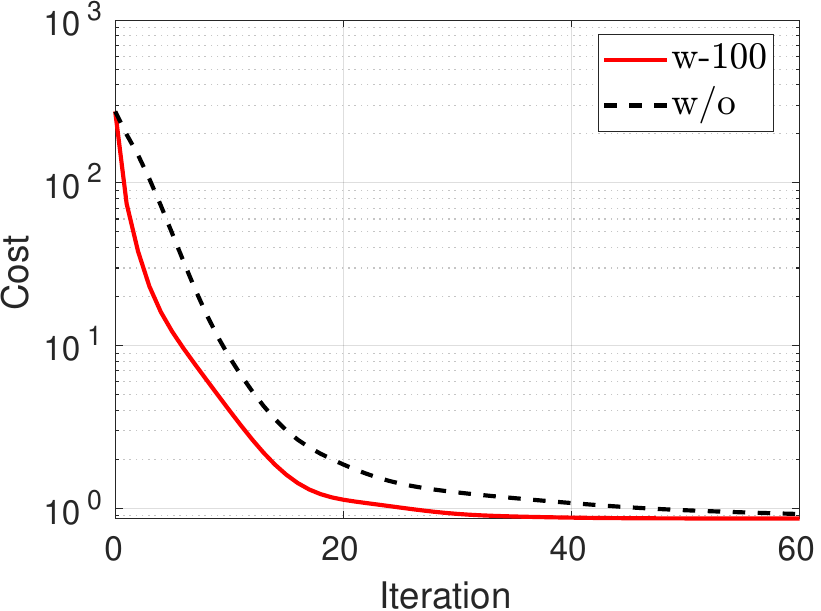}}
    \subfigure[]{
 \includegraphics[width=0.41\linewidth]{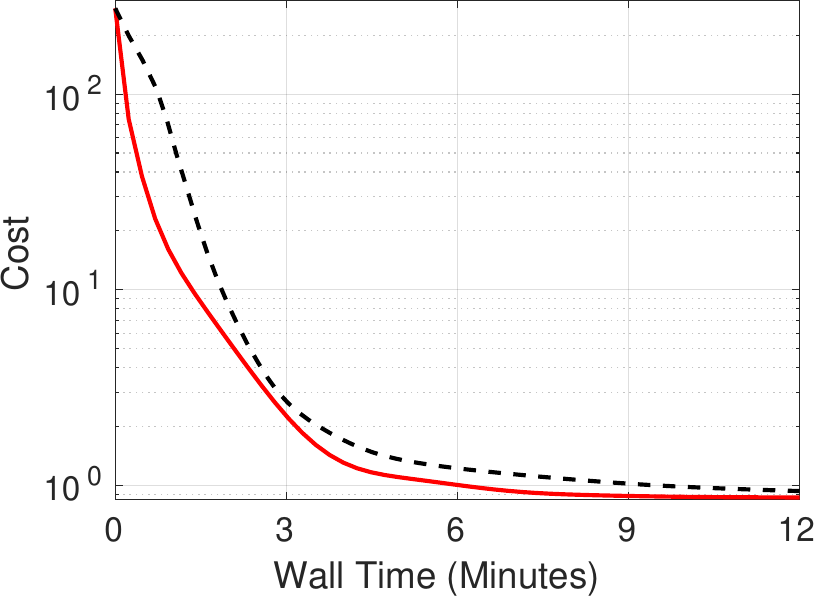}}

    \subfigure[]{
 \includegraphics[width=0.42\linewidth]{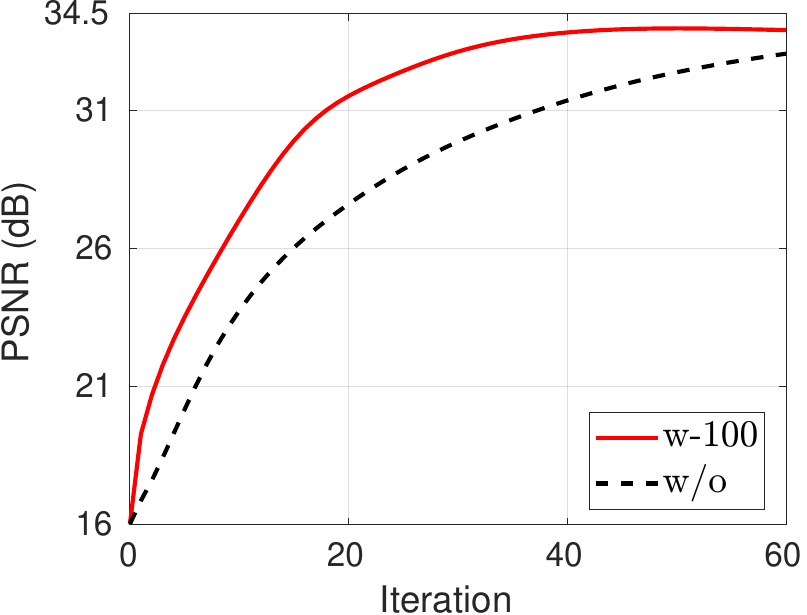}}
    \subfigure[]{
 \includegraphics[width=0.41\linewidth]{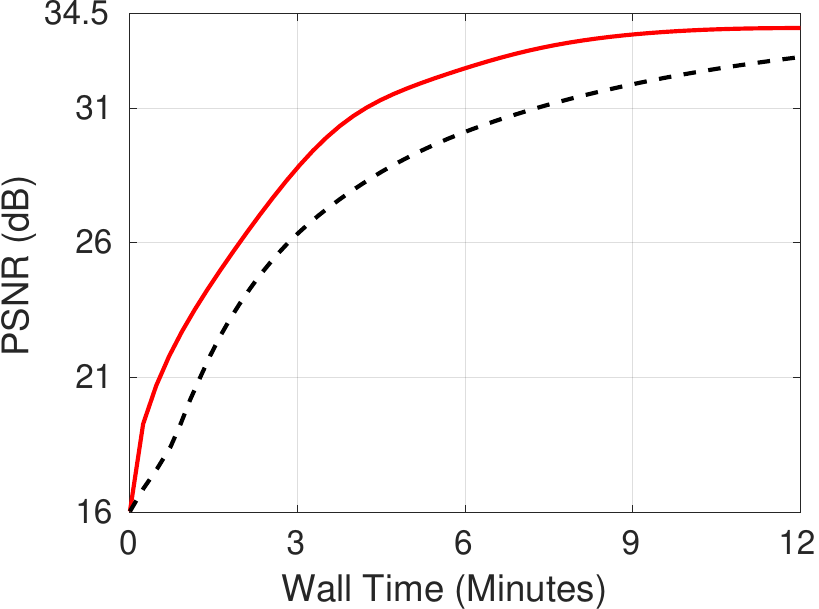}}
    \caption{Comparison of using \RNP for HS$_1$ based parallel-beam CT reconstruction for Fig. S.\ref{fig:CT:GT:a}. w/o denotes the one without using \RNP.}
    \label{fig:supp:fan:im:a:HS1}
\end{figure}

\Cref{fig:supp:fan:im:a:wav,fig:supp:fan:im:a:TV,fig:supp:fan:im:a:HS1} present the cost and PSNR values versus iteration for wavelet, total variation, and HS$_1$ based fan-beam CT reconstruction for Fig. S.\ref{fig:CT:GT:a}. Still, we observed that using \RNP accelerated the convergence in terms of iterations. \Cref{fig:CTFan:RecoImages:a} presents the reconstructed images at different iterations, where we observed that using \RNP yielded much clearer images with fewer iterations.

\subsection{\texorpdfstring{HS$_2$ and HS$_\infty$}{} Based Reconstruction}
\begin{figure}
    \centering
    \subfigure[HS$_2$]{
    \includegraphics[width=0.42\linewidth]{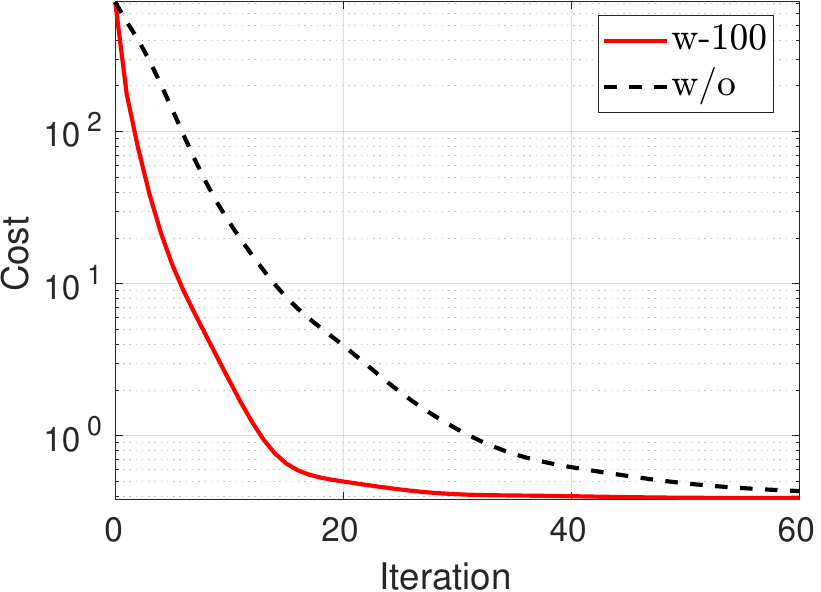}}
    \subfigure[HS$_2$]{
    \includegraphics[width=0.42\linewidth]{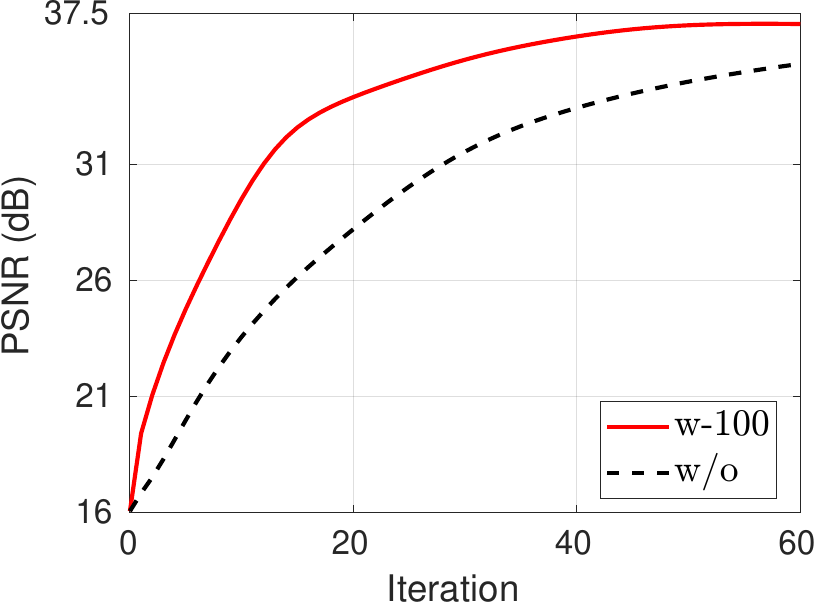}}

    \subfigure[HS$_\infty$]{
    \includegraphics[width=0.42\linewidth]{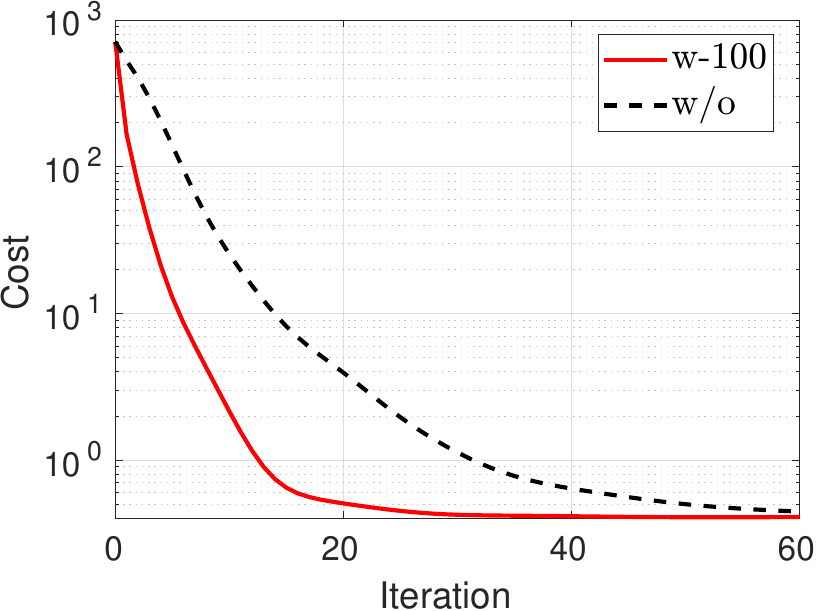}}
    \subfigure[HS$_\infty$]{
    \includegraphics[width=0.42\linewidth]{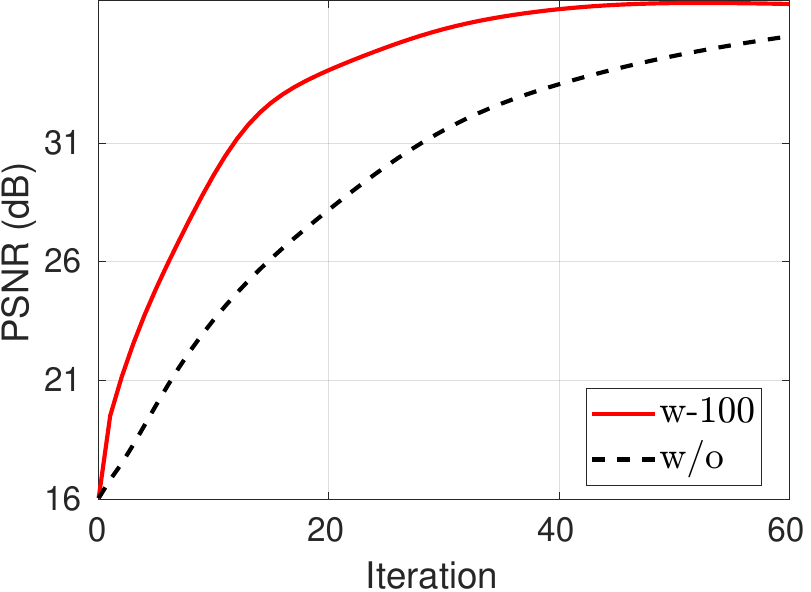}}
    \caption{Comparison of using \RNP for HS$_2$ and H$_\infty$ based parallel-beam CT reconstruction for Fig. S.\ref{fig:CT:GT:a}.}
    \label{fig:supp:parallel:im:a:HS2infty}
\end{figure}

\Cref{fig:supp:parallel:im:a:HS2infty} presents the results of using HS$_2$ and HS$_\infty$ based reconstruciton that we also saw same thrends as before demonstraing the effectiveness of using \RNP to accelerate APG for CT reconstruction.

